\documentclass[11pt,english]{article}
\usepackage[T1]{fontenc}
\usepackage[latin9]{inputenc}
\usepackage{color}
\usepackage{amsmath}
\usepackage{amssymb}
\usepackage{geometry}
\makeatletter
\usepackage{amsmath}
\usepackage{amsthm}
\usepackage{amssymb}

\theoremstyle{plain}
\newtheorem{theorem}{Theorem}[section]

\theoremstyle{plain}
\newtheorem{theorem-seq}{Theorem}

\newenvironment{myproof}[1][Proof.]{\par
	\pushQED{\qed}%
	\normalfont \topsep6\p@\@plus6\p@\relax
	\trivlist
		\item[\hskip\labelsep
		\bfseries
	#1]\ignorespaces
	}{%
	\popQED\endtrivlist\@endpefalse
}


\newcommand{\eqdef}{\stackrel{\rm{def}}{=}}

\newcommand{\N}{\mathbb{N}}

\newcommand{\R}{\mathbb{R}}
\newcommand{\F}{\mathbb{F}}

\ifx \C \undefined
\newcommand{\C}{\mathbb{C}}
\else
\renewcommand{\C}{\mathbb{C}}
\fi

\newcommand{\B}{\left\{ 0,1 \right \}}

\renewcommand{\P}{\mathbf{P}}

\renewcommand{\L}{\mathbf{L}}

\newcommand{\NC}{\mathbf{NC}}
\newcommand{\NCo}{\NC^1}

\usepackage{crossreftools}
\usepackage{cleveref}
\usepackage{aliascnt}
\crefalias{theorem-seq}{theorem}

\let \ref \Cref
\crefname{enumi}{}{}
\crefname{enumii}{}{}
\crefname{enumiii}{}{}

\newaliascnt{conjecture-cnt}{theorem}
    \theoremstyle{plain}    
    \newtheorem{conjecture}[conjecture-cnt]{Conjecture} 
\crefname{conjecture-cnt}{conjecture}{conjectures}
\newaliascnt{proposition-cnt}{theorem}
    \theoremstyle{plain}    
    \newtheorem{proposition}[proposition-cnt]{Proposition} 
\crefname{proposition-cnt}{proposition}{propositions}
\newaliascnt{remark-cnt}{theorem}
    \theoremstyle{definition}
    \newtheorem{remark}[remark-cnt]{Remark}
\crefname{remark-cnt}{remark}{remarks}
\newaliascnt{definition-cnt}{theorem}
\theoremstyle{definition}
\newtheorem{definition}[definition-cnt]{Definition}
\crefname{definition-cnt}{definition}{definitions}
\newaliascnt{lemma-cnt}{theorem}
\theoremstyle{plain}    
\newtheorem{lemma}[lemma-cnt]{Lemma}  
	\crefname{lemma-cnt}{lemma}{lemmas}
\newaliascnt{fact-cnt}{theorem}
    \theoremstyle{plain}    
    \newtheorem{fact}[fact-cnt]{Fact}
\crefname{fact-cnt}{fact}{facts}
\newaliascnt{notation-cnt}{theorem}
    \theoremstyle{definition}    
    \newtheorem{notation}[notation-cnt]{Notation}
\crefname{notation-cnt}{notation}{notations}
\newaliascnt{example-cnt}{theorem}
\theoremstyle{definition}
\newtheorem{example}[example-cnt]{Example}
\crefname{example-cnt}{example}{examples}
\theoremstyle{plain}
\newtheorem*{restated@theorem}{\rep@title}

\newenvironment{restated}[1]{%
	 \def\rep@title{#1}%
	 \begin{restated@theorem}}%
	{\end{restated@theorem}}

\theoremstyle{definition}    
\newtheorem*{acknowledgement*}{Acknowledgement} 

\newcommand{\hastad}{{H{\aa}stad}}
\newcommand{\torma}{{T{\"{o}}rm{\"{a}}}}

\makeatother

\usepackage{babel}
\begin{document}
\title{Toward Better Depth Lower Bounds: A KRW-like theorem for Strong Composition\thanks{A preliminary version of this work appeared in FOCS 2023.}}
\author{Or Meir\thanks{Department of Computer Science, University of Haifa, Haifa 3498838,
Israel. \texttt{ormeir@cs.}\texttt{haifa}\texttt{.ac.}\texttt{il}.
Partially supported by the Israel Science Foundation (grant No. 716/20).}}
\maketitle
\begin{abstract}
\global\long\def\NCo{\mathbf{NC}^{1}}%
\global\long\def\cl{\omega}%
\global\long\def\B{\left\{  0,1\right\}  }%
\global\long\def\d{\diamond}%
\global\long\def\c{\circledast}%
\global\long\def\fg{f\d g}%
\global\long\def\D{\mathsf{D}}%
\global\long\def\pnc{\P\not\subseteq\NCo}%
\global\long\def\KW{\textit{KW}}%
\global\long\def\mKW{\mathit{mKW}}%
\global\long\def\MX{\textit{MUX}}%
\global\long\def\C{\mathsf{CC}}%
\global\long\def\Ch{\C^{\mathrm{hd}}}%
\global\long\def\Cp{\C^{\mathrm{phd}}}%
\global\long\def\NC{\mathsf{NCC}}%
\global\long\def\cNC{\mathrm{co}\mathsf{NCC}}%
\global\long\def\L{\mathsf{L}}%
\global\long\def\cI{\mathcal{I}}%
\global\long\def\cZ{\mathcal{Z}}%
\global\long\def\po{\pi_{1}}%
\global\long\def\KdM{\KW_{f}\d\MX}%
\global\long\def\KcM{\KW_{f}\c\MX}%
\global\long\def\KdMn{\KW_{f}\d\MX_{n}}%
\global\long\def\KcM{\KW_{f}\c\MX}%
\global\long\def\KcMn{\KW_{f}\c\MX_{n}}%
\global\long\def\cA{\mathcal{A}}%
\global\long\def\cB{\mathcal{B}}%
\global\long\def\cE{\mathcal{E}}%
\global\long\def\cF{\mathcal{F}}%
\global\long\def\cG{\mathcal{G}}%
\global\long\def\cX{\mathcal{X}}%
\global\long\def\cY{\mathcal{Y}}%
\global\long\def\cU{\mathcal{U}}%
\global\long\def\cW{\mathcal{W}}%
\global\long\def\cV{\mathcal{V}}%
\global\long\def\cS{\mathcal{S}}%
\global\long\def\Xp{\cX_{\po}}%
\global\long\def\Yp{\cY_{\po}}%
\global\long\def\Ap{\cA_{\po}}%
\global\long\def\Bp{\cB_{\po}}%
\global\long\def\Vp{\cV_{\po}}%
\global\long\def\Gp{\cG_{\po}}%
\global\long\def\XI{\cX^{*}}%
\global\long\def\YI{\cY^{*}}%
\global\long\def\Vb{\cV_{0}}%
\global\long\def\pg{\pi_{1,g}}%
\global\long\def\Xpg{\cX_{\pg}}%
\global\long\def\Ypg{\cY_{\pg}}%

\global\long\def\Apg{\cA_{\pg}}%
\global\long\def\Bpg{\cB_{\pg}}%
\global\long\def\rc{\texttt{receive}}%
\global\long\def\sd{\texttt{send}}%
\global\long\def\rcz{\rc(0)}%
\global\long\def\rco{\rc(1)}%
\global\long\def\sdz{\sd(0)}%
\global\long\def\sdo{\sd(1)}%
\global\long\def\eq{\textsc{Eq}}%
\global\long\def\ineq{\textsc{Ineq}}%
\global\long\def\geq{\textsc{GraphEq}}%
\global\long\def\gineq{\textsc{GraphIneq}}%
One of the major open problems in complexity theory is proving super-logarithmic
lower bounds on the depth of circuits (i.e., $\mathbf{P}\not\subseteq\mathbf{NC}^{1}$).
Karchmer, Raz, and Wigderson (Computational Complexity 5(3/4), 1995)
suggested approaching this problem by proving that the depth complexity
of a composition of functions~$f\diamond g$ is roughly the sum of
the depth complexities of~$f$ and~$g$. They showed that the validity
of this conjecture would imply that $\mathbf{P}\not\subseteq\mathbf{NC}^{1}$.

The intuition that underlies the KRW conjecture is that the composition~$f\diamond g$
should behave like a ``direct-sum problem'', in a certain sense,
and therefore the depth complexity of~$f\diamond g$ should be the
sum of the individual depth complexities. Nevertheless, there are
two obstacles toward turning this intuition into a proof: first, we
do not know how to prove that $f\diamond g$ must behave like a direct-sum
problem; second, we do not know how to prove that the complexity of
the latter direct-sum problem is indeed the sum of the individual
complexities.

In this work, we focus on the second obstacle. To this end, we study
a notion called ``strong composition'', which is the same as~$f\diamond g$
except that it is forced to behave like a direct-sum problem. We prove
a variant of the KRW conjecture for strong composition, thus overcoming
the above second obstacle. This result demonstrates that the first
obstacle above is the crucial barrier toward resolving the KRW conjecture.
Along the way, we develop some general techniques that might be of
independent interest.
\end{abstract}
\newpage{}

\tableofcontents{}

\newpage{}

\section{\label{sec:Introduction}Introduction}

A major frontier of the research on circuit complexity is proving
super-logarithmic lower bounds on the depth complexity of an explicit
function, i.e., proving that $\pnc$. This question is an important
milestone toward proving lower bounds on general circuits, and also
captures the natural question of whether there are tractable computational
tasks that cannot be parallelized. The state of the art is the work
of \hastad~\cite{H93}, who proved a lower bound of $(3-o(1))\cdot\log n$,
following a long line of work~\cite{S61,K72,A87,PZ93,IN93}. This
lower bound has not been improved for three decades except for the
lower order terms~\cite{T14}, and it is an important problem to
break this barrier.

Karchmer, Raz, and Wigderson~\cite{KRW91} proposed approaching this
problem by studying the composition of Boolean functions, defined
as follows: if $f:\B^{m}\to\B$ and $g:\B^{n}\to\B$ are Boolean functions,
then their composition $\fg$ takes inputs in $\left(\B^{n}\right)^{m}$
and is defined by
\begin{equation}
\fg(x_{1},\ldots,x_{m})=f\left(g(x_{1}),\ldots,g(x_{m})\right).\label{eq:composition}
\end{equation}
Let us denote by $\D(f)$ the minimal depth of a circuit with fan-in~$2$
that computes~$f$. The circuit that computes $\fg$ using \ref{eq:composition}
has depth $\D(f)+\D(g)$. Karchmer et al.~\cite{KRW91} conjectured
that this upper bound is roughly optimal:
\begin{conjecture}[The KRW conjecture]
\label{KRW}Let $f:\B^{m}\to\B$ and $g:\B^{n}\to\B$ be non-constant
functions. Then 
\begin{equation}
\D(\fg)\approx\D(f)+\D(g).\label{eq:KRW}
\end{equation}
\end{conjecture}

\noindent Karchmer et al. observed that their conjecture, if proved,
would imply that $\pnc$. The meaning of ``approximate equality''
in \ref{eq:KRW} is intentionally left vague, since there are many
variants that would imply the separation. In particular, the conjecture
would imply that $\pnc$ even if it only holds \emph{for} \emph{some
hard} function~$g$ rather than for \emph{every }function~$g$:
\begin{conjecture}[weak KRW conjecture]
\label{weak-KRW}For every non-constant function $f:\B^{m}\to\B$
and every $n\in\N$ there exists a function $g:\B^{n}\to\B$ such
that 
\begin{equation}
\D(\fg)\ge\D(f)+n-O\left(\log(m\cdot n)\right).\label{eq:weak-KRW}
\end{equation}
\end{conjecture}

\begin{proposition}[{folklore, see, e.g., \cite[Prop. 2.10]{M20}}]
The weak KRW conjecture implies that $\pnc$.
\end{proposition}

\noindent The KRW conjecture has been studied extensively in \cite{KRW91,EIRS91,H93,HW93,GMWW14,DM16,KM18,RMNPR20,M20,FMT21,MS21}.
These works succeeded in proving several special cases of the conjecture,
and in identifying some variants of particular interest. And yet,
the full KRW conjecture, or even just \ref{weak-KRW}, seem beyond
our reach.

\paragraph*{The communication complexity approach.}

It is useful to study the KRW conjecture through the lens of communication
complexity, and specifically, using the framework of \emph{Karchmer-Wigderson
relations}. Let us denote the (deterministic) communication complexity
of a problem~$R$ by $\C(R)$. The \emph{Karchmer-Wigderson relation}
of a function~$f:\B^{n}\to\B$, denoted $\KW_{f}$, is the communication
problem in which the inputs of Alice and Bob are $x\in f^{-1}(1)$
and $y\in f^{-1}(0)$ respectively, and their goal is to find a coordinate~$i$
such that $x_{i}\ne y_{i}$. Karchmer and Wigderson~\cite{KW88}
observed that~$\D(f)=\C(\KW_{f})$. This connection allows us to
study the depth complexity of functions using techniques from communication
complexity.

Now, let $f:\B^{m}\to\B$ and $g:\B^{n}\to\B$ be non-constant functions.
For convenience, we define the notation $\KW_{f}\d\KW_{g}$ to denote
the KW relation~$\KW_{\fg}$ of the composed function. In this relation,
Alice and Bob get $X\in(\fg)^{-1}(1)$ and $Y\in(\fg)^{-1}(0)$, viewed
as $m\times n$ matrices, and their goal is to find an entry $(i,j)$
such that $X_{i,j}\ne Y_{i,j}$. The KRW conjecture can be restated
as:
\[
\C(\KW_{f}\d\KW_{g})\approx\C(\KW_{f})+\C(\KW_{g}).
\]
It is worth noting the obvious protocol for solving $\KW_{f}\d\KW_{g}$:
Let $a=g(X)$ and~$b=g(Y)$ be the column vectors that are obtained
by applying $g$ to the rows of~$X$ and~$Y$ respectively. Observe
that they constitute an instance of~$\KW_{f}$, since by definition~$f(g(X))=1$
and~$f(g(Y))=0$. The players begin by solving $\KW_{f}$ on $a$
and~$b$, thus obtaining a coordinate~$i\in\left[m\right]$ such
that $a_{i}\ne b_{i}$. This implies that $g(X_{i})\ne g(Y_{i})$,
and therefore the rows~$X_{i}$ and~$Y_{i}$ constitute an instance
of~$\KW_{g}$. The players now solve $\KW_{g}$ on the rows~$X_{i}$
and~$Y_{i}$, thus obtaining a coordinate $j\in\left[n\right]$ where
$X_{i,j}\ne Y_{i,j}$. The communication complexity of this protocol
is $\C(\KW_{f})+\C(\KW_{g})$, and the KRW conjecture says that this
obvious protocol is roughly optimal.
\begin{remark}
The text of the introduction up to this point borrows heavily from
\cite{RMNPR20}, with the permission of the authors.
\end{remark}

\subsection{\label{subsec:Our-result}Our result}

In this work, we make progress toward the weak KRW conjecture by decoupling
the two major obstacles toward proving it, and tackling one of them
separately. Specifically, we prove the conjecture for a simpler notion
of composition, called ``strong composition'', which captures one
of the barriers to proving the conjecture, 

In order to motivate the notion of strong composition, recall that
the KRW conjecture says that the above ``obvious protocol'' for
$\KW_{f}\d\KW_{g}$ is essentially optimal. There is a very good intuition
for why this should be the case: First, recall that in the problem~$\KW_{f}\d\KW_{g}$,
the players are looking for an entry~$(i,j)$ such that $X_{i,j}\ne Y_{i,j}$.
It seems obvious that the players have to look for a solution~$(i,j)$
in a row where $a_{i}\ne b_{i}$, since in any other row there is
no guarantee that a solution even exists. Now, observe that finding
such a row~$i$ is equivalent to solving~$\KW_{f}$ on $a$ and~$b$.
Moreover, given such a row~$i$, finding a solution~$(i,j)$ is
equivalent to solving~$\KW_{g}$ on~$X_{i}$ and~$Y_{i}$. Hence,
it seems that in order to solve $\KW_{f}\d\KW_{g}$, the players have
to solve both~$\KW_{f}$ and~$\KW_{g}$, and therefore have to transmit
roughly $\C(\KW_{f})+\C(\KW_{g})$ bits.

Nevertheless, when attempting to turn the above intuition into a formal
proof, one immediately encounters two significant obstacles:
\begin{itemize}
\item While it seems obvious that the players should look for a solution~$(i,j)$
in a row~$i$ where~$a_{i}\ne b_{i}$, it is difficult to prove
that they \emph{must} do so.
\item Even if we could prove that the players have to solve both $\KW_{f}$
and~$\KW_{g}$, it would still not necessarily imply that they must
transmit at least $\C(\KW_{f})+\C(\KW_{g})$ bits. It could be the
case that the players solve both relations using less bits by employing
some clever strategy. For example, perhaps they could somehow ``recycle''
the bits that were used to solve~$\KW_{f}$ in order to solve~$\KW_{g}$.
In fact, showing that this cannot be done is an instance of the well-known
``direct-sum question'' in communication complexity.
\end{itemize}
Indeed, proofs of special cases of the KRW conjecture employ highly
non-trivial arguments to deal with those obstacles. In particular,
these arguments are tailored for specific cases at hand, and do not
seem applicable to the general KRW conjecture or even its weak variant.
In this work, we focus on tackling the second obstacle above, while
avoiding the first obstacle. To this end, we consider the following
strong composition operation, which requires Alice and Bob to find
a solution~$(i,j)$ in rows~$i$ where $a_{i}\ne b_{i}$.
\begin{definition}
Let $f:\B^{m}\to\B$ and $g:\B^{n}\to\B$ be non-constant functions.
The \emph{strong composition} of $\KW_{f}$ and~$\KW_{g}$, denoted
$\KW_{f}\c\KW_{g}$, is the following communication problem: Alice
and Bob take as inputs $X\in(\fg)^{-1}(1)$ and $Y\in(\fg)^{-1}(0)$.
Let $a=g(X)$ and~$b=g(Y)$. The goal of Alice and Bob is to find
an entry $(i,j)$ such that both $a_{i}\ne b_{i}$ and~$X_{i,j}\ne Y_{i,j}$.
\end{definition}

\noindent Observe that this notion of composition is indeed stronger
than the standard notion, in the sense that every protocol that solves
$\KW_{f}\c\KW_{g}$ also solves $\KW_{f}\d\KW_{g}$. This, in turn,
means that \emph{proving the KRW conjecture for strong composition
is necessary for proving the original KRW conjecture} (and the same
goes for the weak variant). Informally, our main result is the following
analogue of the weak KRW conjecture for strong composition (see \textcolor{red}{\ref{main-thm-formal}
}for the formal statement).
\begin{theorem}[Main theorem, informal]
\label{main-thm-informal}There exists a constant $\gamma>0.04$
such that the following holds: for every non-constant function $f:\B^{m}\to\B$
and for every $n\in\N$ there exists a function $g:\B^{n}\to\B$ such
that 
\begin{equation}
\C(\KW_{f}\c\KW_{g})\ge\C(\KW_{f})+n-(1-\gamma)\cdot m-O\left(\log(m\cdot n)\right).\label{eq:main-thm-informal}
\end{equation}
\end{theorem}

\noindent A few remarks are in order:
\begin{itemize}
\item The strong composition $\KW_{f}\c\KW_{g}$ is not a KW relation, and
hence our result does not imply depth lower bounds.
\item The loss of $(1-\gamma)\cdot m$ means that \ref{main-thm-informal}
is considerably weaker quantitatively compared to the weak KRW conjecture.
Yet, the lower bound we get is far from trivial: if this theorem is
proved for standard composition rather than strong composition, it
will imply a new depth lower bound of~$3.04\cdot\log n$. Such a
result would be the first significant improvement in depth lower bounds
in three decades, and hence very exciting. Nevertheless, such a result
would not be strong enough to imply~$\pnc$.
\item We believe that \ref{main-thm-informal} can be strengthened to hold
even when $g$~is a random function (in which case the lower bound
holds with high probability). Nevertheless, we did not verify it.
\item The notion of strong composition $\KW_{f}\c\KW_{g}$ was defined in
passing, without a name, by de Rezende et al. \cite{RMNPR20}. Such
a definition was considered even earlier in private discussions among
researchers: for example, Sajin Koroth suggested this notion to us
in 2019 (private communication).
\item In the setting of monotone circuits, strong composition is equivalent
to standard composition~\cite{KRW91,RMNPR20} (essentially, this
holds because the min-terms and max-terms of $\fg$ force the condition
of strong composition; see \cite[Sec. 3.1.1]{RMNPR24} for a proof).
Since our proof of \ref{main-thm-informal} can be adapted to monotone
circuits in a straightforward manner, this means that our main result
actually holds \emph{for standard composition} in the monotone setting.
Again, we believe this is true even when $g$~is a random (slice)
function. Such a result was not known prior to our work. Unfortunately,
the lower bounds that are implied by this result are weaker than the
state-of-the-art lower bounds in the monotone setting.
\item The only difference between the informal statement of our main theorem
and the formal one is that the formal statement refers to the logarithm
of the formula complexity of~$f$ rather than to~$\C(\KW_{f})$.
\item We did not attempt to optimize the value of the constant~$\gamma$
in \ref{main-thm-informal}. Nevertheless, improving~$\gamma$ to
a value of $0.64$ or larger would require significant new ideas (see
\ref{sec:barrier} for details).
\item We chose the name ``strong composition'' for the relation $\KW_{f}\c\KW_{g}$
since we view both $\KW_{f}\c\KW_{g}$ and $\KW_{f}\d\KW_{g}$ as
ways to compose the \emph{KW relations} $\KW_{f}$ \emph{and~$\KW_{g}$},
where the former is stronger than the latter since it requires the
output to satisfy a stronger requirement. Note, however, that $\KW_{f}\d\KW_{g}$
also corresponds to the composition of \emph{the functions $f$ and~$g$},
whereas $\KW_{f}\c\KW_{g}$ does not correspond to a composition of
\emph{the functions} in any obvious way.
\end{itemize}

\paragraph*{Previous work.}

In their original paper, Karchmer, Raz, and Wigderson~\cite{KRW91}
defined the universal relation~$U$, a simplification of KW relations,
and proposed to prove their conjecture for the composition~$U\d U$
as a step toward the full conjecture. This challenge was met by Edmonds
et al.~\cite{EIRS91}, and an alternative proof was given later by
\hastad~and Wigderson~\cite{HW93}. These results were extended
to compositions of the form~$\KW_{f}\d U$ in the work Gavinsky et
al. \cite{GMWW14}, and their result was improved quantitatively by
Koroth and Meir~\cite{KM18}.

\hastad~\cite{H93} proved the KRW conjecture implicitly for the
composition~$\KW_{f}\d\KW_{\mathrm{parity}}$ for every function~$f$
using the method of random restrictions. Dinur and Meir~\cite{DM16}
provided an alternative proof of that result using the communication
complexity approach. Filmus, Meir, and Tal~\cite{FMT21} generalized
the result of~\cite{H93} to compositions~$\KW_{f}\d\KW_{g}$ where
$f$~is an any function, and $g$~is any function for which the
soft quantum adversary bound is tight. Unfortunately, the result of~\cite{FMT21}
does not imply the weak KRW conjecture since the only functions~$g$
that satisfy the latter condition are relatively easy to compute ($\D(g)\le2\log n$).

De Rezende et al. \cite{RMNPR20} proved the KRW conjecture in the
\emph{monotone} setting for every outer function~$f$ and for a large
family of inner functions~$g$: namely, those functions~$g$ for
which there is a reduction from a lifted problem to~the monotone
KW relation~$\mKW_{g}$. They also introduced a new notion of a ``semi-monotone
composition'', and proved a semi-monotone KRW conjecture for~$U\d\mKW_{g}$
for a similar family of functions~$g$.

Most relevant to the current paper is the work of Mihajlin and Smal~\cite{MS21},
who proved the weak KRW conjecture for compositions of the form~$U\d\KW_{g}$.
Our main result is incomparable to theirs: on the one hand, our result
applies to a KW relation~$\KW_{f}$ rather than the universal relation~$U$;
on the other hand, their result applies to standard composition whereas
our result applies only to strong composition.

\subsection{\label{subsec:Our-techniques}Our techniques}

In order to prove our main result, we develop several new techniques
that we believe to be no less important than the result itself. In
particular, we believe that these techniques will be useful for attacking
the weak KRW conjecture (with standard composition). In what follows,
we provide an overview of the the proof of our main result, including
a detailed discussion of those new techniques.

\subsubsection{Composition of multiplexors}

The bulk of our proof is a lower bound for the ``multiplexor composition''~$\KcMn$:
this communication problem is defined similarly to~$\KW_{f}\c\KW_{g}$,
but now the function~$g$ is given to the players as part of the
input. More specifically, in the problem $\KcMn$, Alice takes as
input a function $g:\B^{n}\to\B$ and a matrix~$X\in(f\d g)^{-1}(1)$,
and Bob takes \emph{the same function}~$g$ and a matrix~$Y\in(f\d g)^{-1}(0)$.
As before, we let~$a=g(X)$ and~$b=g(Y)$. The goal of Alice and
Bob is to find an entry $(i,j)$ such that both $a_{i}\ne b_{i}$
and~$X_{i,j}\ne Y_{i,j}$.

Multiple works have observed that lower bounds for such multiplexor
composition problems can be used to derive depth lower bounds (see,
e.g., \cite{EIRS91,HW93,M20}). Most relevant to this paper is the
following observation of Mihajlin and Smal~\cite{MS21}: for every
function~$f:\B^{m}\to\B$ and every~$n\in\N$, there exists a function~$g:\B^{n}\to\B$
such that
\[
\C(\KW_{f}\d\KW_{g})\ge\Cp(\KdMn)-O\left(\log(m\cdot n)\right),
\]
where $\Cp$ denotes communication complexity in a model called ``partially
half-duplex protocols''. Such a result holds just as as well for
strong composition. Therefore, to prove our main result, it suffices
to prove that for every~$f:\B^{m}\to\B$ and~$n\in\N$ it holds
that
\[
\Cp(\KcMn)\ge\C(\KW_{f})+n-(1-\gamma)\cdot m-O\left(\log(m\cdot n)\right).
\]
In the rest of this section, we describe the proof of the latter lower
bound. In order to streamline the presentation, we ignore the differences
between standard and partially half-duplex protocols in this overview.

\subsubsection{\label{subsec:intro-structure-theorem}Our structure theorem}

Recall that the obvious protocol for~$\KW_{f}\d\KW_{g}$ works in
two stages: the players first solve~$\KW_{f}$ on~$a,b$, thus obtaining
a row~$i\in\left[m\right]$, and then solve~$\KW_{g}$ on the rows~$X_{i},Y_{i}$.
Our goal is to prove that this protocol is roughly optimal. Earlier
works \cite{EIRS91,DM16,KM18,RMNPR20} employed the following general
strategy: Given any protocol~$\Pi$, we break it into two parts,
which roughly correspond to the two stages of the obvious protocol.
We then show that $\Pi$~transmits roughly $\C(\KW_{f})$~bits in
the first part, and roughly $\C(\KW_{g})$~bits in the second part.
This implies that $\Pi$ must transmit roughly $\C(\KW_{f})+\C(\KW_{g})$~bits
overall, as required.

We follow this general strategy in our proof as well. Specifically,
we define a notion of ``live transcripts'', which correspond to
the first stage of the obvious protocol, and prove two claims:
\begin{itemize}
\item There exists a live transcript of length $\C(\KW_{f})-(1-\gamma)\cdot m$.
\item \textbf{The structure theorem:} after transmitting a live transcript,
the protocol must transmit at least $n-O\left(\log(m\cdot n)\right)$
more bits.
\end{itemize}
Together, the two claims imply our lower bound. The first claim is
relatively easy to prove, and the bulk of this paper is devoted to
proving the structure theorem.

Intuitively, a live transcript is a partial transcript~$\po$ in
which the protocol~$\Pi$ has not finished solving~$\KW_{f}$ on~$a$
and~$b$ yet, and has not revealed too much information about $g$,
$X$, and~$Y$. We now describe this notion in detail. To this end,
we begin with setting up some notation. Fix a protocol~$\Pi$ and
a partial transcript~$\pi_{1}$ of~$\Pi$. As usual in communication
complexity, we associate with~$\pi_{1}$ a set $\Xp$ that consists
of all the inputs $(g,X)$ of Alice that are consistent with~$\pi_{1}$.
Similarly, we denote by $\Yp$ the set of all inputs~$(g,Y)$ of
Bob that are consistent with~$\po$. For every function~$g:\B^{n}\to\B$
and strings $a\in f^{-1}(1)$ and $b\in f^{-1}(0)$, we denote
\begin{align*}
\Xp(g) & =\left\{ X\in\B^{m\times n}:(g,X)\in\Xp\right\}  & \Yp(g) & =\left\{ Y\in\B^{m\times n}:(g,Y)\in\Yp\right\} \\
\Xp(g,a) & =\left\{ X\in\Xp(g):g(X)=a\right\}  & \Yp(g,b) & =\left\{ Y\in\Yp(g):g(Y)=b\right\} \\
 & =\Xp(g)\cap g^{-1}(a) &  & =\Yp(g)\cap g^{-1}(b).
\end{align*}
We also denote by $\cA_{\pi_{1}}(g)$ and $\cB_{\pi_{1}}(g)$ the
sets of all strings~$a\in f^{-1}(1)$ and $b\in f^{-1}(0)$ such
that $\Xp(g,a)$ and $\Yp(g,b)$ are non-empty, respectively. Finally,
we denote by~$\Vp$ the set of functions~$g:\B^{n}\to\B$ such that
both~$\Ap(g)$ and~$\Bp(g)$ are non-empty. For convenience, we
focus only on \emph{balanced}\textbf{\emph{ }}functions~$g$ (i.e.,
functions that take the value~$1$ on exactly half of the inputs).
Informally, we say that $\po$ is \emph{alive} if there exists a set~$\cV\subseteq\Vp$
of balanced functions that satisfies the following conditions: 
\begin{itemize}
\item The set $\cV$ is large, i.e., it consists of at least $2^{-O(m)}$~fraction
of all balanced functions. Intuitively, the transcript~$\po$ does
not reveal too much information about~$g$.
\item For every $g\in\cV$, it is still quite hard to solve $\C(\KW_{f})$
on inputs from the set $\Ap(g)\times\Bp(g)$. Specifically, we require
that any protocol that solves~$\KW_{f}$ on~$\Ap(g)\times\Bp(g)$
has to transmit at least $(1-\gamma)\cdot m$~bits.
\item For every $g\in\cV$ and strings $a\in\Ap(g)$ and~$b\in\Bp(g)$,
the sets $\Xp(g,a)$ and~$\Yp(g,b)$ are large. Specifically, the
sets should consist of at least $2^{-\gamma\cdot m}$~fraction of
the sets~$g^{-1}(a)$ and~$g^{-1}(b)$. Intuitively, the transcript~$\po$
does not reveal too much information about~$X,Y$ even conditioned
on~$g$, $a$, and~$b$.
\end{itemize}
For the rest of this overview, we focus on proof of the above structure
theorem.

\subsubsection{Lower bounds using graph coloring}

Our starting point for proving the structure theorem is the following
result.
\begin{lemma}[implicit in \cite{MS21}]
\label{MS-lemma}Let $\Pi$~be a protocol that solves $\KdMn$,
and let $\po$ be a transcript of~$\Pi$. Suppose there is a subset
of functions~$\cV_{1}\subseteq\Vp$ such that, for every two distinct
functions~$g_{A},g_{B}\in\cV_{1}$, the following intersection property
holds:
\begin{itemize}
\item \textbf{Intersection Property:} either $\Xp(g_{A})\cap\Yp(g_{B})\ne\emptyset$
or $\Xp(g_{B})\cap\Yp(g_{A})\ne\emptyset$.
\end{itemize}
Then, after transmitting~$\pi_{1}$, the protocol~$\Pi$ must transmit
at least 
\[
\log\log\left|\cV_{1}\right|-\log\log\log\left|\cV_{1}\right|-O(1)
\]
more bits.
\end{lemma}

\noindent Essentially, \cite{MS21}~prove their lower bound for~$U\d\MX$
by constructing a sufficiently long transcript~$\po$ with such a
set~$\cV_{1}$ of size roughly~$2^{2^{n}}$. \ref{MS-lemma} then
implies a lower bound of roughly~$n$ on the number of bits that
$\Pi$~must communicate after~$\pi_{1}$. 

\ref{MS-lemma} is a powerful tool for proving lower bounds, but the
requirement that the intersection property holds for \emph{every }two
functions~$g_{A},g_{B}\in\cV_{1}$ is quite difficult to satisfy.
In this work, we show that the latter requirement can be weakened
substantially. To this end, we take a graph-theoretic perspective
on \ref{MS-lemma}. We define a graph~$\Gp$ over~$\Vp$ in which
two distinct functions~$g_{A},g_{B}\in\Vp$ are neighbors if and
only if they satisfy the intersection property. Taking this view,
\ref{MS-lemma} gives a lower bound of roughly~$\log\log(\cl(\Gp))$,
where $\cl(\Gp)$ is the \emph{clique number} of~$\Gp$ (i.e., the
size of the largest clique in~$\Gp$).

Our first main technical contribution is the following strengthening
of \ref{MS-lemma}. Let $\chi(\Gp)$ denote the \emph{chromatic number}
of~$\Gp$, i.e., the minimum number of colors required to color the
vertices of~$\Gp$ such that no two adjacent vertices share the same
color, and recall that $\chi(\Gp)\ge\cl(\Gp)$.
\begin{lemma}[\ref{chromatic-number-composition}, informal]
\label{chromatic-number-informal}Let $\Pi$~be a protocol that
solves $\KdM$, and let $\po$ be a transcript of~$\Pi$. Then, after
transmitting~$\pi_{1}$, the protocol~$\Pi$ must transmit at least
\[
\log\log\chi(\Gp)-\log\log\log\chi(\Gp)-O(1)
\]
more bits.
\end{lemma}

\noindent Moving from $\cl(\Gp)$ to~$\chi(\Gp)$ gives us a lot
more flexibility. In particular, it means that instead of \emph{lower
bounding} the size of the largest \emph{clique}, we can \emph{upper
bound} the size of the largest \emph{independent set}. In fact, it
suffices to upper bound the size of independent sets in a large sub-graph
of~$\Gp$ of our choice. This task turns out to be more manageable.
Indeed, the crux of our proof amounts to finding an appropriate sub-graph
and upper bounding the size of its independent sets.

We stress that our \ref{chromatic-number-informal} is proved for
\emph{standard} composition, so it could potentially be used to attack
the weak KRW conjecture in its original form. Nevertheless, both \ref{MS-lemma,chromatic-number-informal}
can be adapted to strong composition as well. The only difference
is that the above intersection property is replaced with the following,
easier to satisfy, intersection property.
\begin{itemize}
\item \textbf{Weak Intersection Property:} there exist matrices $X\in\cX_{\pi_{1}}(g_{A})$
and $Y\in\cY_{\pi_{1}}(g_{B})$ such that $X_{i}=Y_{i}$ for every
$i\in\left[m\right]$ for which $a_{i}\ne b_{i}$, where $a=g_{A}(X)$
and~$b=g_{B}(Y)$ (or the same statement holds when exchanging $g_{A}$
with~$g_{B}$).
\end{itemize}
To see that this property is indeed weaker, observe that the original
intersection property can be rephrased similarly, but requires that
$X_{i}=Y_{i}$ for \emph{every} $i\in\left[m\right]$, rather than
just those rows~$i$ for which $a_{i}\ne b_{i}$.
\begin{remark}
\label{lower-bounding-clique-number}After writing the first version
of this paper, we realized that we could also prove our main theorem
by lower bounding the clique number of~$\Gp$ and applying \ref{MS-lemma}.
Nevertheless, the proof that uses \ref{chromatic-number-informal}
is somewhat simpler. Moreover, the graph-theoretic perspective is
still crucial even when using \ref{MS-lemma} instead of \ref{chromatic-number-informal}.
See \ref{lower-bounding-clique-number-proof} for more details.
\end{remark}

\subsubsection{\label{subsec:Intro-prefix-thick-sets}Prefix-thick sets}

Given \ref{chromatic-number-informal}, the main challenge in our
proof is to upper bound the size of independent sets in the graph~$\Gp$,
where $\po$~is a live transcript of a protocol~$\Pi$. In particular,
this requires us to prove the existence of edges in~$\Gp$. To this
end, we introduce a notion of ``prefix-thick sets'' and prove the
existence of (many) such sets.

As motivation for this notion, fix two functions $g_{A},g_{B}\in\Vp$,
and suppose we would like to prove that they are neighbors in the
graph~$\Gp$. In order to do so, we should construct matrices~$X,Y$
and strings~$a,b$ as in the definition of the weak intersection
property. We will see later that we can construct strings~$a,b$
that agree on $(1-\gamma)$~fraction of the rows. Thus, in order
to show that~$g_{A}$ and~$g_{B}$ are neighbors, it remains to
construct matrices $X\in\cX_{\pi_{1}}(g_{A},a)$ and $Y\in\cY_{\pi_{1}}(g_{B},b)$
that agree on the remaining~$\gamma$~fraction of the rows where~$a_{i}\ne b_{i}$.
To this end, recall that the sets $\cX_{\pi_{1}}(g_{A},a)$ and $\cY_{\pi_{1}}(g_{B},b)$
are relatively large, i.e., they have density at least~$2^{-\gamma\cdot m}$
(since $\po$~is alive).

We now take a step back and consider the following simpler combinatorial
question: given two large sets of $m\times n$ matrices~$\cX$ and~$\cY$,
can we prove the existence of matrices $X\in\cX$ and $Y\in\cY$ that
agree on many rows? More generally, let $\Sigma$ be a finite alphabet,
and let $\cX,\cY\subseteq\Sigma^{m}$ be sets of strings of density
at least~$2^{-\gamma\cdot m}$. Can we prove the existence of strings
$X\in\cX$ and $Y\in\cY$ that agree on an at least $\gamma$~fraction
of their coordinates?

This natural question leads us to the notion of prefix-thick sets.
Recall that the \emph{prefix tree} of a set of strings~$\cX\subseteq\Sigma^{m}$
is the rooted tree of depth~$m$ that is defined as follows: the
vertices of depth~$i$ are the prefixes of length~$i$ of the strings
in~$\cX$; and a string~$x$ of depth~$i$ is the parent of a string
$y$ of depth~$(i+1)$ if and only if $x$ is a prefix of~$y$.
We say that the set~$\cX$ is \emph{prefix thick} if its prefix tree
has minimum degree greater than~$\frac{1}{2}\left|\Sigma\right|$
(or if $\cX$~contains a set with this property). We have the following
easy observation.
\begin{proposition}
\label{prefix-thick-sets-intersect}Let $\cX,\cY\subseteq\Sigma^{m}$.
If $\cX$ and $\cY$ are both prefix thick, then they intersect.
\end{proposition}

In particular, if we can find a large set of coordinates~$I\subseteq\left[m\right]$
such that both $\cX|_{I}$ and~$\cY|_{I}$ are prefix thick, then
$\cX|_{I}$ intersects~$\cY|_{I}$, and hence there exist $X\in\cX$
and $Y\in\cY$ that agree on many coordinates. Our second main technical
contribution is to deduce the following lemma from a result of Salo
and \torma~\cite{ST14}:
\begin{lemma}[informal version of \ref{prefix-thick-lemma}]
\label{prefix-thick-informal}If a set of strings~$\cX\subseteq\Sigma^{m}$
is sufficiently large, then there exist many sets $I\subseteq\left[m\right]$
such that $\cX|_{I}$ is prefix thick.
\end{lemma}

\noindent Using a simple counting argument, this result can be used
to find a set of $\gamma\cdot m$ coordinates~$I\subseteq\left[m\right]$
such that both $\cX|_{I}$ and~$\cY|_{I}$ are prefix thick, as required.

\subsubsection{\label{subsec:intro-proof-of-the-structure-theorem}The proof of
the structure theorem}

Finally, we return to the structure theorem, and explain how to prove
it using the above machinery. Let $f:\B^{m}\to\B$ and let $n\in\N$.
Fix a protocol~$\Pi$ that solves $\KcMn$, and let $\pi_{1}$ be
a live transcript. We would like to prove that after transmitting~$\pi_{1}$,
the protocol~$\Pi$ must still transmit at least $n-O(\log(mn))$
additional bits. Let $\Gp$ be the graph defined over~$\Vp$ according
to the weak intersection property. That is, two functions~$g_{A},g_{B}\in\Vp$
are neighbors in~$\Gp$ if and only if they satisfy the following
property:
\begin{itemize}
\item there exist matrices $X\in\cX_{\pi_{1}}(g_{A})$ and $Y\in\cY_{\pi_{1}}(g_{B})$
such that $X_{i}=Y_{i}$ for every $i\in\left[m\right]$ for which
$a_{i}\ne b_{i}$, where $a=g_{A}(X)$ and~$b=g_{B}(Y)$ (or the
same statement holds when exchanging $g_{A}$ with~$g_{B}$).
\end{itemize}
In our proof, we construct a certain sub-graph~$\cG'$ of~$\Gp$
and prove that every independent set contains at most $2^{-\left(\Omega(2^{n})-O(m\cdot n)\right)}$~fraction
of the vertices of that sub-graph. This implies that the chromatic
number of~$\cG'$ is at least~$2^{\Omega(2^{n})-O(m\cdot n)}$,
and hence the protocol must transmit at least 
\[
\log\log\chi(\Gp)\ge\log\log\chi(\cG')\ge n-O(\log(mn))
\]
additional bits, as required.

\paragraph*{The construction of~$\protect\cG'$.}

Let $\cV\subseteq\Vp$~be the set of balanced functions that is associated
with~$\po$. In order to construct~$\cG'$, we first show that,
for every $g\in\cV$, there exist strings $a_{g}\in\Ap(g)$ and $b_{g}\in,\Bp(g)$,
and a set of rows $I_{g}\subseteq\left[m\right]$, such that:
\begin{itemize}
\item The sets $\Xp(g,a_{g})|_{I_{g}}$ and $\Yp(g,b_{g})|_{I_{g}}$ are
prefix thick.
\item $a_{g}|_{\left[m\right]-I_{g}}=b_{g}|_{\left[m\right]-I_{g}}$.
\end{itemize}
In order to construct such a triplet~$(a_{g},b_{g},I_{g})$, we apply
\ref{prefix-thick-informal} to show that, for every $a$ and~$b$,
there are \emph{many }subsets~$I\subseteq\left[n\right]$ such that
$\Xp(g,a)|_{I}$ and $\Yp(g,b)|_{I}$ are prefix thick. Here, we use
the assumption that $\Xp(g,a)$ and $\Yp(g,b)$ are large. We then
deduce the existence of $a_{g},b_{g},I_{g}$ that satisfy the above
properties using the assumption that it is hard to solve $\C(\KW_{f})$
on inputs from the set $\Ap(g)\times\Bp(g)$, combined with the fortification
theorem of~\cite{DM16}.

Having found such a triplet~$(a_{g},b_{g},I_{g})$ for each~$g\in\cV$,
we construct~$\cG'$ as follows: Fix $(a,b,I)$ to be the most popular
triplet among all the triplets $(a_{g},b_{g},I_{g})$. We define~$\cV'$
to be the set of all functions~$g\in\cV$ whose triplet~$(a_{g},b_{g},I_{g})$
is equal to~$(a,b,I)$, and define~$\cG'$ to be the sub-graph of~$\Gp$
induced by~$\cV'$.

It remains to upper bound the size of independent sets in~$\cG'$.
In what follows, we use the convenient abbreviations $\XI_{g}=\Xp(g,a)|_{I}$,
and~$\YI_{g}=\Yp(g,b)|_{I}$. Observe that two functions~$g_{A},g_{B}\in\cV'$
are neighbors in $\cG'$ if $\XI_{g_{A}}\cap\YI_{g_{B}}\ne\emptyset$
or $\XI_{g_{B}}\cap\YI_{g_{A}}\ne\emptyset$: To see it, observe that
if $\XI_{g_{A}}\cap\YI_{g_{B}}\ne\emptyset$ then there exist matrices~$X\in\cX_{g_{A}}$
and $Y\in\cY_{g_{B}}$ such that $X_{i}=Y_{i}$ for every $i\in I$.
Moreover, for every $i\in\left[m\right]-I$ it holds that $a_{i}=b_{i}$.
Hence, the strings~$a,b$ and the matrices~$X,Y$ satisfy the requirements
of the weak intersection property, so $g_{A}$ and~$g_{B}$ are neighbors
in~$\cG'$.

In addition, in order to streamline the notation, from now on we make
the simplifying assumption that $a$~is the all-ones vector and that
$b$~is the all-zeroes vector. In particular, it holds that $g^{-1}(a)=g^{-1}(1)^{m}$
and $g^{-1}(b)=g^{-1}(0)^{m}$. In making this assumption, we allow
ourselves to ignore the fact that $a|_{\left[m\right]-I}=b|_{\left[m\right]-I}$,
since we will not need this fact again in this overview. We stress
that the actual proof does not use this simplifying assumption.

\paragraph*{Why is $\protect\cG'$ not a clique?}

At first glance, it may seem that~$\cG'$ is, in fact, a clique,
by the following (flawed) reasoning: For every $g_{A},g_{B}\in\cV'$,
the sets $\XI_{g_{A}}$ and~$\YI_{g_{B}}$ are prefix thick by definition,
so by \ref{prefix-thick-sets-intersect} it holds that $\XI_{g_{A}}\cap\YI_{g_{B}}\ne\emptyset$.
Therefore $g_{A}$ and~$g_{B}$ are neighbors.

The error in the foregoing reasoning is that it deduces that the sets
$\XI_{g_{A}}$ and $\YI_{g_{B}}$ intersect from the fact that they
are prefix thick. To see why this is a mistake, let us consider again
our construction of the triplet~$(a,b,I)$. When we applied \ref{prefix-thick-informal}
to $\Xp(g_{A},a)$ and~$\Yp(g_{B},b)$, we did so based on the assumption
that those sets have large density \emph{within the sets}~$g_{A}^{-1}(a)=g_{A}^{-1}(1)^{m}$
and~$g_{B}^{-1}(b)=g_{B}^{-1}(0)^{m}$ respectively. In particular,
we actually viewed $\Xp(g_{A},a)$ and~$\Yp(g_{B},b)$ as sets of
\emph{strings over the ``alphabets''}~$g_{A}^{-1}(1)$ \emph{and}~$g_{B}^{-1}(0)$
respectively. Therefore, while the sets $\XI_{g_{A}}$ and $\YI_{g_{B}}$
are indeed prefix thick, \emph{they satisfy this property over different
alphabets}. This means that we cannot use \ref{prefix-thick-sets-intersect}
to deduce that these sets intersect. Indeed, if $g_{A}=g_{B}$, then
these sets cannot possibly intersect.

This issue makes it difficult to argue that any two particular vertices
$g_{A}$ and~$g_{B}$ in~$\cG'$ are neighbors. Our third main technical
contribution, described next, is to show how to overcome this difficulty.

\paragraph*{Upper bounding the size of independent sets of~$\protect\cG'$.}

We turn to upper bound the size of independent sets in~$\cG'$. Consider
any set of vertices~$\cS\subseteq\cV'$ that is ``too large''.
Let $g_{A}$ and~$g_{B}$ be uniformly distributed functions in~$\cS$.
We prove that $g_{A}$ and~$g_{B}$ are neighbors with non-zero probability,
and this will imply that $\cS$ cannot be an independent set. 

We denote the alphabets~$\Sigma_{A}=g_{A}^{-1}(1)$ and $\Sigma_{B}=g_{B}^{-1}(0)$,
so $g_{A}^{-1}(a)=\Sigma_{A}^{m}$ and~$g_{B}^{-1}(b)=\Sigma_{B}^{m}$.
We also let $\Sigma_{AB}=\Sigma_{A}\cap\Sigma_{B}$, and let
\[
\cU=g_{A}^{-1}(a)|_{I}\cap g_{B}^{-1}(b)|_{I}=(\Sigma_{AB})^{I}.
\]

\noindent Now, recall that the sets $\XI_{g_{A}}$ and $\YI_{g_{B}}$
are prefix thick over $\Sigma_{A}$ and~$\Sigma_{B}$ respectively.
Unfortunately, as explained above, this prefix thickness does not
immediately imply that these sets intersect, since they are prefix
thick over different alphabets.

The key observation is that, with non-zero probability, \emph{the
sets $\XI_{g_{A}}\cap\cU$ and $\YI_{g_{B}}\cap\cU$ are both prefix
thick over the alphabet~$\Sigma_{AB}$}. Since they are both prefix
thick over the same alphabet, it follows that the sets $\XI_{g_{A}}\cap\cU$
and $\YI_{g_{B}}\cap\cU$ must intersect. Hence, the the sets $\XI_{g_{A}}$
and $\YI_{g_{B}}$ intersect with non-zero probability, implying that
$g_{A}$ and~$g_{B}$ are neighbors.

The reason that $\XI_{g_{A}}\cap\cU$ and $\YI_{g_{B}}\cap\cU$ are
prefix thick over~$\Sigma_{AB}$ with high probability is that, since
$\cS$ is relatively large, the sets~$\Sigma_{A},\Sigma_{B}$ are
good samplers. In other words, for every set~$W\subseteq\Sigma_{A}$,
the set $W\cap\Sigma_{B}$ has roughly the same density inside $\Sigma_{AB}=\Sigma_{A}\cap\Sigma_{B}$
as the set~$W$ inside $\Sigma_{A}$ with high probability, and the
same holds if we exchange~$\Sigma_{A}$ with~$\Sigma_{B}$. We use
this property to show that with sufficiently high probability, for
every vertex~$v$ in the prefix tree of~$\XI_{g_{A}}$, the density
of~$v$'s children in~$\Sigma_{AB}$ is roughly the same as their
density in~$\Sigma_{A}$, and the same goes for~$\YI_{g_{B}}$.
Since the prefix trees of~$\XI_{g_{A}}$ and~$\YI_{g_{B}}$have
minimum degree greater than~$\frac{1}{2}\cdot\left|\Sigma_{A}\right|$
and~$\frac{1}{2}\cdot\left|\Sigma_{B}\right|$ respectively, it follows
that the corresponding trees for \emph{$\XI_{g_{A}}\cap\cU$ }and\emph{
$\YI_{g_{B}}\cap\cU$} have minimum degree greater than $\frac{1}{2}\left|\Sigma_{AB}\right|$,
as required.
\begin{remark}
In the formal proof, we actually require prefix trees of~$\XI_{g_{A}}$
and~$\YI_{g_{B}}$ to have minimum degree greater than $\left(\frac{1}{2}+\varepsilon\right)\cdot\left|\Sigma_{A}\right|$
and $\left(\frac{1}{2}+\varepsilon\right)\cdot\left|\Sigma_{B}\right|$
respectively, in order to guarantee that the minimum degrees remain
larger than $\frac{1}{2}\cdot\left|\Sigma_{AB}\right|$ after the
intersection with~$\cU$.
\end{remark}

\begin{remark}
\label{lower-bounding-clique-number-proof}As mentioned in \ref{lower-bounding-clique-number},
in retrospect we realized that it is also possible to lower bound
the clique number of~$\Gp$. To this end, observe that last proof
actually shows that the graph~$\cG'$ is very dense. In other words,
the graph~$\cG'$ has a very large average degree. We now remove
the vertices whose degree is slightly smaller than the average degree,
and obtain a sub-graph~$\cG''$ that has a very large minimum degree.
Finally, we construct a clique in~$\cG''$ using an iterative greedy
algorithm: in each iteration, we add an arbitrary vertex~$v$ of~$\cG''$
to the clique, and remove from~$\cG''$ all the vertices that are
not neighbors of~$v$. The clique that is constructed in this way
is sufficiently large to imply the structure theorem.
\end{remark}

\subsubsection{\label{subsec:technical-comparison-with-previous-works}Comparison
with previous works}

We conclude this section by comparing our techniques with the techniques
of the previous works on the subject. As explained above, our proof
shares a common proof strategy with several previous works \cite{EIRS91,DM16,KM18,RMNPR20}.
Recall that in this strategy, we first define a notion of a ``live''
transcript, in which the protocol is still sufficiently far from solving
$\KW_{f}$ on~$a$ and~$b$, and in which not too much information
has been revealed on the inputs of the players We then prove that
every protocol must have a live transcript of length $\approx\C(\KW_{f})$.
Finally, we prove a structure theorem, which says that when the protocol
reaches a live transcript, it still has to communicate $\approx\C(\KW_{g})$
bits. Together, the two last items imply that every protocol must
communicate at least $\approx\C(\KW_{f})+\C(\KW_{g})$.

Our work also shares with the latter works the following general strategy
for proving the structure theorem:
\begin{enumerate}
\item Given a live transcript $\pi_{1}$, we measure the amount of information
that $\pi_{1}$~reveals on each row of $X$ and~$Y$ individually.
\item We partition the rows to ``revealed rows'', on which $\pi_{1}$
reveals a lot of information, and to ``unrevealed rows'', on which
$\pi_{1}$ reveals only a little information.
\item \label{enu:proof-strategy-bounding-revealed-rows}We argue that since
the total \emph{amount} of information that $\pi_{1}$~reveals on
$X$ and~$Y$ is not too large, the number of revealed rows must
be relatively small.
\item \label{enu:proof-strategy-forcing-agreement-on-revealed-rows}We restrict
the protocol to inputs in which the column vectors~$a$ and~$b$
agree on the revealed rows. Showing that we can afford to do it is
non-trivial, and relies on the assumption that in~$\pi_{1}$, the
protocol is still far from solving~$\KW_{f}$, as well as on the
fact that there are not too many revealed rows (as we showed in the
previous step).
\item \label{enu:proof-strategy-preventing-solution-on-revealed-rows}We
force the protocol to output a solution~$(i,j)$ for which $a_{i}\ne b_{i}$.
This implies in particular that the solution must belong to the unrevealed
rows (since we forced the revealed rows to satisfy $a_{i}=b_{i}$
in the previous step).
\item \label{enu:proving-hardness-on-unrevealed-rows}We argue that since
$\pi_{1}$ reveals only little information on the unrevealed rows,
the protocol must spend $\approx\C(\KW_{g})$ bits in order to find
a solution in such rows.
\end{enumerate}
In particular, Steps~\ref{enu:proof-strategy-preventing-solution-on-revealed-rows}
and \ref{enu:proving-hardness-on-unrevealed-rows} correspond to what
we referred to in \ref{subsec:Our-result} as ``the first and second
obstacles'' respectively. The main differences between the various
proofs are the ways in which they measure information and implement
the above steps. We now describe the analogues of the above steps
in our proof:
\begin{itemize}
\item The information revealed on individual rows is measured using the
notion of ``prefix-thickness''. In particular, the unrevealed rows
are those in which the sets $\cX$ and~$\cY$ are prefix-thick. More
specifically, the set of unrevealed rows is the set we denoted by~$I$
in the construction of~$\cG'$ in \ref{subsec:intro-proof-of-the-structure-theorem}.
\item Step~\ref{enu:proof-strategy-bounding-revealed-rows} is implemented
via \ref{prefix-thick-informal}, which implies that there is a large
prefix-thick set of rows (in other words, there are many unrevealed
rows).
\item Step~\ref{enu:proof-strategy-forcing-agreement-on-revealed-rows}
is implemented in the construction of~$\cG'$ in \ref{subsec:intro-proof-of-the-structure-theorem}.
Specifically, when we define~$\cG'$ to be the induced subgraph of
vertices that correspond to a fixed triplet~$(a,b,I)$, with $I$~being
the set of unrevealed rows, we in fact restrict the protocol to inputs
satisfying $a|_{\left[m\right]-I}=b|_{\left[m\right]-I}$.
\item Step~\ref{enu:proof-strategy-preventing-solution-on-revealed-rows}
is given to us for free by the definition of strong composition. Indeed,
as discussed in \ref{subsec:Our-result}, the whole motivation for
the notion of strong composition is that we would like to avoid dealing
with this obstacle for now.
\item Step~\ref{enu:proving-hardness-on-unrevealed-rows} is implemented
by upper bounding the size of independent sets in~$\cG'$ (as described
in \ref{subsec:intro-proof-of-the-structure-theorem}), and then deducing
a lower bound on the communication complexity from the chromatic number
of~$\cG'$ (using \ref{chromatic-number-informal}).
\end{itemize}
Taking the above perspective, we believe that the this work makes
three important contributions to the literature on the KRW conjecture,
which we discuss in detail below.

\paragraph*{Implementing Step~\ref{enu:proving-hardness-on-unrevealed-rows}
for $\protect\MX$.}

The first contribution is developing a technique for implementing
the above Step~\ref{enu:proving-hardness-on-unrevealed-rows} for
$\KW_{f}\d\MX$. In the aforementioned works, the implementation of
this step is tailored to the choice of the inner relation~$\KW_{g}$
and its specific properties, whether it was the universal relation~\cite{EIRS91,KM18},
the KW relation of the parity function~\cite{DM16}, or a monotone
KW relation with a certain ``lifted'' structure~\cite{RMNPR20}.
In our case, the inner relation is the multiplexor relation~$\MX$,
and for years, it has not been clear how we can implement Step~\ref{enu:proving-hardness-on-unrevealed-rows}
for it. 

While some earlier works suggested ideas for how such implementation
might be carried out \cite{EIRS91,M20}, the first tangible progress
on this question was made by~\cite{MS21} in their proof of the KRW
conjecture for $U\diamond\MX$ (where $U$ is the universal relation).
Unfortunately, their proof as a whole seemed tailored to the case
where the outer relation is the universal relation, and it was not
clear how to adapt it to the case where the outer relation is a KW
relation~$\KW_{f}$. In particular, their proof did not use the above
proof strategy for proving the structure theorem.

Our contribution in this regard consists of both extracting an important
technique that was implicit in the proof \cite{MS21} (namely, the
``intersection property'', \ref{MS-lemma}), and strengthening it
to work with the chromatic number rather than the clique number (\ref{chromatic-number-informal}).
This technique is crucial for carrying out Step~\ref{enu:proving-hardness-on-unrevealed-rows}
in the above proof strategy for~$\MX$, and we believe it might be
useful for future works.

\paragraph*{Bypassing the limitations of existing information measures.}

The previous works \cite{EIRS91,KM18,RMNPR20} measure the information
that $\pi_{1}$ reveals on individual rows of~$X$ and~$Y$ using
a notion called \emph{average degree} or \emph{unpredictability~}\cite{EIRS91,RM97}.
Using this measure of information, \cite{KM18} showed that it is
possible to guarantee that $\pi_{1}$~reveals as little as $2$~bits
of information per unrevealed row. In fact, if one is willing to prove
a lower bound that is quantitatively weaker than theirs, this can
be pushed down to $(1+\varepsilon)$~bits of information per unrevealed
row. Unfortunately, this technique suffers from two limitations that
prevent us from using it in our context:
\begin{itemize}
\item Having $(1+\varepsilon)$ bits of information per unrevealed row is
too much: we can only afford less than one bit of information per
unrevealed row. To see why, recall that in order to establish the
weak intersection property, we want to show that there exist matrices~$X$
and~$Y$ that agree on the unrevealed rows. In other words, we would
like to show that the supports of these rows intersect. Now, in order
to make sure, for example, that the supports of~$X_{1}$ and~$Y_{1}$
intersect, we need to guarantee that each of these supports contains
more than half of all the strings --- that is, that less than one
bit of information was revealed about~$X_{1}$ and~$Y_{1}$.
\item The notion of average degree is inherently an ``average-case'' notion
of information, while our proof seems to require a ``worst-case''
notion. To see why we seem to need a worst-case notion, recall again
that our goal is to force equality in the unrevealed rows of~$X$
and~$Y$. If the unrevealed rows are, say, the first two rows, then
in order to force them both to be equal in~$X$ and~$Y$, we a guarantee
that $\pi_{1}$ reveals little information about $X_{2}$ and~$Y_{2}$
\emph{even conditioned on $X_{1}=Y_{1}$}. The notion of average degree
cannot provide us with such a guarantee.
\end{itemize}
With regard to the second limitation, we note that worst-case notions
of information exist in the literature. For start, \cite{RM97} introduced
the notion the ``thick'' sets, which is a worst-case version of
average degree (the name ``prefix thick'' was chosen to allude to
their notion). Moreover, \cite{DM16} measured information using min-entropy,
which is a worst-case notion. Nevertheless, the two notions lead to
much weaker quantitative statements: they only provide a bound of
$O(\log m)$ bits of information per revealed row, which is far from
what we can afford.

Our contribution here is identifying a new way to measure information
that does provide what we need, namely, prefix-thick sets. These sets
guarantee that the support of every row covers more than half of all
the strings (i.e. less than one bit revealed), conditioned on every
assignment to the previous rows (i.e., a worst case guarantee). This
is exactly the guarantee we were looking for, and we believe it might
be of use to future works as well.

Equally importantly, in proving \ref{prefix-thick-informal}, we show
that this notion of information can indeed be used --- that is, we
show that we can indeed find a large prefix-thick set of rows. While
this lemma follows rather easily from a result of \cite{ST14} (\ref{ST-result}),
we note that this result was discovered in another community (namely,
the community of dynamical systems), and to the best of our knowledge
it was not known in the complexity community. Thus, an additional
contribution of this work is importing a new tool from a different
community.

\paragraph*{Conditioning on $a$ and~$b$.}

The notion of prefix-thick sets is important because it gives us a
\emph{worst-case} measure of information, but it is insufficient on
its own. While, by definition, $\pi_{1}$ reveals less than one bit
of information about each row in the prefix-thick set, we want this
property to hold simultaneously with the property that $a_{i}=b_{i}$
holds for each revealed row. Nevertheless, if we force the equality
$a_{i}=b_{i}$ on the revealed rows, this may leak additional information
on the unrevealed rows, and violate the prefix-thick guarantee. Indeed,
dealing with this issue is exactly where the argument of~\cite{KM18}
loses the additional one bit of information.

In order to avoid this loss, we change our measure of information
once more: instead of measuring the information revealed on~$X$
and~$Y$ on their own, we measure this information \emph{conditioned
on $a$ and~$b$}. This conditioning guarantees that when we force
the equality $a_{i}=b_{i}$ on the unrevealed rows, this forcing does
not leak additional information about the unrevealed rows (since the
information on the unrevealed rows was measured conditioned on the
whole values of~$a$ and~$b$ anyway). Formally, this conditioning
is implemented in the fact that we work with different sets $\cX_{\pi_{1}}(a)$
and~$\cY_{\pi_{1}}(b)$ for every choice of~$a$ and~$b$, whereas
previous works worked with single sets~$\cX_{\pi_{1}}$ and~$\cY_{\pi_{1}}$
(without dependence on~$a$ and~$b$) --- see also the third item
in the definition of a live transcript in \ref{subsec:intro-structure-theorem}.
Using this conditioning to avoid the loss in information is another
new idea of this work.

Unfortunately, this conditioning on~$a$ and~$b$ creates a new
problem: now the unrevealed rows of~$X$ and~$Y$ are prefix-thick
only \emph{when conditioned on~$a$ and~$b$}. In other words, the
support of each unrevealed row~$X_{i}$ contains more than half of
all the strings~$x$ \emph{such that $g(x)=a_{i}$} (rather than
half of all the strings in~$\B^{n}$). It is therefore unclear how
to ensure that $X$ and~$Y$ agree on the unrevealed rows when we
are only given this weaker guarantee.

Our third main contribution is showing how to bypass this obstacle
using the sampling properties of the sets $g^{-1}(1),g^{-1}(0)$,
as described in \ref{subsec:intro-proof-of-the-structure-theorem}
in the part on upper bounding the size of independent sets. While
this idea is somewhat technical, it is a key idea that allows us to
work with the information \emph{conditioned on~$a$ and~$b$}, and
therefore allows us to break the barrier of $(1+\varepsilon)$ bits
per row. We therefore believe that this idea, too, may be of use to
future works on the KRW conjecture.

\paragraph*{Summary.}

Wrapping up, the key contributions of this work to the literature
on the KRW conjecture are the abstraction and strengthening of the
intersection technique of~\cite{MS21}; identifying the notion of
prefix-thick sets as a useful way to define unrevealed rows and showing
that they can be found; and showing how to use conditioning on~$a$
and~$b$ to reduce the amount of information per unrevealed row,
while using sampling to mitigate the issues caused by this conditioning.
We believe that these new techniques would be of use for future works
on the subject.
\begin{remark}
The above description of the general proof strategy for the structure
theorem is a bit inaccurate in the case of~\cite{EIRS91,KM18}. Specifically,
in \ref{enu:proof-strategy-preventing-solution-on-revealed-rows},
they do not force the protocol to output a solution~$(i,j)$ such
that~$a_{i}\ne b_{i}$, but use a somewhat different argument of
similar flavor.
\end{remark}

\subsection{The organization of the paper}

We cover the required preliminaries in \ref{sec:Preliminaries}. Then,
in \ref{sec:Main-Theorem}, we state our structure theorem and use
it to prove our main theorem (\ref{main-thm-informal}). We explain
our method for proving multiplexor lower bounds using the chromatic
number in \ref{sec:Graph-Coloring}, and introduce the notion of prefix-thick
sets in \ref{sec:Prefix-thick-sets}. We then prove the structure
theorem in \ref{sec:Structure-Theorem}. We conclude by showing, in
\ref{sec:barrier}, an example that demonstrates that improving the
constant~$\gamma$ to~$0.64$ or above would require new ideas.

\section{\label{sec:Preliminaries}Preliminaries}

All the logarithms in this paper are of base~$2$. For any $n\in\N$,
we denote by $\left[n\right]$ the set $\left\{ 1,\ldots,n\right\} $.
If $S$ is a subset of some universe~$\cU$, we refer to the fraction~$\left|S\right|/\left|\cU\right|$
as \emph{the density of}~$S$ \emph{(within~$\cU$)}.

Given an alphabet~$\Sigma$ and a set~$I\subseteq\left[n\right]$,
we denote by $\Sigma^{I}$ the set of strings of length~$\left|I\right|$
whose coordinates are indexed by~$I$. Given a string~$w\in\Sigma^{n}$
and a set $I\subseteq\left[n\right]$, we denote by $w|_{I}\in\Sigma^{I}$
the projection of~$w$ to the coordinates in~$I$. Given a set of
strings $\cW\subseteq\Sigma^{n}$ and a set $I\subseteq\left[n\right]$,
we denote by $\cW|_{I}$ the set of projections of strings in~$\cW$
to~$I$. We denote by $\circ$ the concatenation operator of strings,
and denote by~$\left|w\right|$ the length of the string~$w$.

We denote by $\B^{m\times n}$ the set of Boolean $m\times n$~matrices,
and for a set $I\subseteq\left[m\right]$, we denote by $\B^{I\times n}$
the set of $\left|I\right|\times n$ matrices whose entries are indexed
by $I\times\left[n\right]$. Given a matrix $X\in\B^{m\times n}$
and a set~$I\subseteq\left[m\right]$, we denote by $X|_{I}\in\B^{I\times n}$
the projection of~$X$ to the rows in~$I$. Here, too, we extend
this notation to sets of matrices~$\cW\subseteq\B^{m\times n}$.
We denote by $X_{i}\in\B^{n}$ the $i$-th row of~$X$.

For a function~$g:\B^{n}\to\B$ and a matrix~$X\in\B^{m\times n}$,
we denote by~$g(X)$ the column vector that is obtained by applying~$g$
to every row of~$X$. Given a column vector~$a\in\B^{m}$, we denote
by $g^{-1}(a)$ the set of all matrices~$X\in\B^{m\times n}$ such
that $g(X)=a$.

The \emph{clique number} of a graph~$G$, denoted $\cl(G)$, is the
size of the largest clique in~$G$. The \emph{independence number}
of a graph~$G$, denoted $\alpha(G)$, is the size of the largest
independent set in~$G$. The \emph{chromatic number} of a graph~$G$,
denoted $\chi(G)$, is the the least number of colors required to
color the vertices of~$G$. It is easy to prove that if a graph~$G$
has $n$~vertices, then $\chi(G)$ is lower bounded by both~$\cl(G)$
and~$n/\alpha(G)$.

We use the following standard corollaries of the Chernoff-Hoeffding
bounds for independent random variables and negatively-associated
random variables (see, e.g., Theorems~1.1 and~3.1 in Sections 1.6
and 3.1 of \cite{DP09} respectively for the relevant versions of
the Chernoff-Hoeffding bounds):
\begin{fact}
\label{random-set-size}Let $S$ be a uniformly distributed subset
of~$\left[n\right]$. For every $\beta>0$ it holds that 
\begin{align*}
\Pr\left[\left|S\right|<\left(\frac{1}{2}-\beta\right)\cdot n\right] & <2^{-2\log e\cdot\beta^{2}\cdot n}
\end{align*}
\end{fact}

\begin{fact}
\label{sampling}Let $T$ be a subset of some universe~$\cU$, let
$k\in\N$, and let $S$ be a uniformly distributed subset of~$\cU$
of size~$k$. For every $\beta>0$ it holds that
\[
\Pr\left[\frac{\left|S\cap T\right|}{\left|S\right|}<\frac{\left|T\right|}{\left|\cU\right|}-\beta\right]<2^{-2\log e\cdot\beta^{2}\cdot k}
\]
and
\[
\Pr\left[\frac{\left|S\cap T\right|}{\left|S\right|}>\frac{\left|T\right|}{\left|\cU\right|}+\beta\right]<2^{-2\log e\cdot\beta^{2}\cdot k}
\]
\end{fact}

We also use the binary entropy function, and the following standard
approximation that it gives for the binomial coefficients.
\begin{notation}
The \emph{binary entropy function} $H_{2}:\left[0,1\right]\to\left[0,1\right]$
is the function that maps every $0<p<1$ to 
\[
H_{2}(p)=p\cdot\log\frac{1}{p}+\left(1-p\right)\cdot\log\frac{1}{1-p},
\]
and maps $0$~and~$1$ to~$0$.
\end{notation}

\begin{fact}[{see, e.g., \cite[Example 12.1.3]{CT91}}]
\label{binomial-entropy-approximation}For every $k,n\in\N$ such
that $k\le n$ it holds that
\[
\frac{1}{n+1}\cdot2^{H_{2}(\frac{k}{n})\cdot n}\le\binom{n}{k}\le2^{H_{2}(\frac{k}{n})\cdot n}.
\]
\end{fact}

\subsection{\label{subsec:Depth-complexity}Depth complexity and formula complexity}

In this section, we cover the basics of depth complexity. In the introduction,
we defined the depth complexity~$\D(f)$ of a function~$f$ as the
minimum depth of a circuit computing~$f$ with fan-in~$2$. For
our purposes in this paper, however, it is more convenient to use
an equivalent definition: namely, the minimum depth of a (de Morgan)
formula that computes~$f$.
\begin{definition}
A \emph{(de Morgan) formula}\textsf{~$\phi$} is a rooted binary
tree, whose leaves are identified with literals of the forms $x_{i}$
and $\neg x_{i}$, and whose internal vertices are labeled as AND
($\wedge$) or OR ($\vee$) gates. Here, the same literal can be associated
with more than one leaf. Such a formula $\phi$ over variables~$x_{1},\ldots,x_{n}$
computes a function from~$\B^{n}$ to $\B$ in the natural way. The
\emph{size} of a formula is the number of its \emph{leaves} (which
is the same as the number of its gates up to a factor of~$2$). The
\emph{depth}~of a formula is the depth of the tree.
\end{definition}

\begin{definition}
\label{formulas-computing-functions} The \emph{formula complexity}
of a Boolean function~$f:\B^{n}\to\B$, denoted $\L(f)$, is the
minimal size of a formula that computes~$f$. The \emph{depth complexity}
of~$f$, denoted~$\D(f)$, is the minimal depth of a formula that
computes $f$. If $f$ is a constant function, then we define $\L(f)=0$
and~$\D(f)=-\infty$.
\end{definition}

In what follows, we generalize the latter definition from functions
to promise problems, which will be useful when we discuss Karchmer-Wigderson
relations later. 
\begin{definition}
\label{formulas-separating-sets}Let $\cX,\cY\subseteq\B^{n}$ be
disjoint sets. We say that a formula~$\phi$ \emph{separates} $\cX$
and $\cY$ if $\phi(\cX)=1$ and $\phi(\cY)=0$. The \emph{formula
complexity of the rectangle $\cX\times\cY$}, denoted~$\L(\cX\times\cY)$,
is the size of the smallest formula that separates $\cX$ and~$\cY$.
The \emph{depth complexity} \emph{of} \emph{the rectangle}~$\cX\times\cY$,
denoted~$\D(\cX\times\cY)$, is the smallest depth of a formula that
separates $\cX$ and~$\cY$. If $\cX$ or~$\cY$ are empty, we define
$\L(\cX\times\cY)=0$ and $\D(\cX\times\cY)=-\infty$.
\end{definition}

Note that \ref{formulas-computing-functions} is indeed a special
case of \ref{formulas-separating-sets} where $\cX=f^{-1}(1)$ and~$\cY=f^{-1}(0)$.
Formula complexity has the following useful sub-additivity property.
\begin{proposition}
\label{complexity-subadditivity}Let $\cX,\cY$ be disjoints subsets
of~$\B^{n}$. Then, for every two partitions $\cX=\cX_{0}\cup\cX_{1}$
and $\cY=\cY_{0}\cup\cY_{1}$, it holds that
\begin{eqnarray*}
\L(\cX\times\cY) & \le & \L(\cX_{0}\times\cY)+\L(\cX_{1}\times\cY)\\
\L(\cX\times\cY) & \le & \L(\cX\times\cY_{0})+\L(\cX\times\cY_{1}).
\end{eqnarray*}
\end{proposition}

We also use the following standard upper bound about the formula complexity
of parity, which is obtained by computing the function in the natural
way.
\begin{proposition}
\label{complexity-parity}The formula complexity of the parity function
over $n$~bits is at most $4n^{2}$.
\end{proposition}

\begin{myproof}
First, observe that the parity of two bits can be computed by a formula
of size~$4$ at follows:
\[
x_{1}\oplus x_{2}=(x_{1}\wedge\neg x_{2})\vee(\neg x_{1}\wedge x_{2}).
\]
Next, observe that for every $k\in\N$, the parity of $2^{k}$ variables
can be computed recursively, by first computing the parity of the
first and last $2^{k-1}$ variables separately, and then computing
the parity of the resulting two bits. This idea leads to the following
recursive formula:
\[
\bigoplus_{i=1}^{2^{k}}x_{i}=\left(\bigoplus_{i=1}^{2^{k-1}}x_{i}\wedge\neg\bigoplus_{i=2^{k-1}+1}^{2^{k}}x_{i}\right)\vee\left(\neg\bigoplus_{i=1}^{2^{k-1}}x_{i}\wedge\bigoplus_{i=2^{k-1}+1}^{2^{k}}x_{i}\right).
\]
It can be easily proved by induction that the size of this formula
is at most $\left(2^{k}\right)^{2}$. Finally, for every natural number~$n$,
we can compute the parity of $n$~variables by adding dummy variables
(which are fixed to~$0$) so the total number of variables is the
smallest power of two that is at least~$n$. Since this power of
two is at most~$2n$, the size of the resulting formula is at most
$(2n)^{2}=4n^{2}$.
\end{myproof}

\subsection{\label{subsec:Communication-complexity}Communication complexity}

Let $\cX$, $\cY$, and $\cZ$ be sets, and let $R\subseteq\cX\times\cY\times\cZ$
be a relation. The \emph{communication problem}~\cite{Y79} that
corresponds to $R$ is the following: two players, Alice and Bob,
get inputs $x\in\cX$ and $y\in\cY$, respectively. They would like
to find~$z\in\cZ$ such that $(x,y,z)\in R$. To this end, they send
bits to each other until they find~$z$, but they would like to send
as few bits as possible. The \emph{communication complexity} of $R$
is the minimal number of bits that is transmitted by a protocol that
solves~$R$. More formally, we define a protocol as a binary tree,
in which every vertex represents a possible state of the protocol,
and every edge represents a message that moves the protocol from one
state to another: 
\begin{definition}
\label{def:protocol}A \emph{(deterministic) protocol} from $\cX\times\cY$
to~$\cZ$ is a rooted binary tree with the following structure:
\begin{itemize}
\item Every vertex~$v$ of the tree is labeled by a rectangle $\cX_{v}\times\cY_{v}$
where $\cX_{v}\subseteq\cX$ and $\cY_{v}\subseteq\cY$. The root
is labeled by the rectangle $\cX\times\cY$. The rectangle $\cX_{v}\times\cY_{v}$
is the set of pairs of inputs that lead the players to the vertex~$v$. 
\item Each internal vertex~$v$ is \emph{owned} by Alice or by Bob. Informally,
$v$ is owned by Alice if it is Alice's turn to speak at state~$v$,
and same for Bob. 
\item The two outgoing edges of every internal vertex are labeled by $0$
and~$1$ respectively. 
\item For every internal vertex $v$ that is owned by Alice, the following
holds: Let $v_{0}$ and $v_{1}$ be the children of $v$ associated
with the out-going edges labeled with $0$ and~$1$, respectively.
Then,

\begin{itemize}
\item $\cX_{v}=\cX_{v_{0}}\cup\cX_{v_{1}}$, and $\cX_{v_{0}}\cap\cX_{v_{1}}=\emptyset$. 
\item $\cY_{v}=\cY_{v_{0}}=\cY_{v_{1}}$. 
\end{itemize}
Informally, when the players are at the vertex~$v$, Alice sends
$0$ to~Bob if her input is in $\cX_{v_{0}}$ and $1$ if her input
is in~$\cX_{v_{1}}$. An analogous property holds for vertices owned
by Bob, while exchanging the roles of $\cX$ and~$\cY$.
\item Each leaf is labeled with a value~$z_{\ell}\in\cZ.$ The value $z_{\ell}$
is the output of the protocol at~$\ell$. 
\end{itemize}
We say that the protocol solves a relation $R\subseteq\cX\times\cY\times\cZ$
if, for every leaf~$\ell$, it holds that $\cX_{\ell}\times\cY_{\ell}\times\left\{ z_{\ell}\right\} \subseteq R$. 
\end{definition}

\begin{definition}
Given a protocol $\Pi$ and a vertex $v$ of~$\Pi$, the \emph{transcript
of~$v$} is the string that is obtained by concatenating the labels
of the edges on the path from the root to~$v$. Intuitively, this
string consists of the messages that Alice and Bob sent in their conversation
until they got to~$v$. Observe that the transcript determines the
vertex~$v$ uniquely and vice versa, so we often identify the transcript
with~$v$. If $v$ is a leaf of the protocol, we say that the transcript
is a \emph{full transcript}, and otherwise we say that it is a \emph{partial
transcript}.
\end{definition}

\begin{definition}
The \emph{communication complexity} of a (deterministic) protocol
$\Pi$, denoted $\C(\Pi)$, is the the depth of the protocol tree.
In other words, it is the maximum number of bits that can be sent
in an execution of the protocol on any pair of inputs~$(x,y)$. The
\emph{(deterministic) communication complexity} of a relation~$R$,
denoted~$\C(R)$, is the minimal communication complexity of a (deterministic)
protocol that solves $R$. 
\end{definition}

\begin{definition}
We define the \emph{size} of a protocol~$\Pi$ to be its number of
leaves. The \emph{protocol size} of a relation~$R$, denoted $\L(R)$,
is the minimal size of a protocol that solves it (this is also known
as the \emph{protocol partition number} of~$R$). 
\end{definition}

\subsubsection{Non-deterministic protocols}

Let $f:\cX\times\cY\to\B$, and consider the communication problem
in which Alice and Bob compute~$f(x,y)$ on inputs $x\in\cX$ and~$y\in\cY$
respectively. In a non-deterministic protocol for~$f$, there is
an untrusted prover who knows both~$x$ and~$y$, and whose goal
is to convince Alice and Bob that $f(x,y)=1$. The prover attempts
to do so by sending a witness~$w$ from some witness set~$\cW$
to both Alice and Bob. We require that the prover succeeds in convincing
both Alice and Bob to accept if and only if it indeed holds that $f(x,y)=1$,
and define the complexity of the protocol as the number of bits that
the prover sends.
\begin{definition}
\label{non-deterministic-protocol}Let $f:\cX\times\cY\to\B$, and
let $\cW$ be a set. A \emph{non-deterministic protocol~$\Pi$ for~$f$}
is a pair of functions $a_{\Pi}:\cX\times\cW\to\B$ and $b_{\Pi}:\cY\times\cW\to\B$
such that for every $(x,y)\in\cX\times\cY$ the following holds: $f(x,y)=1$
if and only if there exists a witness~$w\in\cW$ such that both $a_{\Pi}(x,w)=1$
and $b_{\Pi}(y,w)=1$. The \emph{complexity }of the protocol is $\log\left|\cW\right|$.
\end{definition}

\begin{definition}
\label{non-deterministic-complexity}Let $f:\cX\times\cY\to\B$. The
\emph{non-deterministic communication complexity} of~$f$, denoted
$\NC(f)$, is the minimal complexity of a non-deterministic protocol
for~$f$. The \emph{co-non-deterministic communication complexity}
of~$f$, denoted $\cNC(f)$, is the non-deterministic communication
complexity of the negation of~$f$.
\end{definition}

The following relationship between the non-deterministic communication
complexities is an easy observation.
\begin{proposition}
\label{NCC-vs-coNCC}Let $f:\cX\times\cY\to\B$. Then,
\begin{align*}
\NC(f) & \le2^{\cNC(f)}.
\end{align*}
\end{proposition}

Let $\cX$ be a any set. A standard example in communication complexity
is the equality function~$\eq_{\cX}:\cX\times\cX\to\B$, which outputs~$1$
if and only if~$x=y$. We denote its negation, the inequality function,
by $\ineq_{\cX}$. It is well known that $\NC(\eq_{\cX})=\log\left|\cX\right|$
and that 
\[
\log\log\left|\cX\right|\le\cNC(\eq_{\cX})\le\log\log\left|\cX\right|+1.
\]

\begin{remark}
The above presentation of non-deterministic communication complexity
is somewhat non-standard. Usually, non-deterministic communication
complexity is defined as the number of monochromatic rectangles required
to cover the inputs in~$f^{-1}(1)$. Nevertheless, it is not hard
to verify that the two definitions are equivalent.
\end{remark}

\subsection{Karchmer-Wigderson and multiplexor relations}

In what follows, we provide some background on Karchmer-Wigderson
relations and the related multiplexor relation. We first define KW
relations for general rectangles, and then specialize the definition
for functions.
\begin{definition}
\label{KW-relations-rectangles}Let $\cX,\cY\subseteq\B^{n}$ be disjoint
sets. The \emph{KW relation} $\KW_{\cX\times\cY}$ is the communication
problem in which Alice's input is~$x\in\cX$, Bob's input is~$y\in\cY$,
and they would like to find a coordinate $i\in\left[n\right]$ such
that $x_{i}\ne y_{i}$. Note that such a coordinate~$i$ always exists,
since $x\ne y$.
\end{definition}

\begin{definition}
\label{KW-relations-functions}Let $f:\B^{n}\to\B$ be a non-constant
function. The \emph{KW relation}\textsf{ }\emph{of}\textsf{~$f$},
denoted $\KW_{f}$, is defined by $\KW_{f}\eqdef\KW_{f^{-1}(1)\times f^{-1}(0)}$. 
\end{definition}

KW relations are related to depth and formula complexity in the following
way.
\begin{theorem}[\cite{KW88}, see also \cite{R90,KKN92,GMWW14}]
\label{KW-connection}For every two disjoints sets $\cX,\cY\subseteq\B^{n}$
it holds that $\D(\cX\times\cY)=\C(\KW_{\cX\times\cY})$, and $\L(\cX\times\cY)=\L(\KW_{\cX\times\cY})$.
In particular, for every non-constant $f:\B^{n}\to\B$, it holds that
$\D(f)=\C(\KW_{f})$, and $\L(f)=\L(\KW_{f})$.
\end{theorem}

\subsubsection{Fortification}

Given a protocol solving a KW relation, we sometimes want to relate
the amount of information that Alice and Bob transmit about their
inputs to the decrease in the complexity of the KW relation. For example,
we may want to argue that if Alice transmitted only one bit of information
about her input, then the complexity of the KW relation decreased
by only one bit. Unfortunately, this is not true in general. Nevertheless,
\cite{DM16} showed that every rectangle can be ``fortified'' such
that it satisfies this property. Intuitively, we say that a rectangle~$\cX\times\cY$
is fortified if when Alice and Bob speak, the complexity decreases
in proportion to the amount of information transmitted. Formally,
we have the following definition and result.
\begin{definition}
\label{fortified-rectangle}Let $\rho>0$. We say that a rectangle
$\cX\times\cY$ is \emph{$\rho$-fortified on Alice's side} if for
every $\tilde{\cX}\subseteq\cX$ it holds that 
\[
\frac{\L(\tilde{\cX}\times\cY)}{\L(\cX\times\cY)}\ge\rho\cdot\frac{\bigl|\tilde{\cX}\bigr|}{\bigl|\cX\bigr|}.
\]
Similarly, we say that $\cX\times\cY$ is \emph{$\rho$-fortified
on Bob's side} if the same holds for subsets $\tilde{\cY}\subseteq\cY$. 
\end{definition}

\begin{theorem}[fortification theorem, \cite{DM16}]
\label{foritification}Let $\cX,\cY\subseteq\B^{n}$ be disjoint
sets. There exists a subset $\cX'\subseteq\cX$ such that $\cX'\times\cY$
is $\frac{1}{4n}$-fortified on Alice's side, and such that $\L(\cX'\times\cY)\ge\frac{1}{4}\cdot\L(\cX\times\cY)$.
An analogous statement holds for Bob's side.
\end{theorem}

\subsubsection{Multiplexor relations}

As discussed in the introduction, for the most part of the paper we
focus on proving lower bounds for a certain ``multiplexor composition''.
Informally, the multiplexor relation~\cite{EIRS91} is a KW relation~$\KW_{g}$
in which the function~$g$ is given to the players as part of the
input. It is often more convenient, however, to allow Alice and Bob
to take as inputs (possibly distinct) functions~$g_{A},g_{B}:\B^{n}\to\B$
respectively, and allow them to reject if they detect that the promise~$g_{A}=g_{B}$
was violated. It is not hard to see that modifying the relation in
this way changes its complexity by at most a constant term.
\begin{definition}[\cite{EIRS91}]
\label{multiplexor}Let $n\in\N$. The \emph{(same-function) multiplexor
relation}, denoted $\MX_{n}$, is the following communication problem:
Alice gets a function~$g_{A}:\B^{n}\to\B$ and a string~$x\in g_{A}^{-1}(1)$,
and Bob gets a function~$g_{B}:\B^{n}\to\B$ and a string~$y\in g_{B}^{-1}(0)$.
Their goal is to find a coordinate $i\in\left[n\right]$ such that
$x_{i}\ne y_{i}$, and they are allowed to output the special symbol~$\bot$
if $g_{A}\ne g_{B}$.
\end{definition}

It can be shown that $\C(\MX_{n})=n+\Theta(1)$. The upper bound follows
from a protocol of Tardos and Zwick~\cite{TZ97}, and the lower bound
follows from the techniques of~\cite{MS21} (see Lemma~33 there).

\subsection{\label{subsec:Half-duplex-protocols}Half-duplex protocols}

One of the important properties of protocols in the standard model
of communication complexity is that, at any given point, the players
know whose turn it is to speak. This is captured in \ref{def:protocol},
for example, in the assumption that every vertex is owned by one of
the players. Hoover et al.~\cite{HIMS18} considered a different
model, called ``half-duplex channels'', in which the players do
not necessarily agree on whose turn it is to speak. Intuitively, a
half-duplex channel models a walkie-talkie device: such a device has
a ``push to talk'' button, where the user has to push the button
in order to speak, and to release it in order to listen. In particular,
if both sides try to speak simultaneously, the communication is lost.

A bit more formally, the communication in a half-duplex channel is
executed in rounds. At each round, Alice should choose whether she
would like to send a bit or to receive, and the same goes for Bob.
If Alice chooses to receive and Bob chooses to send a bit~$\sigma\in\B$,
then Alice receives the bit~$\sigma$, as in standard protocols (and
vice versa). Such rounds are called \emph{classical}. If both Alice
and Bob choose to send bits, then the bits are lost, with Alice and
Bob being none the wiser. Such rounds are called \emph{wasted}. If
both Alice and Bob choose to receive, the round is called \emph{silent},\emph{
}and there are multiple ways to define the behavior of the channel
in such a round (see \cite{HIMS18}). As in \cite{MS21}, we use the
definition of a ``half-duplex channel with adversary'': in silent
rounds, Alice and Bob receive bits that are chosen by an adversary.

An interesting property of half-duplex protocols with adversary is
that Alice and Bob do not necessarily agree on the ``state'' of
the protocol. Put differently, each player has its own view of what
has transpired so far, and these views are not necessarily consistent.
Therefore, when defining such protocols formally, we define them as
a pair of trees rather than one: each tree defines the behavior and
viewpoint of one of the players.
\begin{definition}[Half-duplex protocols with adversary \cite{HIMS18}]
\label{half-duplex-protocol}A half-duplex protocol~$\Pi$ from
$\cX\times\cY$ to~$\cZ$ is a pair of full $4$-ary trees~$\Pi_{A}$
and~$\Pi_{B}$ with the following structure.
\begin{itemize}
\item Each vertex~$v$ of~$\Pi_{A}$ (respectively,~$\Pi_{B}$) is labeled
with a set~$\cX_{v}\subseteq\cX$ (respectively, $\cY_{v}\subseteq\cY$).
Intuitively, $\cX_{v}$ is the set of inputs that can reach the vertex~$v$.
\item The four outgoing edges of every internal vertex are labeled by $\rcz$,
$\rco$, $\sdz$, and $\sdo$ respectively. Intuitively, this label
consists of the action that the player chooses to take and its outcome.
For every vertex~$v$ of~$\Pi_{A}$, denote by $v_{\rcz}$, $v_{\rco}$,
$v_{\sdz}$, and~$v_{\sdo}$ the children of~$v$ that are associated
with each of its outgoing edges.
\item For every vertex~$v$ of~$\Pi_{A}$, the following holds:
\begin{itemize}
\item $\cX_{v_{\rcz}}=\cX_{v_{\rco}}$.
\item $\cX_{v}=\cX_{v_{\rcz}}\cup\cX_{v_{\sdz}}\cup\cX_{v_{\sdo}}$.
\item The sets $\cX_{v_{\rcz}}$, $\cX_{v_{\sdz}}$, and $\cX_{v_{\sdo}}$are
pairwise disjoint.
\end{itemize}
Intuitively, when Alice is at the vertex~$v$, Alice sends~$0$
to~Bob if her input is in $\cX_{v_{\sd(0)}}$, sends~$1$ if her
input is in~$\cX_{v_{\sdo}}$, or receives if her input is in~$\cX_{v_{\rcz}}=\cX_{v_{\rco}}$.
An analogous property holds for the vertices of~$\Pi_{B}$, while
exchanging the roles of $\cX$ and~$\cY$.
\item Each leaf~$\ell$ is labeled with a value~$z_{\ell}\in\cZ.$ Intuitively,
$z_{\ell}$ is the output of the player at~$\ell$. 
\end{itemize}
When given an input~$(x,y)\in\cX\times\cY$, the protocol executes
the following algorithm. We initialize a pair of vertices~$(u,v)$
to the roots of~$\Pi_{A}$ and~$\Pi_{B}$ respectively. Then, at
each round, the algorithm performs the following steps: 
\begin{itemize}
\item For each $\sigma\in\B$, if $x\in\cX_{u_{\sd(\sigma)}}$ and $y\in\cY_{v_{\rc(0)}}=\cY_{v_{\rco}}$,
then the pair $(u,v)$ is updated to $(u_{\sd(\sigma)},v_{\rc(\sigma)})$.
In this case, the round is called \emph{classical}.
\item The same as the previous step, but exchanging the roles of $x,\cX,u$
with $y,\cY,v$. In this case, too, the round is called \emph{classical}.
\item For every $\sigma,\tau\in\B$, if $x\in\cX_{u_{\sd(\sigma)}}$ and
$y\in\cY_{v_{\sd(\tau)}}$, then the pair $(u,v)$ is updated to $(u_{\sd(\sigma)},v_{\sd(\tau)})$.
In this case the round is called \emph{wasted}.
\item If $x\in\cX_{u_{\rc(0)}}=\cX_{u_{\rc(1)}}$ and $y\in\cY_{v_{\rc(0)}}=\cY_{v_{\rco}}$,
then the pair $(u,v)$ is updated to $(u_{\rc(\sigma)},v_{\rc(\tau)})$
for some bits~$\sigma,\tau\in\B$. We think of the choice of the
bits~$\sigma,\tau$ as being taken by an adversary. In this case
the round is called \emph{silent}.
\end{itemize}
We require that for every~$(x,y)\in\cX\times\cY$ and for every possible
choices of the adversary, the vertices~$u$ and~$v$ reach leaves
at the same round, at which point the algorithm halts. We also require
that when the algorithm halts, both leaves are labeled with the same
output. If, in a particular execution of the protocol, the algorithm
reaches a pair of leaves~$(u,v)$ that are both labeled by an output~$z\in\cZ$,
then we define~$z$ to be the output of that particular execution.
We stress that the protocol may output different outputs in different
executions on the same input, depending on the choices of the adversary.
We say that the protocol \emph{solves} a relation~$R\subseteq\cX\times\cY\times\cZ$
if for every input~$(x,y)$, and in every possible execution of~$\Pi$
on~$(x,y)$, the output~$z$ at that execution satisfies $(x,y,z)\in R$.
\end{definition}

\begin{remark}
Note that the requirement that the vertices~$(u,v)$ always reach
leaves at the same round implies that the trees~$\Pi_{A},\Pi_{B}$
have the same depth.
\end{remark}

\begin{definition}
The \emph{communication complexity} of a half-duplex protocol~$\Pi$,
denoted~$\Ch(\Pi)$, is the depth of the trees~$\Pi_{A},\Pi_{B}$.
The \emph{half-duplex communication complexity} of a relation~$R\subseteq\cX\times\cY\times\cZ$,
denoted~$\Ch(R)$, is the minimal communication complexity of a half-duplex
protocol that solves it.
\end{definition}

Observe that every standard deterministic protocol can be viewed as
a half-duplex protocol, so~$\Ch(R)\le\C(R)$. On the other hand,
as noted in~\cite{HIMS18}, every half-duplex protocol can be simulated
by a standard protocol while incurring a factor of~$2$ in the complexity,
and hence $\C(R)\le2\cdot\Ch(R)$. In short, $\C(R)$ and $\Ch(R)$
are always within a factor of~$2$ of each other. Nevertheless, as
in~\cite{HIMS18,MS21}, saving this factor of~$2$ will be important
in our application.

We have seen that the players in a half-duplex protocol may have different
views. In particular, it follows that there is no well-defined transcript
for the execution of the protocol, since Alice and Bob may see different
transcripts. Nevertheless, given a transcript~$\pi\in\B^{*}$ and
an input~$x$ of Alice, we can still say whether $\pi$~matches
Alice's view in the protocol when given the input~$x$. Informally,
we say that $x\in\cX$~is \emph{consistent} with~$\pi\in\B^{*}$
if there exists some execution of the protocol in which Alice gets
the input~$x$ such that, for every round~$i$:
\begin{itemize}
\item if Alice decides to send a bit on input~$x$, then that bit is~$\pi_{i}$; 
\item and if Alice decides to receive on input~$x$, then she receives
the bit~$\pi_{i}$.
\end{itemize}
More formally, we can define this notion as follows.
\begin{definition}
\label{consistent-inputs}Fix a half-duplex protocol~$\Pi$ from
$\cX\times\cY$ to~$\cZ$, and let $\pi\in\B^{*}$. We say that a
vertex~$v$ of depth~$\left|\pi\right|$ in the tree~$\Pi_{A}$
\emph{is consistent} with $\pi$ if, for every $i\in\left[\left|\pi\right|\right]$,
the $i$-th edge in the path from the root to~$v$ is labeled with
either $\rc(\pi_{i})$ or $\sd(\pi_{i})$. We say that an input~$x\in\cX$
\emph{is consistent with }the~$\pi$ if there exists a vertex~$v$
of the tree~$\Pi_{A}$ that is consistent with~$\pi$ such that
$x\in\cX_{v}$. We define the notion that a vertex~$v$ in the tree~$\Pi_{B}$
or an input~$y\in\cY$ are consistent with~$\pi$ similarly.
\end{definition}

\begin{notation}
\label{consistent-inputs-set}Let $\Pi$ be a half-duplex protocol
from $\cX\times\cY$ to~$\cZ$, and let $\pi\in\B^{*}$. We denote
by $\cX_{\pi}$ (respectively, $\cY_{\pi}$) the set of inputs~$x\in\cX$
(respectively $y\in\cY$) that are consistent with~$\pi$. We say
that $\pi$~is a \emph{transcript} of~$\Pi$ if and only if both
$\cX_{\pi}$ and~$\cY_{\pi}$ are non-empty. 
\end{notation}

\begin{remark}
\label{half-duplex-transcript-vs-vertex}Observe that a transcript~$\pi$
contains the bits that were sent during the execution of the protocol,
but does not register who sent them. Hence, there could be inputs
$x,y$ of Alice and Bob respectively that are consistent with the
same transcript~$\pi$, but for which Alice and Bob still have ``different
beliefs'' about who sent each bit. For example, it could be the case
that on inputs~$x,y$ both Alice and Bob send the bit~$\pi_{1}$
in the first round.

In particular, recall that in the standard model of communication
complexity, the transcript~$\pi$ determines a vertex~$v$ of the
protocol, and vice versa. Moreover, the rectangle~$\cX_{\pi}\times\cY_{\pi}$
consists of exactly those pairs of inputs~$(x,y)$ on which the protocol
reaches~$v$. In the half-duplex model, on the other hand, different
pairs $(x,y)$ in the rectangle~$\cX_{\pi}\times\cY_{\pi}$ may lead
the protocol to different pairs of vertices~$(u,v)$. In fact, depending
on the choices of the adversary, even the same pair~$(x,y)$ may
lead the players to different pairs of vertices~$(u,v)$. This means
that some standard arguments in communication complexity require a
lot more care when executed in the half-duplex model.

Nevertheless, it is important to note that a transcript~$\pi$ and
a consistent input of Alice $x\in\cX_{\pi}$ do determine together
a vertex in~$\Pi_{A}$. More formally, for every $x\in\cX_{\pi}$,
the vertex~$v$ corresponding to~$x$ and~$\pi$ in \ref{consistent-inputs}
is unique. To see it, observe that the input~$x$ and the transcript~$\pi$
together determine the actions of the above algorithm on the vertex~$u$,
and hence determine the vertex that $u$~will reach after~$\left|\pi\right|$~steps.
\end{remark}

\subsubsection{Partially half-duplex protocol and multiplexors}

Lower bounds for multiplexors in the half-duplex model imply the existence
of hard KW relations in the standard model:
\begin{lemma}[{\cite[Lemma 31]{MS21}}]
For every~$n\in\N$ there exists a function~$g:\B^{n}\to\B$ such
that
\[
\C(\KW_{g})\ge\Ch(\MX_{n})-O(\log n).
\]
\end{lemma}

\noindent This result may not seem so impressive on its own, since
the existence of a hard KW relation follows immediately from counting
arguments. Nevertheless, the importance of this result is that similar
lemmas can also be proved for compositions of the form~$\KW_{f}\d\KW_{g}$
--- and this is exactly the kind of lower bounds we need in order
to prove the weak KRW conjecture. In fact, this was the original motivation
for introducing the half-duplex model (see discussion in~\cite{HIMS18}).

Unfortunately, the half-duplex model is rather complicated. In particular,
the lower bounds we have for multiplexors in this model are worse
than the ones we have in the standard model. In order to remedy this
issue, \cite{MS21} introduced a weaker variant of the model, called
``partially half-duplex protocols'', which is easier to analyze.
This model is only applicable to a particular kind of communication
problems, but the important thing is that it is applicable to multiplexors.
\begin{definition}[{\cite[Lemma 32]{MS21}}]
Let $R\subseteq\cX\times\cY\times\cZ$. Suppose that there exists
a set~$\cV$ such that the inputs in~$\cX$ and~$\cY$ are pairs
of the forms $(g_{A},x)$ and~$(g_{B},y)$ respectively where $g_{A},g_{B}\in\cV$.
Let $\Pi$ be a half-duplex protocol that solves~$R$. We say that
$\Pi$ is \emph{partially half-duplex} if whenever it is given an
input $\left((g_{A},x),(g_{B},y)\right)$ such that $g_{A}=g_{B}$,
all the rounds in the execution of~$\Pi$ on the input are classical.
The \emph{partially half-duplex communication complexity} of~$R$,
denoted~$\Cp(R)$, is the minimal communication complexity of a partially
half-duplex protocol that solves~$R$.
\end{definition}

\cite{MS21} observed that lower bounds against partially half-duplex
protocols are sufficient to imply the existence of hard KW relations.
\begin{lemma}[\cite{MS21}]
For every~$n\in\N$ there exists a function~$g:\B^{n}\to\B$ such
that
\[
\C(\KW_{g})\ge\Cp(\MX_{n})-O(\log n).
\]
\end{lemma}

\noindent In \ref{sec:Main-Theorem}, we prove a similar lemma for
the strong composition~$\KcM$, using exactly the same ideas as~\cite{MS21}.

\subsection{Linear codes and the Varshamov bound}

Let $C$ be a linear subspace of~$\B^{n}$ of dimension~$k$ (where
we identify $\B$ with the field~$\F_{2}$), and let $d\in\left[n\right]$.
We say that $C$ is a \emph{(linear) code} with distance~$d$ if
every non-zero vector~$c\in C$ has at least~$d$ non-zero coordinates
(i.e., its \emph{Hamming weight} is at least~$d$). The following
theorem, due to Varshamov, establishes the existence of linear codes
with a good trade-off between the distance and the dimension.
\begin{theorem}[\label{varshamov}Varshamov's bound \cite{V57}]
 For every $0<\delta<\frac{1}{2}$, $\varepsilon>0$, and for every
sufficiently large~$n\in\N$, there exists a linear code~$C\subseteq\B^{n}$
with distance~$\delta\cdot n$ and dimension at least $(1-H(\delta)-\varepsilon)\cdot n$.
\end{theorem}

\section{\label{sec:Main-Theorem}Main theorem}

In this section, we prove our main theorem, stated formally next.
\begin{theorem}
\label{main-thm-formal}There exists a constant $\gamma>0.04$ such
that the following holds: for every non-constant function $f:\B^{m}\to\B$
and for every $n\in\N$ there exists a function $g:\B^{n}\to\B$ such
that 
\begin{equation}
\C(\KW_{f}\c\KW_{g})\ge\log\L(\KW_{f})-(1-\gamma)\cdot m+n-O\left(\log(m\cdot n)\right).\label{eq:main-thm-formal}
\end{equation}
\end{theorem}

The section is organized as follows: We start with reducing our task
to proving lower bounds on the multiplexor composition~$\KcM$ in
\ref{subsec:Reduction-to-multiplexor}. Then, in \ref{subsec:Structure-theorem},
we define the notion of live transcripts and state our structure theorem.
Finally, we derive our main theorem from the structure theorem in
\ref{subsec:Proof-of-main-theorem}.

\subsection{\label{subsec:Reduction-to-multiplexor}Reduction to multiplexor
lower bounds}

Informally, the multiplexor composition $\KcM$ is a variant of the
composition~$\KW_{f}\c\KW_{g}$ in which the function~$g$ is given
to the players as part of the input. In the formal definition, however,
it is more convenient to allow Alice and Bob to take as inputs (possibly
distinct) functions $g_{A}$ and~$g_{B}$ respectively, and allow
them to reject if they detect that~$g_{A}\ne g_{B}$. We now define
the multiplexor compositions~$\KcM$ and~$\KdM$ formally.
\begin{definition}
Let $f:\B^{m}\to\B$ be a non-constant function, and let $n\in\N$.
The composition~$\KdMn$ is the following communication problem:
Alice gets as an input a function~$g_{A}:\B^{n}\to\B$ and a matrix~$X\in\B^{m\times n}$
such that $(f\d g_{A})(X)=1$. Bob gets as an input a function~$g_{B}:\B^{n}\to\B$
and a matrix~$Y\in\B^{m\times n}$ such that $(f\d g_{B})(Y)=0$.
Their goal is to find an entry~$(i,j)\in\left[m\right]\times\left[n\right]$
such that $X_{i,j}\ne Y_{i,j}$, and they are also allowed to output
a special symbol~$\bot$ if~$g_{A}\ne g_{B}$.

The strong composition $\KcM_{n}$ is defined similarly, except that
the entry~$(i,j)$ is also required to satisfy that $a_{i}\ne b_{i}$,
where $a=g_{A}(X)$ and~$b=g_{B}(Y)$.
\end{definition}

\noindent The following lemma allows us to focus, for the rest of
the paper, on proving lower bounds for~$\KcM$ in the partially half-duplex
model.
\begin{lemma}
\label{reduction-to-multiplexor}Let $f:\B^{m}\to\B$ be a non-constant
function, and let $n\in\N$. Then, there exists a function~$g:\B^{n}\to\B$
such that
\[
\C(\KW_{f}\d\KW_{g})\ge\Cp(\KdMn)-\log(m\cdot n)-3.
\]
Similarly, there exists a function~$g:\B^{n}\to\B$ such that
\[
\C(\KW_{f}\c\KW_{g})\ge\Cp(\KcMn)-\log(m\cdot n)-4.
\]
\end{lemma}

\begin{myproof}
Let $f:\B^{m}\to\B$ be a non-constant function, and let $n\in\N$.
We prove the statement for~$\KW_{f}\c\KW_{g}$, and the statement
for $\KW_{f}\d\KW_{g}$ can be proved similarly. For every function~$g:\B^{n}\to\B$,
fix $\Pi_{g}$ to be some optimal (standard) protocol that solves~$\KW_{f}\c\KW_{g}$,
and let $c$~denote the maximal communication complexity of~$\Pi_{g}$
over all choices of the function~$g$. We show that there exists
a partially half-duplex protocol~$\Pi$ that solves $\KcMn$ with
complexity~$c+\left\lceil \log(m\cdot n)\right\rceil +3$, and this
will imply the required result.

The protocol~$\Pi$ works as follows. On input $(g_{A},X)$, Alice
simulates the protocol~$\Pi_{g_{A}}$ on the matrix~$X$: specifically,
Alice sends a bit whenever it is her turn to send a bit according
to~$\Pi_{A}$, and she receives whenever it is Bob's turn to send
a bit. If Alice finishes in less than $c$~rounds, then she receives
until $c$~rounds have passed. Similarly, on input~$(g_{B},Y)$,
Bob simulates the protocol~$\Pi_{g_{B}}$ on the matrix~$Y$, and
then receives. Finally, after $c$ rounds have passed, Alice and Bob
have solutions ($i_{A},j_{A}$) and~$(i_{B},j_{B})$ to their respective
relations. Now, Alice sends $(i_{A},j_{A})$, $a_{i_{A}},$ and~$X_{i_{A},j_{A}}$
to Bob (and Bob receives). Finally, Bob checks that $(i_{A},j_{A})=(i_{B},j_{B})$,
$a_{i_{A}}\ne b_{i_{A}}$, and $X_{i_{A},j_{A}}\ne Y_{i_{A},j_{A}}$.
If this is indeed the case, then Bob sends the bit~$1$ to Alice,
and otherwise he sends~$0$ to Alice (meanwhile, Alice receives).
If Bob sent~$1$, then the output of the protocol is $(i_{A},j_{A})$.
If Bob sent~$0$, then the output of the protocol is~$\bot$: the
reason is that this case can only happen if Alice and Bob were simulating
different protocols, meaning that $g_{A}\ne g_{B}$. Clearly, the
complexity of~$\Pi$ is $c+\left\lceil \log(m\cdot n)\right\rceil +3$.
This concludes the proof.
\end{myproof}
\begin{remark}
The foregoing proof is a straightforward adaptation of the proofs
in~\cite{MS21}.
\end{remark}

\begin{remark}
The composition $\KdMn$ was defined by the author in~\cite{M20}.
The definition here is a bit different from the definition there:
specifically, in the definition of~\cite{M20}, each player gets
$m$~inner functions $g_{1},\ldots,g_{m}$ rather than just one.
Moreover, Alice and Bob are promised to receive the same inner functions
(rather than allowing them to output~$\bot$ if they got different
functions). This difference is not substantial, and the definition
that we use in this paper is more suitable for our purposes.
\end{remark}

\subsection{\label{subsec:Structure-theorem}The structure theorem}

Intuitively, our structure theorem says that every efficient protocol
that solves $\KcMn$ has roughly the same structure as the obvious
protocol. More specifically, this theorem says that if, at a given
point of the execution of the protocol, not much information was transmitted
about $g_{A},g_{B},X,Y$, and the protocol is still sufficiently far
from solving~$\KW_{f}$ on~$a$ and~$b$, then the protocol must
transmit at least $n-O(\log(m\cdot n))$ more bits. A point in the
execution that satisfies the latter requirements is called a \emph{live
transcript}. In what follows, we define live transcripts and state
our structure theorem formally. We start by recalling some notation
from the introduction. 
\begin{notation}
\label{V-X-Y-A-B}Let $f:\B^{m}\to\B$, let $n\in\N$, and let $\Pi$
be a partially half-duplex protocol that solves~$\KcMn$. Let $\pi$~be
a (possibly partial) transcript of~$\Pi$, and let $\cX_{\pi}$ and~$\cY_{\pi}$
be the sets of inputs that are consistent with~$\pi$ as in \ref{consistent-inputs-set}.
For every function $g:\B^{n}\to\B$ and strings $a\in f^{-1}(1)$
and $b\in f^{-1}(0)$, we denote
\begin{align*}
\cX_{\pi}(g) & =\left\{ X\in\B^{m\times n}:(g,X)\in\cX_{\pi}\right\}  & \cY_{\pi}(g) & =\left\{ Y\in\B^{m\times n}:(g,Y)\in\cY_{\pi}\right\} \\
\cX_{\pi}(g,a) & =\left\{ X\in\cX_{\pi}:g(X)=a\right\}  & \cY_{\pi}(g,b) & =\left\{ Y\in\Yp:g(Y)=b\right\} \\
 & =\cX_{\pi}(g)\cap g^{-1}(a) &  & =\cY_{\pi}(g)\cap g^{-1}(b).
\end{align*}
We also denote by $\cA_{\pi}(g)$ and $\cB_{\pi}(g)$ the sets of
all strings~$a\in f^{-1}(1)$ and $b\in f^{-1}(0)$ such that $\cX_{\pi}(g,a)$
and $\cY_{\pi}(g,b)$ are non-empty, respectively. Finally, we denote
by~$\cV_{\pi}$ the set of all functions~$g:\B^{n}\to\B$ such that
both~$\cA_{\pi}(g)$ and~$\cB_{\pi}(g)$ are non-empty.
\end{notation}

\begin{notation}
We denote by $\Vb$ be the set of all \emph{balanced} functions from~$\B^{n}$
to~$\B$ (i.e., the functions that take the value~$1$ on exactly
half of the inputs).
\end{notation}

\begin{definition}
\label{live-transcript}Let $\kappa\in\N$ be a universal constant
to be fixed later ($\kappa\ge8$ is enough), and let~$\gamma>0$.
Let $f$, $n$, and~$\Pi$ be as in \ref{V-X-Y-A-B}. We say that
a (possibly partial) transcript~$\pi_{1}$ of $\Pi$ is $\gamma$\emph{-alive}
if and only if there exists a set~$\cV\subseteq\Vp\cap\Vb$ of \emph{balanced
}functions that satisfies the following conditions:
\begin{itemize}
\item $\left|\cV\right|\ge2^{-m}\cdot\left|\Vb\right|$.
\item For every~$g\in\cV$, it holds that $\log\L\left(\Ap(g)\times\Bp(g)\right)\ge(1-\gamma)\cdot m+\kappa\log m+\kappa$.
\item For every $g\in\cV$, $a\in\Ap(g)$, and $b\in\Bp(g)$, it holds that
$\left|\Xp(g,a)\right|\ge2^{-\gamma\cdot m+1}\cdot\left|g^{-1}(a)\right|$
and $\left|\Yp(g,b)\right|\ge2^{-\gamma\cdot m+1}\cdot\left|g^{-1}(b)\right|$.
\end{itemize}
When $\gamma$~is clear from the context, we drop it and just write
that $\pi_{1}$ is alive.
\end{definition}

We are now ready to state our structure theorem.
\begin{theorem}[structure theorem]
\label{structure-theorem}There exists a constant~$\gamma>0.04$
such that the following holds: Let $f:\B^{m}\to\B$, let $n\in\N$,
and let $\Pi$ be a partially half-duplex protocol that solves~$\KcMn$.
For every $\gamma$-live transcript $\po$ of~$\Pi$, the complexity
of~$\Pi$ is at least 
\[
\left|\po\right|+n-O(\log(m\cdot n)).
\]
\end{theorem}

\subsection{\label{subsec:Proof-of-main-theorem}Proof of main theorem from structure
theorem}

Let $\gamma>0.04$ be the constant from \ref{structure-theorem},
and let $\kappa$~be the universal constant from \ref{live-transcript}.
Let $f:\B^{m}\to\B$ and let $n\in\N$. We may assume without loss
of generality that $\log\L(\KW_{f})>(1-\gamma)\cdot m$, since otherwise
the desired result follows easily from a counting argument. By \ref{reduction-to-multiplexor},
it suffices to prove that
\[
\Cp(\KcMn)\ge\log\L(\KW_{f})-(1-\gamma)\cdot m+n-O\left(\log(m\cdot n)\right),
\]
and this will imply the desired result. Fix a partially half-duplex
protocol $\Pi$ that solves~$\KcMn$. We prove that the communication
complexity of~$\Pi$ is at least the right-hand side of the last
equation. To this end, we construct a sufficiently long $\gamma$-live
transcript~$\po$, and then apply the structure theorem to~$\pi_{1}$.

In order to construct the live transcript~$\po$, we first construct,
for each $g\in\Vb$, a ``candidate transcript''~$\pg$ that satisfies
the following properties:
\begin{enumerate}
\item \label{enu:candidate-length}$\left|\pg\right|=\log\L(\KW_{f})-(1-\gamma)\cdot m-\kappa\log m-\kappa$.
\item \label{enu:candidate-g}$g\in\cV_{\pg}$.
\item \label{enu:candidate-a-b}$\log\L\left(\Ap(g)\times\Bp(g)\right)\ge(1-\gamma)\cdot m+\kappa\log m+\kappa$.
\item \label{enu:candidate-X-Y}For every $a\in\Ap(g)$ and $b\in\Bp(g)$,
it holds that $\left|\Xp(g,a)\right|\ge2^{-\gamma\cdot m+1}\cdot\left|g^{-1}(a)\right|$
and $\left|\Yp(g,b)\right|\ge2^{-\gamma\cdot m+1}\cdot\left|g^{-1}(b)\right|$.
\end{enumerate}
We then choose~$\po$ to be the most popular value of~$\pg$ over
all functions~$g\in\Vb$, and choose~$\cV$ to be the set of functions~$g\in\Vb$
such that $\pg=\po$. A simple counting argument then shows that $\left|\cV\right|\ge2^{-m}\cdot\left|\Vb\right|$,
and hence $\po$~is alive.

We turn to describing the construction of a single candidate transcript~$\pg$.
Fix a function~$g\in\Vb$. In what follows, we abbreviate and write
$\Apg=\Apg(g)$ and $\Xpg(a)=\Xpg(g,a)$, and similarly for $\Bpg$
and~$\Ypg$. Let $\Pi_{g}$ be the protocol that is obtained from~$\Pi$
by hard-wiring the inputs of the players such that~$g_{A}=g_{B}=g$.
Observe that since the original protocol~$\Pi$ is \emph{partially}
half-duplex, it holds that all the rounds in the hardwired protocol~$\Pi_{g}$
are classical. In other words, we can think of~$\Pi_{g}$ as a standard
deterministic protocol rather than as a half-duplex protocol. We choose
the candidate transcript~$\pg$ to be a transcript of~$\Pi_{g}$,
and construct it iteratively, bit-by-bit. Intuitively, at each iteration
we choose the next bit of~$\pg$ such that it transmits at most one
bit of information about~$X,Y$ and decreases the complexity of solving
$\KW_{f}$ on~$a$ and~$b$ by at most one bit. Formally, we initialize
$\pg$ to the empty transcript, and then, in each iteration, we perform
the following steps:
\begin{itemize}
\item Without loss of generality, assume that it is Alice's turn to speak
next in $\pg$.
\item For every $a\in\Apg$ and $X\in\Xpg(a)$, let $\sigma_{a,X}$ be the
bit that Alice sends in~$\Pi_{g}$ at~$\pi_{1,g}$ on input~$X$.
\item For every $a\in\Apg$, let $\sigma_{a}$ be majority value of~$\sigma_{a,X}$
over all $X\in\Xpg(a)$.
\item For each $\sigma\in\B$, let $\cA_{\sigma}=\left\{ a\in\Apg\mid\sigma_{a}=\sigma\right\} $.
By the sub-additivity property of formula complexity, it holds that
\[
\L(\Apg\times\Bpg)\le\L(\cA_{0}\times\Bpg)+\L(\cA_{1}\times\Bpg).
\]
In particular, there exists a bit~$\sigma\in\B$ such that $\L(\cA_{\sigma}\times\Bpg)\ge\L(\Apg\times\Bpg)/2$.
\item Append $\sigma$ to~$\pg$.
\item If it is Bob's turn to speak next in~$\pg$, perform the foregoing
steps while exchanging $a\in\Apg$ with~$b\in\Bpg$ and $X\in\Xpg(a)$
with~$Y\in\Ypg(b)$.
\end{itemize}
We repeat these iterations until $\pg$ is either of length 
\begin{equation}
\log\L(f)-(1-\gamma)\cdot m-\kappa\log m-\kappa\label{first-stage-length}
\end{equation}
or a leaf of~$\Pi_{g}$. It remains to show that $\pi_{1,g}$ satisfies
the above properties of a candidate transcript.

First, observe that the transcript~$\pg$ satisfies Property~\ref{enu:candidate-g}
(i.e., $g\in\cV_{\pg}$) since it is a transcript of~$\Pi_{g}$.
Next, we show that $\pg$~satisfies Property~\ref{enu:candidate-a-b}.
At the beginning of the construction, we have $\Apg=f^{-1}(1)$ and
$\Bpg=f^{-1}(0)$. In particular, it holds that $\L(\Apg\times\Bpg)=\L(f)$.
Observe that at each iteration of the construction, the formula complexity
$\L(\Apg\times\Bpg)$ is decreased by at most a factor of~$2$. Hence,
at the end of the construction, it holds that
\[
\L(\Apg\times\Bpg)\ge\L(f)\cdot2^{-\left|\pg\right|}\ge\L(f)\cdot2^{-\left(\log L(f)-(1-\gamma)\cdot m-\kappa\log m-\kappa\right)}\ge2^{(1-\gamma)\cdot m+\kappa\log m+\kappa},
\]
so~$\pg$ indeed satisfies Property~\ref{enu:candidate-a-b}.

We now show that the transcript~$\pg$ satisfies Property~\ref{enu:candidate-length}.
To this end, it suffices to show that $\pg$~cannot be a leaf of~$\Pi_{g}$.
Suppose for the sake of contradiction that $\pg$ is a leaf of~$\Pi_{g}$.
By definition, $\Pi_{g}$ is a protocol that solves~$\KW_{f}\c\KW_{g}$,
and therefore $\pg$~is labeled with a solution~$(i,j)$ for that
relation. This means that for every $a\in\Apg$ and~$b\in\Bpg$,
and for every $X\in\cX_{\pi_{g}}(a)$ and~$Y\in\Ypg(a)$, it should
hold that $a_{i}\ne b_{i}$ and $X_{i,j}\ne Y_{i,j}$. Nevertheless,
the requirement that $a_{i}\ne b_{i}$ for every $a\in\Apg$ and~$b\in\Bpg$
implies that $i$~is a solution for~$\KW_{f}$ on $\Apg\times\Bpg$.
In other words, the relation~$\KW_{f}$ is solved on~$\Apg\times\Bpg$,
and thus $\L(\Apg\times\Bpg)=1$. This, however, contradicts the above
lower bound on $\L(\Apg\times\Bpg)$. It follows that $\pg$ cannot
be a leaf.

Finally, we prove that $\pg$~satisfies Property~\ref{enu:candidate-X-Y}.
Observe that at the beginning of the construction, for every $a\in\Apg$
and~$b\in\Bpg$ it holds that $\Xpg(a)=g^{-1}(a)$ and $\Ypg(b)=g^{-1}(b)$.
Furthermore, observe that in each iteration of the construction, the
sizes of the sets $\Xpg(a)$ and~$\Ypg(b)$ decrease by at most a
factor of~$2$ for every $a\in\Apg$ and~$b\in\Bpg$. Moreover,
the construction performs at most~$\gamma\cdot m-\kappa\log m-\kappa$
iterations, since $L(f)\le2^{m}$ for every function~$f$. It follows
that at the end of the construction we have 
\begin{align*}
\left|\Xpg(a)\right| & \ge2^{-\gamma\cdot m+1}\cdot\left|g^{-1}(a)\right|\\
\left|\Ypg(b)\right| & \ge2^{-\gamma\cdot m+1}\cdot\left|g^{-1}(b)\right|.
\end{align*}

We conclude the proof by choosing the transcript~$\pi_{1}$ of~$\Pi$
to be the most popular transcript~$\pg$ over all the functions~$g\in\Vb$.
We prove that $\pi_{1}$ is $\gamma$-alive. First, we choose the
set~$\cV$ that is associated with~$\po$ to be the set of functions~$g\in\cV_{0}$
for which $\pg=\po$. Since the transcripts~$\pg$ are of the same
length, and since this length is smaller than~$m$, the number of
those transcripts is at most~$2^{m}$. By an averaging argument,
it follows that~$\left|\cV\right|\ge2^{-m}\cdot\left|\Vb\right|$.
Furthermore, since $\po$~satisfies Properties~\ref{enu:candidate-a-b}
and~\ref{enu:candidate-X-Y} for every~$g\in\cV$, it also satisfies
the other requirements of the definition of live transcripts. Hence,
$\pi_{1}$ is $\gamma$-alive. By the structure theorem, the complexity
of the protocol~$\Pi$ is at least
\[
\left|\po\right|+n-O(\log(m\cdot n))\ge\log\L(f)+n-(1-\gamma)\cdot m-O(\log(m\cdot n)),
\]
as required.

\section{\label{sec:Graph-Coloring}Multiplexor lower bounds via graph coloring}

In this section, we develop a general method for proving lower bounds
on multiplexor-composition problems, by extending a previous method
of Mihajlin and Smal~\cite{MS21}. We first develop the method for
the composition~$\KdMn$, and then for the strong composition~$\KcMn$.
We note that in this paper we only apply the method for $\KcMn$.
Nevertheless, we hope that the method will also be useful in the future
for proving lower bounds on $\KdMn$, and therefore state and prove
it for this relation as well.
\begin{definition}
\label{characteristic-graph-composition}Let $\Pi$ be a partially
half-duplex protocol that solves a relation $\KdMn$ for some $f:\B^{m}\to\B$
and $n\in\N$, and let $\po$ be a (possibly partial) transcript of~$\Pi$
such that $\Vp\ne\emptyset$. The \emph{characteristic graph of~$\po$},
denoted~$\Gp$, is the graph whose vertices are the functions in~$\Vp$,
and whose edges are defined as follows: two functions $g_{A},g_{B}\in\Vp$
are neighbors in~$\Gp$ if and only if either $\Xp(g_{A})\cap\Yp(g_{B})\ne\emptyset$
or $\Xp(g_{B})\cap\Yp(g_{A})\ne\emptyset$.
\end{definition}

\begin{lemma}
\label{chromatic-number-composition}Let $\Pi,f,n$ be as in~\ref{characteristic-graph-composition}.
For every transcript~$\po$ of~$\Pi$ such that $\Vp\ne\emptyset$,
the complexity of~$\Pi$ is at least
\[
\left|\po\right|+\log\log\chi(\Gp)-\log\log\log\chi(\Gp)-4.
\]
\end{lemma}

As discussed in the introduction, \cite{MS21} implicitly proved a
similar result, in which the chromatic number~$\chi(\Gp)$ is replaced
by the clique number~$\cl(\Gp)$. Since it is always the case that
$\chi(\Gp)\ge\cl(\Gp)$, our lemma generalizes their result. We now
state the analogous definition and lemma for strong composition.
\begin{definition}
\label{characteristic-graph-strong-composition}Let $\Pi$ be a partially
half-duplex protocol that solves a relation $\KcMn$ for some $f:\B^{m}\to\B$
and $n\in\N$, and let $\po$ be a (possibly partial) transcript of~$\Pi$
such that $\Vp\ne\emptyset$. The \emph{characteristic graph of~$\po$},
denoted~$\Gp$, is the graph whose vertices are the functions in~$\Vp$,
and in which two functions $g_{A},g_{B}\in\Vp$ are neighbors if and
only if they satisfy the following property:
\begin{itemize}
\item \textbf{Weak Intersection Property:} there exist matrices $X\in\cX_{\pi_{1}}(g_{A})$
and $Y\in\cY_{\pi_{1}}(g_{B})$ such that $X_{i}=Y_{i}$ for every
$i\in\left[m\right]$ for which $a_{i}\ne b_{i}$, where $a=g_{A}(X)$
and~$b=g_{B}(Y)$ (or the same statement holds when exchanging $g_{A}$
and~$g_{B}$).
\end{itemize}
\end{definition}

\begin{lemma}
\label{chromatic-number-strong-composition}Let $\Pi,f,n$ be as in~\ref{characteristic-graph-strong-composition}.
For every transcript~$\po$ of~$\Pi$ such that $\Vp\ne\emptyset$,
the complexity of~$\Pi$ is at least
\[
\left|\po\right|+\log\log\chi(\Gp)-\log\log\log\chi(\Gp)-4.
\]
\end{lemma}

\begin{example}
If $\po$ is the empty transcript, then the set $\Vp$ consists of
all functions from~$\B^{n}$ to~$\B$. In this case, it can be verified
that a sufficient condition for two functions~$g_{A},g_{B}\in\Vp$
to be neighbors in~$\Gp$ is that~$g_{A}^{-1}(\sigma)\cap g_{B}^{-1}(\tau)\ne\emptyset$
for every $\sigma,\tau\in\B$ (according to both \ref{characteristic-graph-composition,characteristic-graph-strong-composition}).
Therefore, the set of all balanced functions~$g\in\Vp$ such that
$g(\overline{1})=1$ is a clique in~$\Gp$. This clique has at least
$2^{2^{n}-O(n)}$~vertices, and therefore \ref{chromatic-number-composition,chromatic-number-strong-composition}
yield a lower bound of $n-O(\log n)$ in this case.
\end{example}

\begin{example}
Suppose that $\Pi$ is a standard protocol that solves the relation~$\KdMn$
(the case of $\KcMn$ is similar). Let $\po$ be a full transcript
of~$\Pi$. Then, $\po$~is labeled with an output that is either~$\bot$
or an entry~$(i,j)$. We consider the two cases separately:
\end{example}

\begin{itemize}
\item If $\po$ is labeled with $\bot$, then $\Vp=\emptyset$, and therefore
\ref{chromatic-number-composition,chromatic-number-strong-composition}
are inapplicable. The reason is that, in this case, for every function~$g:\B^{n}\to\B$,
either $\Xp(g)$ or~$\Yp(g)$ must be empty: otherwise, there would
have existed pairs $(g,X)\in\Xp$ and~$(g,Y)\in\Yp$, and the protocol
cannot output~$\bot$ when given these inputs.
\item Suppose that $\po$ is labeled with an entry~$(i,j)$. We claim that
in this case the graph $\Gp$ contains no edges. To see why, observe
that there must exist a bit~$\sigma\in\B$ such that for every $(g_{A},X)\in\Xp$
and $(g_{B},Y)\in\Yp$ it holds that $X_{i,j}=\sigma$ and $Y_{i,j}=1-\sigma$:
otherwise, there would have existed pairs $(g_{A},X)\in\Xp$ and~$(g_{B},Y)\in\Yp$
for which $X_{i,j}=Y_{i,j}$, and on those inputs the protocol cannot
output~$(i,j)$. Nevertheless, this implies that $\Xp(g_{A})\cap\Yp(g_{B})=\emptyset$
for every two functions $g_{A},g_{B}\in\Vp$, and therefore the graph~$\Gp$
does not contain any edges. It follows that $\chi(\Gp)=1$, and hence
\ref{chromatic-number-composition} does not give a meaningful lower
bound
\end{itemize}
In the rest of this section, we prove \ref{chromatic-number-composition,chromatic-number-strong-composition}.
Mihajlin and Smal~\cite{MS21} proved their result by reduction from
the co-non-deterministic communication complexity of the equality
function~$\eq$. We generalize their result by considering a certain
``graph equality'' problem, and using it as the starting point of
the reduction. We define the graph equality problem and lower bound
its co-non-deterministic complexity in \ref{subsec:graph-equality}.
Then, we prove the above lemmas by reducing from that problem in \ref{subsec:reduction-to-graph-equality}.

\subsection{\label{subsec:graph-equality}The graph equality problem}

De Wolf~\cite{dW_phd} studied the following promise version of the
equality problem.
\begin{definition}
Fix a graph~$G=(V,E)$. The \emph{graph equality problem of}~$G$,
denoted $\geq_{G}$, is the following promise problem: Alice and Bob
get as inputs vertices $v_{A},v_{B}\in V$ respectively. Their goal
is to output~$1$ if $v_{A}=v_{B}$, and $0$ if $v_{A}$ and~$v_{B}$
are neighbors. If neither is the case, then the output of Alice and
Bob may be either~$0$ or~$1$. We denote the negation of the problem
by $\gineq_{G}$.
\end{definition}

Ilango, Loff, and Oliveira~\cite{ILO20} proved the following result
on the non-deterministic complexity of~$\geq_{G}$.
\begin{proposition}[\cite{ILO20}]
\label{graph-equality-nc}For every graph~$G=(V,E)$, it holds that
$\NC(\geq_{G})=\log\chi(G)$.
\end{proposition}

Using the latter result, we deduce the following bounds on the co-non-deterministic
complexity of the graph equality problem.
\begin{proposition}
\label{graph-equality-cnc}For every graph~$G=(V,E)$, it holds that
\[
\log\log\chi(G)\le\NC(\gineq_{G})\le\log\log\chi(G)+1.
\]
\end{proposition}

\begin{myproof}
The lower bound follows from \ref{graph-equality-nc} by using the
fact that the non-deterministic communication complexity of a problem
is at most exponential in its co-non-deterministic communication complexity
(\ref{NCC-vs-coNCC}). We prove the upper bound by reduction from~$\gineq_{G}$
to $\ineq_{\left[\chi(G)\right]}$.

Consider the following non-deterministic protocol for~$\gineq_{G}$:
Suppose that Alice and Bob get vertices $v_{A}$ and~$v_{b}$ that
are neighbors in~$G$, and Merlin wants to convince them that $v_{A}\ne v_{B}$.
We fix in advance some optimal coloring of~$G$ that is known to
both players. Since $v_{A}$ and~$v_{B}$ are neighbors, they must
be colored with different colors. Merlin now proves to Alice and Bob
that the colors of~$v_{A}$ and~$v_{B}$ are different using an
optimal non-deterministic protocol for~$\ineq_{\left[\chi(G)\right]}$.
Clearly, if $v_{A}$ and~$v_{B}$ are neighbors then Merlin can always
convince Alice and Bob to accept, whereas if $v_{A}=v_{B}$ then Merlin
would never succeed in convincing them that $v_{A}$ and~$v_{B}$
have different colors. The complexity of this protocol is $\NC(\ineq_{\left[\chi(G)\right]})\le\log\log\chi(G)+1$.
\end{myproof}

\paragraph*{Related work.}

The graph equality problem was defined by de Wolf~\cite{dW_phd}
in the context of quantum fingerprinting. He showed that the one-round
deterministic communication complexity of this problem is $\log\chi(G)$.
This result was extended to the case of unbounded rounds by Bri{\"{e}}t
et al.~\cite{BBLPS15}, and to non-deterministic communication complexity
by Ilango et al.~\cite{ILO20}.

We note that the graph equality problem is also related to the works
of Alon and Orlitsky on dual-source coding~\cite{AO95,AO96}. The
results on the non-deterministic and co-non-deterministic complexities
of this problem are also similar in spirit to the work of Alon on
edge coloring~\cite{A87_LineGraphColoring}. A few additional similar
results are the characterization of the communication complexity of
the $\textsc{Exactly-N}$ problem in the Number-on-Forehead model
in terms of a certain chromatic number~\cite{CFL83}, a related observation
of~\cite{LY93}, and a variant of \ref{graph-equality-nc} for $3$-regular
hypergraphs~\cite[Thm. 9]{AB23}.

\subsection{\label{subsec:reduction-to-graph-equality}Proof of \ref{chromatic-number-composition,chromatic-number-strong-composition}}

We turn to proving \ref{chromatic-number-composition,chromatic-number-strong-composition}.
Our proof follows the main ideas of~\cite{MS21}, with the following
main difference: The reduction of~\cite{MS21} starts from the problem~$\ineq_{\mathcal{C}}$,
where $\mathcal{C}$~is some clique in the characteristic graph~$\Gp$.
Our reduction, on the other hand, starts from the problem~$\gineq_{\Gp}$.

As a warm-up, we first prove a simpler result that only holds for
standard protocols, rather than partially half-duplex ones.
\begin{proposition}
\label{chromatic-composition-classical}Let $\Pi$ be a (standard)
deterministic protocol that solves a relation $\KdMn$ for some $f:\B^{m}\to\B$
and $n\in\N$. For every transcript~$\pi_{1}$ of~$\Pi$, the complexity
of~$\Pi$ is at least $\left|\po\right|+\log\log\chi(\Gp)-2$.
\end{proposition}

\begin{myproof}
Let $\Pi,f,n$ be as in the proposition, and let $\po$ be a transcript
of~$\Pi$. Let $c$~be the maximal length of a string~$\pi_{2}$
such that the concatenation~$\pi_{1}\circ\pi_{2}$ is a transcript
of~$\Pi$. We construct a non-deterministic protocol for $\gineq_{\Gp}$
with complexity~$c+1$, and this will imply that 
\[
c+1\ge\cNC(\geq_{\Gp})\ge\log\log\chi(\Gp)-1.
\]
and this will yield the desired result. Consider the following protocol
for $\gineq_{\Gp}$: Suppose that Alice and Bob get as inputs~$g_{A},g_{B}\in\Vp$
respectively. Merlin would like to convince Alice and Bob that the
functions $g_{A},g_{B}$ are neighbors in~$\Gp$. If this is indeed
the case, then by the definition of~$\Gp$, it holds that either
$\Xp(g_{A})\cap\Yp(g_{B})\ne\emptyset$ or $\Xp(g_{B})\cap\Yp(g_{A})\ne\emptyset$.
Merlin begins by telling Alice and Bob which is the case, at the cost
of one bit. Without loss of generality, assume that we are in the
first case. 

Next, Merlin takes a matrix~$X\in\Xp(g_{A})\cap\Yp(g_{B})$, and
computes the transcript of~$\Pi$ when Alice and Bob get $(g_{A},X)$
and $(g_{B},X)$ respectively. This transcript must be of the form
$\po\circ\pi_{2}$ (by definition of $\Xp(g_{A})$ and $\Yp(g_{B})$).
Furthermore, the transcript~$\po\circ\pi_{2}$ must output~$\bot$,
as the protocol cannot find an entry~$(i,j)$ where the matrices
of Alice and Bob differ (since they are the same matrix). Merlin now
sends the string~$\pi_{2}$ to Alice and Bob as the rest of the witness.
Observe that the complexity of the protocol is~$c+1$.

When Alice receives such a witness from Merlin, she accepts if and
only if $\po\circ\pi_{2}$ is indeed a transcript of~$\Pi$ that
outputs~$\bot$, and there exists a matrix~$X$ such that the input
$(g_{A},X)$ is consistent with $\po\circ\pi_{2}$. Bob does the same
on his side. Clearly, if $g_{A}$ and~$g_{B}$ are indeed neighbors
in~$\Gp$, then Merlin can convince Alice and Bob that this is the
case. 

We turn to prove the soundness of the protocol. Suppose that Alice
and Bob accept some witness~$w$ from Merlin. We prove that $g_{A}\ne g_{B}$.
Let $\pi_{2}$~be the string contained in the witness~$w$. Then,
$\pi_{1}\circ\pi_{2}$ is a transcript of~$\Pi$ that outputs~$\bot$,
and there exist matrices~$X,Y\in\B^{m\times n}$ such that the inputs
$(g_{A},X)$ and $(g_{B},Y)$ are consistent with~$\pi_{1}\circ\pi_{2}$.
In particular, this implies that when given inputs $(g_{A},X)$ and
$(g_{B},Y)$, the protocol outputs~$\bot$. It follows that $g_{A}\ne g_{B}$,
since the protocol~$\Pi$ is only allowed to output~$\bot$ on the
inputs $(g_{A},X)$ and $(g_{B},Y)$ if $g_{A}\ne g_{B}$.
\end{myproof}
In order to prove \ref{chromatic-number-composition}, we would like
to apply a similar argument to partially half-duplex protocols. In
this setting, however, we encounter a new issue. Recall that in the
above proof, the transcript~$\pi_{2}$ was obtained by executing
the protocol~$\Pi$ on the inputs $(g_{A},X)$ and $(g_{B},X)$.
In the half-duplex setting, Alice and Bob do not share the same view
of the protocol, so there may not be a single transcript~$\pi_{2}$
that they both agree on. In principle, we could replace~$\pi_{2}$
in the above argument with two transcripts~$\pi_{2,A}$ and~$\pi_{2,B}$,
corresponding to the views of Alice and Bob respectively. Unfortunately,
that would cost us a factor of~$2$ in the lower bound, and we are
not willing to lose such a factor. 

\cite{MS21} resolve this issue using the following nice observation:
the viewpoints of Alice and Bob on~$\pi_{2}$ diverge only if non-classical
rounds took place in the execution of the protocol. Nevertheless,
since the protocol is \emph{partially} half-duplex, such non-classical
rounds can take place only if~$g_{A}\ne g_{B}$. Hence, the presence
of non-classical rounds is, on its own, a convincing proof that $g_{A}\ne g_{B}$.
This leads to a natural strategy for Merlin: if the execution of the
protocol on $(g_{A},X)$ and $(g_{B},X)$ encountered non-classical
rounds, use them as a proof that $g_{A}\ne g_{B}$, and otherwise,
send the transcript~$\pi_{2}$ as in the proof \ref{chromatic-composition-classical}.
We now provide the formal proof.
\begin{restated}{\ref{chromatic-number-composition}}
Let $\Pi$ be a partially half-duplex protocol that solves a relation
$\KdMn$ for some $f:\B^{m}\to\B$ and $n\in\N$. For every transcript~$\po$
of~$\Pi$, the complexity of~$\Pi$ is at least
\[
\left|\po\right|+\log\log\chi(\Gp)-\log\log\log\chi(\Gp)-4.
\]
\end{restated}

\begin{myproof}
Let $\Pi,f,n$ be as in the lemma, and let $\Pi_{A}$ and $\Pi_{B}$
be the protocol trees that are associated with~$\Pi$ according to
\ref{half-duplex-protocol}. Let $\po$ be a transcript of~$\Pi$.
Observe that since $\po$ is a transcript of~$\Pi$, there exists
at least one pair of vertices $(u,v)$ of~$\Pi_{A}$ and~$\Pi_{B}$
that are consistent with~$\pi_{1}$. Let $c$~be the maximum, over
all such pairs, of the number of rounds that the protocol may execute
after reaching~$(u,v)$. We prove that 
\[
c\ge\log\log\chi(\Gp)-\log\log\log\chi(\Gp)-4,
\]
and this will yield the desired result. If $c\ge\log\log\chi(\Gp)$,
then we are done. Suppose otherwise. We construct a non-deterministic
protocol for $\gineq_{\Gp}$ with complexity at most~$c+\log c+3$.
This will imply that 
\[
c+\log c+3\ge\NC(\gineq_{\Gp})\ge\log\log\chi(\Gp)-1,
\]
and hence
\[
c\ge\log\log\chi(\Gp)-\log c-4\ge\log\log\chi(\Gp)-\log\log\log\chi(\Gp)-4,
\]
as required.

Consider the following protocol: Suppose that Alice and Bob get as
inputs~$g_{A},g_{B}\in\Vp$ respectively. Merlin would like to convince
Alice and Bob that the functions $g_{A},g_{B}$ are neighbors in~$\Gp$.
If this is indeed the case, then by the definition of~$\Gp$, it
holds that either $\Xp(g_{A})\cap\Yp(g_{B})\ne\emptyset$ or $\Xp(g_{B})\cap\Yp(g_{A})\ne\emptyset$.
Merlin begins by telling Alice and Bob which is the case, at the cost
of one bit. Without loss of generality, assume that we are in the
first case.

Let $X$ be a matrix in the intersection~$\Xp(g_{A})\cap\Yp(g_{B})$.
By definition, the input $(g_{A},X)$ of Alice is consistent with~$\po$,
and therefore there exists a (unique) vertex~$u$ of $\Pi_{A}$ that
is consistent with~$\pi_{1}$ such that $(g_{A},X)\in\cX_{u}$ (see
\ref{half-duplex-transcript-vs-vertex}). Similarly, there exists
a (unique) vertex~$v$ of $\Pi_{B}$ that is consistent with~$\pi_{1}$
such that $(g_{B},X)\in\cY_{v}$. Merlin continues to execute the
protocol~$\Pi$ on inputs $(g_{A},X)$ and~$(g_{B},X)$ starting
from the pair of vertices~$(u,v)$. Note that this execution performs
at most $c$~rounds. There are now two cases: either this execution
encountered a non-classical round, or not. Merlin sends Alice and
Bob an additional bit saying which is the case.

If we are in the first case, Merlin sends to Alice and Bob the following
information:
\begin{itemize}
\item The index~$i$ of the first non-classical round.
\item The transcript~$\pi_{2}$ of the first $i-1$~rounds. Note that
since these rounds are classical, there is a single transcript that
describes both Alice's and Bob's viewpoints.
\item A bit saying whether the $i$-th round is a wasted or a silent round.
\end{itemize}
Note that this information can be encoded using at most $\left\lceil \log c\right\rceil +(c-1)+1$
bits. Together with the two bits that were used to tell Alice and
Bob the cases in which we are, the complexity of the protocol in this
case is at most $c+\log c+3$.

When Alice receives such a witness from Merlin, she accepts if and
only if there exists a matrix~$X$ such that the following holds:
Let $u$~be the unique vertex of~$\Pi_{A}$ that is determined by~$(g_{A},X)$
and~$\po$. Then, the input~$(g_{A},X)$ is required to be consistent
with the transcript~$\pi_{2}$ in the first $i-1$ rounds when starting
from~$u$. Furthermore, if Merlin claimed that the $i$-th round
is wasted (respectively, silent), then Alice checks that she chooses
to send (respectively, receive) at the $i$-th round when given the
input~$(g_{A},X)$ and having seen the transcript~$\pi_{2}$ after
starting from~$u$. Similarly, Bob accepts if and only if there exists
a matrix~$Y$ that satisfies the same conditions while exchanging
$(g_{A},X)$ and $\Pi_{A}$ with $(g_{B},Y)$ and~$\Pi_{B}$. The
completeness of the protocol in this case is obvious. For the soundness,
observe that if $g_{A}=g_{B}$, then all the rounds of the protocol
are classical regardless of the choice of~$X,Y$ (since it is \emph{partially}
half-duplex), and therefore there is no convincing transcript that
Merlin may send.

If we are in the second case (i.e., the execution did not encounter
a non-classical round), then Merlin proceeds as in the proof of \ref{chromatic-composition-classical}:
Merlin sends the (classical) transcript~$\pi_{2}$ of the execution
to Alice and Bob. The players accepts if and only if the transcript~$\po\circ\pi_{2}$
outputs~$\bot$ and there are matrices~$X,Y$ such that $(g_{A},X)$
and~$(g_{B},Y)$ are consistent with~$\pi_{1}\circ\pi_{2}$. The
analysis of the completeness and the soundness is the same as before,
and the complexity in this case is at most~$c+2$.
\end{myproof}
We turn to proving \ref{chromatic-number-strong-composition}, which
is the analogue of \ref{chromatic-number-composition} for strong
composition. Here, we use exactly the same argument, where the only
difference is that the matrix~$X$ is replaced with two matrices~$X,Y$
that witness the weak intersection property. For completeness, we
provide the full proof below.
\begin{restated}{\ref{chromatic-number-strong-composition}}
Let $\Pi$ be a partially half-duplex protocol that solves a relation
$\KcMn$ for some $f:\B^{m}\to\B$ and $n\in\N$. For every transcript~$\po$
of~$\Pi$, the complexity of~$\Pi$ is at least
\[
\left|\po\right|+\log\log\chi(\Gp)-\log\log\log\chi(\Gp)-4.
\]
\end{restated}

\begin{myproof}
Let $\Pi,f,n$ be as in the lemma, and let $\Pi_{A}$ and $\Pi_{B}$
be the protocol trees that are associated with~$\Pi$ according to
\ref{half-duplex-protocol}. Let $\po$ be a transcript of~$\Pi$.
Observe that since $\po$ is a transcript of~$\Pi$, there exists
at least one pair of vertices $(u,v)$ of~$\Pi_{A}$ and~$\Pi_{B}$
that are consistent with~$\pi_{1}$. Let $c$~be the maximum, over
all such pairs, of the number of rounds that the protocol may execute
after reaching~$(u,v)$. We prove that 
\[
c\ge\log\log\chi(\Gp)-\log\log\log\chi(\Gp)-4,
\]
and this will yield the required result. If $c\ge\log\log\chi(\Gp)$,
then we are done. Suppose otherwise. We construct a non-deterministic
protocol for $\gineq_{\Gp}$ with complexity at most~$c+\log c+3$.
This will imply that 
\[
c+\log c+3\ge\NC(\gineq_{\Gp})\ge\log\log\chi(\Gp)-1,
\]
and hence
\[
c\ge\log\log\chi(\Gp)-\log c-4\ge\log\log\chi(\Gp)-\log\log\log\chi(\Gp)-4,
\]
as required.

Consider the following protocol: Suppose that Alice and Bob get as
inputs~$g_{A},g_{B}\in\Vp$ respectively. Merlin would like to convince
Alice and Bob that the functions $g_{A},g_{B}$ are neighbors in~$\Gp$.
If this is indeed the case, then by the definition of~$\Gp$, one
of the following cases holds:
\begin{itemize}
\item There exist matrices $X\in\cX_{\pi_{1}}(g_{A})$ and $Y\in\cY_{\pi_{1}}(g_{B})$
such that $X_{i}=Y_{i}$ for every $i\in\left[m\right]$ for which
$a_{i}\ne b_{i}$ (where $a=g_{A}(X)$ and~$b=g_{B}(Y)$).
\item The same statement holds when exchanging $g_{A}$ and~$g_{B}$.
\end{itemize}
Merlin begins by telling Alice and Bob which is the case, at the cost
of one bit. Without loss of generality, assume that we are in the
first case.

Let $X,Y,a,b$ be matrices and strings as in the first case above.
By definition, the input $(g_{A},X)$ of Alice is consistent with~$\po$,
and therefore there exists a (unique) vertex~$u$ of $\Pi_{A}$ that
is consistent with~$\pi_{1}$ such that $(g_{A},X)\in\cX_{u}$ (see
\ref{half-duplex-transcript-vs-vertex}). Similarly, there exists
a (unique) vertex~$v$ of $\Pi_{B}$ that is consistent with~$\pi_{1}$
such that $(g_{B},Y)\in\cY_{v}$. Merlin continues to execute the
protocol~$\Pi$ on inputs $(g_{A},X)$ and~$(g_{B},Y)$ starting
from the pair of vertices~$(u,v)$. Note that this execution performs
at most $c$~rounds. There are now two cases: either this execution
encountered a non-classical round, or not. Merlin sends Alice and
Bob an additional bit saying which is the case.

If we are in the first case, Merlin sends to Alice and Bob the following
information:
\begin{itemize}
\item The index~$i$ of the first non-classical round.
\item The transcript~$\pi_{2}$ of the first $i-1$~rounds. Note that
since these rounds are classical, there is a single transcript that
describes both Alice's and Bob's viewpoints.
\item A bit saying whether the $i$-th round is a wasted or a silent round.
\end{itemize}
Note that this information can be encoded using at most $\left\lceil \log c\right\rceil +(c-1)+1$
bits. Together with the two bits that were used to tell Alice and
Bob the cases in which we are, the complexity of the protocol in this
case is at most $c+\log c+3$.

When Alice receives such a witness from Merlin, she accepts if and
only if there exists a matrix~$X$ such that the following holds:
Let $u$~be the vertex of~$\Pi_{A}$ that is determined by~$(g_{A},X)$
and~$\po$. Then, the input~$(g_{A},X)$ is required to be consistent
with the transcript~$\pi_{2}$ in the first $i-1$ rounds when starting
from~$u$. Furthermore, if Merlin claimed that the $i$-th round
is wasted (respectively, silent), then Alice checks that she chooses
to send (respectively, receive) at the $i$-th round, when given the
input~$(g_{A},X)$ and having seen the transcript~$\pi_{2}$ after
starting from~$u$. Similarly, Bob accepts if and only if there exists
a matrix~$Y$ that satisfies the same conditions while exchanging
$(g_{A},X)$ and $\Pi_{A}$ with $(g_{B},Y)$ and~$\Pi_{B}$. The
completeness of the protocol in this case is obvious. For the soundness,
observe that if $g_{A}=g_{B}$, then all the rounds of the protocol
are classical regardless of the choice of~$X,Y$ (since it is \emph{partially}
half-duplex), and therefore there is no convincing transcript that
Merlin may send.

Suppose that we are in the second case, i.e., the execution did not
encounter a non-classical round. In this case, the transcript $\pi_{1}\circ\pi_{2}$
must output~$\bot$, as the protocol cannot find an entry~$(i,j)$
where $a_{i}\ne b_{i}$ and~$X_{i,j}\ne Y_{i,j}$ (by definition
of $X,Y,a,b$). Merlin now sends~$\pi_{2}$ to Alice and Bob as the
rest of the witness. Observe that the complexity of the protocol in
this case is~$c+1$.

When Alice receives such a witness from Merlin, she accepts if and
only if $\po\circ\pi_{2}$ is indeed a transcript of~$\Pi$ that
outputs~$\bot$, and there exists a matrix~$X$ such that the input
$(g_{A},X)$ is consistent with $\po\circ\pi_{2}$. Bob does the same
on his side. Again, the completeness of the protocol in this case
is obvious.

We turn to prove the soundness of the protocol. Suppose that Alice
and Bob accept some witness~$w$ from Merlin. We prove that $g_{A}\ne g_{B}$.
Let $\pi_{2}$~be the string contained in the witness~$w$. c Then,
$\pi_{1}\circ\pi_{2}$ is a transcript of~$\Pi$ that outputs~$\bot$,
and there exist matrices~$X,Y\in\B^{m\times n}$ such that the inputs
$(g_{A},X)$ and $(g_{B},Y)$ are consistent with~$\pi_{1}\circ\pi_{2}$.
In particular, this implies that when given inputs $(g_{A},X)$ and
$(g_{B},Y)$, the protocol may transmit the transcript $\po\circ\pi_{2}$,
and hence may output~$\bot$. It follows that $g_{A}\ne g_{B}$,
as required, since the protocol~$\Pi$ is only allowed to output~$\bot$
on the inputs $(g_{A},X)$ and $(g_{B},Y)$ if $g_{A}\ne g_{B}$.
\end{myproof}

\section{\label{sec:Prefix-thick-sets}Prefix-thick sets}

In this section, we introduce the notion of prefix-thick sets, and
show how to lower bound their number using a result of Salo and \torma~\cite{ST14}.
Let $\Sigma$ be a finite alphabet of size~$q$. We start by recalling
the definition of a prefix tree.
\begin{definition}
\label{prefix-tree}Let $\cX\subseteq\Sigma^{m}$. The \emph{prefix
tree} of~$\cX$ is a rooted tree~$T_{\cX}$ of depth~$m$ that
satisfies the following properties:
\begin{itemize}
\item The vertices of~$T_{\cX}$ at depth~$i$ are all the length-$i$
prefixes of strings in~$\cX$. In particular, the root of~$T$ is
the empty string, and its leaves are the strings in~$\cX$.
\item A vertex $y\in\Sigma^{i+1}$ at depth~$i+1$ is a child of a vertex
$x\in\Sigma^{i}$ at depth~$i$ if and only if $x$~is a prefix
of~$y$. In this case, we label the edge from~$x$ to~$y$ with
the symbol~$\sigma\in\Sigma$ that satisfies $y=x\circ\sigma$ (i.e.,
$\sigma=y_{i+1}$), and say that $y$~is the $\sigma$\emph{-child}
of~$x$.
\end{itemize}
\end{definition}

\begin{definition}
\label{prefix-thick}Let $\cX\subseteq\Sigma^{m}$ be a set of strings,
and let $t\in\R$. We say that $\cX$ is \emph{prefix thick with degree~$t$}
if there is a subset~$\cX'\subseteq\cX$ whose prefix tree has minimum
degree that is greater than~$t$. When $\Sigma$ is clear from the
context, we abbreviate and say that $\cX$ is \emph{prefix thick}
if it is prefix thick with degree~$q/2$.
\end{definition}

As discussed in the introduction, our motivation for using prefix-thick
sets is the following easy observation.
\begin{restated}{\ref{prefix-thick-sets-intersect}}
Let $\cX,\cY\subseteq\Sigma^{m}$. If $\cX$ and $\cY$ are both
prefix thick, then~$\cX\cap\cY\ne\emptyset$.
\end{restated}

\begin{myproof}
Let $\cX,\cY\subseteq\Sigma^{m}$ be as in the proposition. Assume
that $\cX$ and~$\cY$ are both prefix thick with degree~$q/2$,
and let $\cX'$ and~$\cY'$ be the corresponding subsets of~$\cX$
and~$\cY$ respectively. Let $T_{\cX'}$ and~$T_{\cY'}$ be the
prefix trees of~$\cX'$ and~$\cY'$ respectively, and note that
the degree of every vertex in these trees is at most~$q$. Since
the degrees of the roots in both $T_{\cX'}$ and~$T_{\cY'}$ is greater
than~$q/2$, there must be some vertex~$\sigma_{1}\in\Sigma$ that
is a child of the root in both trees. Similarly, since the degree
of~$\sigma_{1}$ is greater than~$q/2$ in both $T_{\cX'}$ and~$T_{\cY'}$,
there must be some vertex~$(\sigma_{1},\sigma_{2})\in\Sigma^{2}$
that is a child of~$\sigma_{1}$ in both trees. Continuing in this
fashion, we obtain a leaf $(\sigma_{1},\ldots.,\sigma_{m})\in\Sigma^{m}$
that belongs to both trees, and this leaf is a string in the intersection~$\cX\cap\cY$. 
\end{myproof}
\begin{remark}
\label{prefix-thick-different-alphabets}In \ref{sec:Structure-Theorem},
we use the following generalization of the above definitions: we consider
$m$~different alphabets~$\Sigma_{1},\ldots,\Sigma_{m}$ of the
same size~$q$, and work with strings in~$\Sigma_{1}\times\cdots\times\Sigma_{m}$.
It is not hard to check that all the definitions and results in this
section work in this setting without a change.

In fact, we could even work with $m$~different alphabets of \emph{different}
sizes. The results in this section would continue to hold in such
a case, but with a somewhat more cumbersome phrasing. For the simplicity
of the presentation, we decided to skip this more general version.\medskip{}
\end{remark}

\noindent Given a set~$\cX\subseteq\Sigma^{m}$, we would like to
lower bound the number of coordinate sets~$I\subseteq\left[m\right]$
for which~$\cX|_{I}$ is prefix thick. To this end, we use a result
of Salo and \torma~\cite{ST14} (see also~\cite{S21}). We first
introduce a few additional terms.
\begin{definition}[{\cite[Def. 5.6]{ST14}}]
Let $\cX\subseteq\Sigma^{m}$ be a set of strings and let $T_{\cX}$
be its prefix tree. We say that a vector~$w\in\left[q\right]^{m}$
is a \emph{branching structure} of~$\cX$ if $T_{\cX}$ has a sub-tree
of depth~$m$ in which, for every $i\in\left[m\right]$, the degree
of all the vertices at depth~$i$ is~$w_{i}$. The \emph{winning
set} of~$\cX$ is the set of all branching structures of~$\cX$.
\end{definition}

\begin{remark}
A ``winning set'' is called that way since \cite{ST14} present
their result in terms of winning possibilities in a certain game over
strings. We describe their result in terms of prefix trees since this
view is more useful for our purposes.
\end{remark}

Next, observe that $\cX|_{I}$ is prefix thick with degree~$t$ if
and only if there is a branching structure~$w$ of~$\cX$ such that
$w_{i}>t$ for every $i\in I$. Thus, we can obtain lower bounds on
the number of such sets~$I$ from the number of branching vectors.
The following result of~\cite{ST14} tells us exactly the size of
the winning set.
\begin{lemma}[{\label{ST-result}\cite[Prop. 5.7]{ST14}}]
\label{winning-set-size}Let $\cX\subseteq\Sigma^{m}$ be a set of
strings and let $\cW(\cX)$ be its winning set. Then, $\left|\cW(\cX)\right|=\left|\cX\right|$.
\end{lemma}

\begin{myproof}
We prove the proposition by induction on~$m$. For base case of $m=1$,
observe that for every $\cX\subseteq\Sigma$ it holds that $\cW(\cX)=\left\{ 1,\ldots,\left|\cX\right|\right\} $.
We assume that the lemma holds for~$m$, and prove it for $m+1$.
Let $\cX\subseteq\Sigma^{m+1}$ be a set of strings and let $\cW(\cX)\subseteq\left[q\right]^{m+1}$
be its winning set. For every symbol $\sigma\in\Sigma$, let 
\[
\cX_{\sigma}=\left\{ x\in\Sigma^{m}:\sigma\circ x\in\cX\right\} ,
\]
and let $\cW(\cX_{\sigma})\subseteq\left[q\right]^{m}$ be the winning
set of~$\cX_{\sigma}$. By the induction assumption, it holds that
$\left|\cW(\cX_{\sigma})\right|=\left|\cX_{\sigma}\right|$ for every
$\sigma\in\Sigma$. For every $k\in\left[q\right]$, let $\cW_{k}$
be the set of vectors $w\in\left[q\right]^{m}$ such that $k\circ w\in\cW(\cX)$,
and for every $w\in\left[q\right]^{m}$, let $c_{w}\in\left[q\right]$
denote the number of symbols~$\sigma\in\Sigma$ such that $w\in\cW(\cX_{\sigma})$.
The key observation is that for every $k\in\left[q\right]$ and $w\in\left[q\right]^{m}$,
it holds that $w\in\cW_{k}$ if and only if $k\le c_{w}$. We prove
this observation next:
\begin{itemize}
\item \textbf{The ``only if'' direction:} Assume that $w\in\cW_{k}$,
so $k\circ w\in\cW(x)$. This implies that the prefix tree~$T_{\cX}$
of~$\cX$ has a sub-tree~$T_{k\circ w}$ that corresponds to the
branching structure~$k\circ w$. In particular, the root of~$T_{k\circ w}$
has $k$~children. Let $\sigma_{1},\ldots,\sigma_{k}$ be the labels
of the outgoing of edges of the root of~$T_{k\circ w}$, and let
$T_{\sigma_{i}}$ be the sub-tree that is rooted at the $\sigma_{i}$-child
of the root. Then, each tree $T_{\sigma_{i}}$ is a sub-tree of the
prefix tree of~$\cX_{\sigma_{i}}$, and has branching structure~$w$.
Therefore, for each~$\sigma_{i}$, it holds that $w\in\cW(\cX_{\sigma_{i}})$,
so~$c_{w}\ge k$, as required.
\item \textbf{The ``if'' direction:} Assume that $k\le c_{w}$. Then,
there exist $k$~symbols~$\sigma_{1},\ldots,\sigma_{k}$ such that
$w\in\cW(\cX_{\sigma_{i}})$ for each~$i\in\left[k\right]$. Therefore,
for each symbol~$\sigma_{i}$, the prefix tree of~$\cX_{\sigma_{i}}$
has a sub-tree $T_{\sigma_{i}}$ that corresponds to the branching
structure~$w$. Now, let $T_{k\circ w}$ be the prefix tree in which
the children of the root are the symbols~$\sigma_{1},\ldots,\sigma_{k}$,
and the sub-tree that is rooted at~$\sigma_{i}$ is~$T_{\sigma_{i}}$.
It is not hard to see that $T_{k\circ w}$ is a sub-tree of the prefix
tree~$T_{\cX}$ of~$\cX$, and has branching structure $k\circ w$.
Hence, $k\circ w\in\cW(\cX)$, or alternatively $w\in\cW_{k}$.
\end{itemize}
Given the above key observation, we prove the lemma as follows:
\begin{align*}
\left|\cW(\cX)\right| & =\sum_{k=1}^{q}\left|\cW_{k}\right|\\
 & =\sum_{k=1}^{q}\sum_{w\in\cW_{k}}1\\
 & =\sum_{w\in\left[q\right]^{m}}\sum_{k\in\left[q\right]:w\in\cW_{k}}1\\
 & =\sum_{w\in\left[q\right]^{m}}\sum_{k=1}^{c_{w}}1 & \text{(the key observation)}\\
 & =\sum_{w\in\left[q\right]^{m}}c_{w}\\
 & =\sum_{w\in\left[q\right]^{m}}\sum_{\sigma\in\Sigma:w\in\cW(\cX_{\sigma})}1 & \text{(definition of \ensuremath{c_{w}})}\\
 & =\sum_{\sigma\in\Sigma}\sum_{w\in\cW(\cX_{\sigma})}1\\
 & =\sum_{\sigma\in\Sigma}\left|\cW(\cX_{\sigma})\right|\\
 & =\sum_{\sigma\in\Sigma}\left|\cX_{\sigma}\right| & \text{(induction hypothesis)}\\
 & =\left|\cX\right|,
\end{align*}
as required.
\end{myproof}
\noindent We now use \ref{winning-set-size} to lower bound the number
of subsets for which $\cX|_{I}$ is prefix thick. Specifically, the
following result says that the fraction of such sets~$I$ is not
much smaller than the density of~$\cX$ inside~$\Sigma^{m}$.
\begin{lemma}
\label{prefix-thick-lemma}Let $\cX\subseteq\Sigma^{m}$, let $\varepsilon\ge0$,
and let $\cF$ be the family of subsets~$I\subseteq\left[m\right]$
such that $\cX|_{I}$ is prefix thick with degree~$(\frac{1}{2}+\varepsilon)\cdot q$.
Then,
\[
\frac{\left|\cF\right|}{2^{m}}\ge2^{-2\log e\cdot\varepsilon\cdot m}\cdot\frac{\left|\cX\right|}{\left|\Sigma^{m}\right|}.
\]
\end{lemma}

\begin{myproof}
Let $\cX$, $\varepsilon$, and~$\cF$ be as in \ref{prefix-thick-lemma},
and let $\cW$ be the winning set of~$\cX$. Let $\phi:\cW\to2^{\left[m\right]}$
be the mapping that maps every branching structure $w\in\cW$ to the
set $I\subseteq\left[m\right]$ of coordinates~$i$ such that $w_{i}>\left(\frac{1}{2}+\varepsilon\right)\cdot q$.
Now, observe that for every~$I\subseteq\left[m\right]$, the preimage
$\phi^{-1}(I)$ is contained in the set of all vectors in~$\left[q\right]^{m}$
whose entries that are greater than~$\left(\frac{1}{2}+\varepsilon\right)\cdot q$
are exactly those in~$I$. Hence, for every $I\subseteq\left[m\right]$,
it holds that
\begin{align*}
\left|\phi^{-1}(I)\right| & =\left[q-\left\lfloor (\frac{1}{2}+\varepsilon)\cdot q\right\rfloor \right]^{\left|I\right|}\left\lfloor (\frac{1}{2}+\varepsilon)\cdot q\right\rfloor ^{m-\left|I\right|}\\
 & \le\left\lfloor (\frac{1}{2}+\varepsilon)\cdot q\right\rfloor ^{m}\\
 & \le\left((\frac{1}{2}+\varepsilon)\cdot q\right)^{m}\\
 & =\left(1+2\varepsilon\right)^{m}\cdot\frac{q^{m}}{2^{m}}\\
 & \le e^{2\varepsilon m}\cdot\frac{q^{m}}{2^{m}} & \text{(since \ensuremath{1+x\le e^{x}})}.
\end{align*}
Now, by \ref{winning-set-size}, it follows that
\begin{align*}
\left|\cX\right| & =\left|\cW\right|\\
 & =\sum_{I\in\cF}\left|\phi^{-1}(I)\right|\\
 & \le\sum_{I\in\cF}\frac{q^{m}}{2^{m}}\cdot e^{2\varepsilon m}\\
 & =\frac{q^{m}}{2^{m}}\cdot2^{2\log e\cdot\varepsilon\cdot m}\cdot\left|\cF\right|.
\end{align*}
The lemma follows from by dividing both sides by $2^{2\log e\cdot\varepsilon\cdot m}\cdot q^{m}$.
\end{myproof}
\begin{remark}
We note that the last proof was suggested to us by Ville Salo (over
MathOverflow), although we also came up with it independently.
\end{remark}

\begin{remark}
\ref{prefix-thick-lemma} can be viewed as a generalization of the
Sauer-Shelah lemma~\cite{Sauer72,Shelah72}, and in particular of
its strengthening due to Pajor~\cite{Pajor85}. Specifically, if
$\Sigma=\B$, then $\cX|_{I}$ is a prefix-thick set if and only if
$I$ is shattered by~$\cX$. If we substitute $q=2$ and~$\varepsilon=0$
in \ref{prefix-thick-lemma}, we get that the number of shattered
sets~$\left|\cF\right|$ is at least~$\left|\cX\right|$, as shown
by \cite{Pajor85}. Furthermore, \ref{winning-set-size} of~\cite{ST14}
itself is a generalization of a result of~\cite{ARS02} on ordered-shattering
sets, which is another strengthening of~\cite{Pajor85}.
\end{remark}

\section{\label{sec:Structure-Theorem}Proof of the structure theorem}

In this section, we prove our structure theorem, restated next.
\begin{restated}{\ref{structure-theorem}}
There exists a constant~$\gamma>0.04$ such that the following holds:
Let $f:\B^{m}\to\B$, let $n\in\N$, and let $\Pi$ be a partially
half-duplex protocol that solves~$\KcMn$. For every $\gamma$-live
transcript $\po$ of~$\Pi$, the complexity of~$\Pi$ is at least
\[
\left|\po\right|+n-O(\log(m\cdot n)).
\]
\end{restated}

\noindent Let $\gamma>0$ be a universal constant that will be fixed
later to a value greater than~$0.04$. Let $f,n,\Pi$ be as in the
structure theorem, and $\po$ be a $\gamma$-live transcript~$\po$
of~$\Pi$, and let $\Gp$~be the characteristic graph of~$\po$
as in \ref{characteristic-graph-strong-composition}. By \ref{chromatic-number-strong-composition},
in order to prove the desired lower bound, it is sufficient to prove
that 
\[
\log\log\chi(\Gp)\ge n-O(\log(m\cdot n)).
\]
We prove the latter bound by proving an upper bound on the independence
number of a sub-graph~$\cG'$ of~$\Gp$. The following two lemmas
construct that sub-graph and bound its independence number respectively.
\begin{lemma}
\label{construction-of-G'}There exist a universal constant $\varepsilon>0$,
a set of balanced functions~$\cV'\subseteq\cV_{0}$, a subset $I\subseteq\left[m\right]$,
and strings $a\in f^{-1}(1)$ and~$b\in f^{-1}(0)$, that satisfy
the following properties:
\begin{itemize}
\item $\left|\cV'\right|\ge2^{-4m}\cdot\left|\cV_{0}\right|$.
\item $a|_{\left[m\right]-I}=b|_{\left[m\right]-I}$.
\item For every $g\in\cV'$, the sets $\Xp(g,a)|_{I}$ and $\Yp(g,a)|_{I}$
are prefix thick with degree~$\left(\frac{1}{2}+\varepsilon\right)\cdot2^{n-1}$.
\end{itemize}
\end{lemma}

\begin{lemma}
\label{independence-number-of-G'}Let $\varepsilon$ and $\cV'$ be
the constant and set from \ref{construction-of-G'}, and let $\cG'$
be the sub-graph of~$\Gp$ induced by~$\cV'$. Then, 
\[
\alpha(\cG')\le2^{-(\frac{\log e}{32}\cdot\varepsilon^{2}\cdot2^{n}-\frac{1}{2}\cdot m\cdot n-4\cdot m-1)}\cdot\left|\cV'\right|.
\]
\end{lemma}

\noindent We prove \ref{construction-of-G',independence-number-of-G'}
in \ref{subsec:Construction-of-G',subsec:Independence-number-of-G'}
respectively. We now derive the structure theorem from those lemmas.
\begin{myproof}[Proof of \ref{structure-theorem}.]
Let $\gamma,f,n,\Pi,\po$ be as above. Let $\varepsilon$ and~$\cV'$
be the constant and set whose existence is guaranteed by \ref{construction-of-G'},
and let $\cG'$ be the sub-graph of~$\Gp$ induced by~$\cV'$. By
\ref{independence-number-of-G'}, it holds that
\[
\chi(\cG')\ge\left|\cV'\right|/\alpha(\cG')\ge2^{\frac{\log e}{32}\cdot\varepsilon^{2}\cdot2^{n}-\frac{1}{2}\cdot m\cdot n-4\cdot m-1}.
\]
Clearly, $\chi(\Gp)\ge\chi(\cG')$, and therefore
\[
\log\log\chi(\Gp)\ge n+\log\left(\frac{\log e}{32}\cdot\varepsilon^{2}\right)-\log(\frac{1}{2}\cdot m\cdot n)-\log(4\cdot m)-1=n-O\left(\log(m\cdot n)\right).
\]
By \ref{chromatic-number-strong-composition}, it follows that the
communication complexity of the protocol~$\Pi$ is at least
\[
\left|\po\right|+\log\log\chi(\Gp)-\log\log\log\chi(\Gp)-4\ge\left|\po\right|+n-O\left(\log(m\cdot n)\right),
\]
as required.
\end{myproof}

\subsection{\label{subsec:Construction-of-G'}The construction of~$\protect\cG'$}

In this section, we prove \ref{construction-of-G'}. Recall that $\po$~is
$\gamma$-alive, and hence there exists a set of balanced functions~$\cV\subseteq\Vp$
that satisfies the following conditions:
\begin{itemize}
\item $\left|\cV\right|\ge2^{-m}\cdot\left|\Vb\right|$.
\item For every~$g\in\cV$, it holds that $\log\L\left(\Ap(g)\times\Bp(g)\right)\ge(1-\gamma)\cdot m+\kappa\log m+\kappa$.
\item For every $g\in\cV$, $a\in\Ap(g)$, and $b\in\Bp(g)$, it holds that
$\left|\Xp(g,a)\right|\ge2^{-\gamma\cdot m+1}\cdot\left|g^{-1}(a)\right|$
and $\left|\Yp(g,b)\right|\ge2^{-\gamma\cdot m+1}\cdot\left|g^{-1}(b)\right|$.
\end{itemize}
The crux of our proof is showing that for every~$g\in\cV$, there
exists a subset~$I_{g}\subseteq\left[m\right]$ and strings~$a_{g}\in f^{-1}(1)$
and~$b_{g}\in f^{-1}(0)$ that satisfy the following properties:
\begin{itemize}
\item $a_{g}|_{\left[m\right]-I_{g}}=b_{g}|_{\left[m\right]-I_{g}}$.
\item The sets $\Xp(g,a_{g})|_{I_{g}}$ and $\Yp(g,a_{g})|_{I_{g}}$ are
prefix thick with degree~$\left(\frac{1}{2}+\varepsilon\right)\cdot2^{n-1}$.
\end{itemize}
Having shown that, we will choose the triplet~$(I,a,b)$ to be the
most popular triplet among all triplets~$(I_{g},a_{g},b_{g})$ for
all~$g\in\cV$, and choose~$\cV'$ to be the set of those functions~$g\in\cV$
whose corresponding triplet is~$(I,a,b)$. By an averaging argument,
it will follow that 
\[
\left|\cV'\right|\ge2^{-3m}\cdot\left|\cV\right|\ge2^{-4m}\cdot\left|\Vb\right|,
\]
thus completing the proof.

For the rest of this section, we focus on proving the existence of
such~$I_{g},a_{g},b_{g}$ for every~$g\in\Vp$. Fix a function~$g\in\Vp$,
and let $0<\beta<\frac{1}{2}$ be some universal constant to be fixed
later. For convenience, we abbreviate $\cA=\Ap(g)$, $\cB=\Bp(g)$,
$\cX(a)=\Xp(g,a)$, and $\cY(b)=\cY(g,b)$. The high-level idea of
our construction of $I_{g},a_{g},b_{g}$ is the following:
\begin{enumerate}
\item \label{enu:choosing-subset-of-A}For every $a\in\cA$, the set~$\cX(a)$
is large (since $\po$ is alive), and hence there are many subsets~$I\subseteq\left[m\right]$
such that $\cX(a)|_{I}$ is prefix thick by \ref{prefix-thick-lemma}.
\item In particular, for every~$a\in\cA$ many of those subsets~$I$ are
large (this follows by a standard concentration bound).
\item Hence, by an averaging argument, there exists a single large set~$I_{1}\subseteq\left[m\right]$
such that~$\cX(a)|_{I_{1}}$ is prefix thick for many strings~$a\in\cA$.
We denote the set of those strings by~$\cA_{1}$.
\item \label{enu:choosing-subset-of-B}Repeating the same argument for~$b\in\cB$
and~$\cY(b)$, but \emph{this time restricting ourselves to coordinates
in~$I_{1}$}, we get that there exists a large set~$I_{2}\subseteq I_{1}$
such that $\cY(b)|_{I_{2}}$ is prefix thick for many strings~$b\in\cB$.
We denote the set of those strings by~$\cB_{1}$.
\item Note that it also holds that $\cX(a)|_{I_{2}}$ is prefix thick for
every $a\in\cA_{1}$ (since $I_{2}\subseteq I_{1}$). We therefore
choose $I_{g}=I_{2}$.
\item \label{enu:dense-sets-formula-complexity-step}Recall that the formula
complexity~$\L(\cA\times\cB)$ is large (since $\po$ is alive).
Using the fact that the sets~$\cA_{1}$ and~$\cB_{1}$ are dense
within $\cA$ and~$\cB$ respectively, we deduce that the formula
complexity~$\L(\cA_{1}\times\cB_{1})$ is large too (this deduction
is non-trivial, as we explain below).
\item Finally, since $I_{g}$ is large and~$\L(\cA_{1}\times\cB_{1})$
is large, we claim that there must exist strings~$a_{g}\in\cA_{1}$
and~$b_{g}\in\cB_{1}$ such that $a_{g}|_{\left[m\right]-I_{g}}=b_{g}|_{\left[m\right]-I_{g}}$:
otherwise, Alice and Bob could have solved $\KW_{\cA_{1}\times\cB_{1}}$
too efficiently by sending their bits in the coordinates of~$\left[m\right]-I_{g}$.
\end{enumerate}
There is one issue in the argument as described above. In Step~\ref{enu:dense-sets-formula-complexity-step},
we would like to deduce that the complexity $\L(\cA_{1}\times\cB_{1})$
is large using the fact that the sets~$\cA_{1}$ and~$\cB_{1}$
are large. Nevertheless, such a deduction is false in general: it
is easy to construct examples of sets~$\cA_{1}$ and~$\cB_{1}$
that are dense in~$\cA$ and~$\cB$ but for which $\L(\cA_{1}\times\cB_{1})$
is small. In order to remedy this issue, we apply the fortification
theorem of~\cite{DM16} (see \ref{foritification}) on Alice's and
Bob's sides before Steps~\ref{enu:choosing-subset-of-A} and~\ref{enu:choosing-subset-of-B}
respectively. This theorem guarantees that the largeness of~$\cA_{1}$
and~$\cB_{1}$ will imply that the formula complexity~$\L(\cA_{1}\times\cB_{1})$
is large, as required by Step~\ref{enu:dense-sets-formula-complexity-step}.
We now execute the above argument formally.

First, we apply \ref{foritification} (the fortification theorem)
to the rectangle $\cA\times\cB$ on Alice's side. This yields a subset
$\cA_{0}\subseteq\cA$ such that the rectangle~$\cA_{0}\times\cB$
is $\frac{1}{4m}$-fortified on Alice's side and satisfies 
\[
\L(\cA_{0}\times\cB)\ge\frac{1}{4}\cdot\L(\cA\times\cB).
\]

Second, let $a\in\cA_{0}$. Since $\po$ is $\gamma$-alive, it holds
that $\left|\cX(a)\right|\ge2\cdot2^{-\gamma\cdot m}\cdot\left|g^{-1}(a)\right|$.
We now apply \ref{prefix-thick-lemma} in order to find sets~$I$
for which~$\cX(a)|_{I}$ is prefix thick. To this end, we view $\cX(a)$
as a subset of strings in $g^{-1}(a)=g^{-1}(a_{1})\times\cdots\times g^{-1}(a_{m})$,
where $g^{-1}(a_{1}),\ldots,g^{-1}(a_{m})$ are alphabets of size~$2^{n-1}$
since $g$~is balanced. Let $\cF_{a}$ denote the family of subsets~$I\subseteq\left[m\right]$
such that $\cX(a)|_{I}$ is prefix-thick with degree~$\left(\frac{1}{2}+\varepsilon\right)\cdot2^{n-1}$.
By \ref{prefix-thick-lemma}, it holds that $\left|\cF_{a}\right|\ge2\cdot2^{-(\gamma+2\log e\cdot\varepsilon)\cdot m}\cdot2^{m}$.

Next, for every string~$a\in\cA_{0}$, let $\cF_{a}'$ denote the
family of subsets~$I\in\cF_{a}$ of size at least $(\frac{1}{2}-\beta)\cdot m$.
By \ref{random-set-size}, the number of subsets $I\subseteq\left[m\right]$
whose size is less than $(\frac{1}{2}-\beta)\cdot m$ is at most $2^{-2\log e\cdot\beta^{2}\cdot m}\cdot2^{m}$,
and therefore
\[
\left|\cF_{a}'\right|\ge\left|\cF_{a}\right|-2^{-2\log e\cdot\beta^{2}\cdot m}\cdot2^{m}\ge\left(2\cdot2^{-(\gamma+2\log e\cdot\varepsilon)\cdot m}-2^{-2\log e\cdot\beta^{2}\cdot m}\right)\cdot2^{m}
\]
We will later choose the constants~$\beta,\gamma,\varepsilon$ such
that they satisfy that $\gamma+2\log e\cdot\varepsilon\le2\log e\cdot\beta^{2}$,
which implies that $\left|\cF_{a}'\right|\ge2^{-(\gamma+2\log e\cdot\varepsilon)\cdot m}\cdot2^{m}$.

It now follows by an averaging argument that there exists a set~$I_{1}\subseteq\left[m\right]$
such that $I_{1}\in\cF_{a}'$ for at least $2^{-(\gamma+2\log e\cdot\varepsilon)\cdot m}$~fraction
of the strings $a\in\cA_{0}$. Let $\cA_{1}$ be the set of those
strings~$a\in\cA_{0}$, so $\left|\cA_{1}\right|\ge2^{-(\gamma+2\log e\cdot\varepsilon)\cdot m}\cdot\left|\cA_{0}\right|$.
Since the rectangle~$\cA_{0}\times\cB$ is $\frac{1}{4m}$-fortified
on Alice's side, it holds that
\[
\L(\cA_{1}\times\cB)\ge\frac{1}{4m}\cdot2^{-(\gamma+2\log e\cdot\varepsilon)\cdot m}\cdot\L(\cA_{0}\times\cB)\ge2^{-(\gamma+2\log e\cdot\varepsilon)\cdot m-\log m-4}\cdot\L(\cA\times\cB).
\]

We turn to apply the same argument to the inputs on Bob's side, but
this time we restrict ourselves to the coordinates in the set~$I_{1}$.
We first apply \ref{foritification} (the fortification theorem) to
the rectangle $\cA_{1}\times\cB$ on Bob's side. This yields a subset
$\cB_{0}\subseteq\cB$ such that the rectangle~$\cA_{1}\times\cB_{0}$
is $\frac{1}{4m}$-fortified on Bob's side and satisfies 
\[
\L(\cA_{1}\times\cB_{0})\ge\frac{1}{4}\cdot\L(\cA_{1}\times\cB)\ge2^{-(\gamma+2\log e\cdot\varepsilon)\cdot m-\log m-6}\cdot\L(\cA\times\cB).
\]
Second, observe that for every $b\in\cB_{0}$, it holds that $\left|\cY(b)\right|\ge2\cdot2^{-\gamma\cdot m}\cdot\left|g^{-1}(b)\right|$,
and hence $\left|\cY(b)|_{I_{1}}\right|\ge2\cdot2^{-\gamma\cdot m}\cdot\left|g^{-1}(b)|_{I_{1}}\right|$.
Following similar steps as before, we denote by~$\cF_{b}$, for every
$b\in\cB$, the family of subsets~$I\subseteq I_{1}$ such that $\cY(b)|_{I}$
is prefix-thick with degree~$\left(\frac{1}{2}+\varepsilon\right)\cdot2^{n-1}$.
Again, \ref{prefix-thick-lemma} implies that
\[
\left|\cF_{b}\right|\ge2\cdot2^{-\gamma\cdot m-2\log e\cdot\varepsilon\cdot\left|I_{1}\right|}\cdot2^{\left|I_{1}\right|}\ge2\cdot2^{-(\gamma+2\log e\cdot\varepsilon)\cdot m}\cdot2^{\left|I_{1}\right|}.
\]
Next, we let $\cF_{b}'$ be the family of subsets~$I\in\cF_{b}$
of size at least 
\[
\left(\frac{1}{2}-\beta\right)\cdot\left|I_{1}\right|\ge\left(\frac{1}{2}-\beta\right)^{2}\cdot m,
\]
and observe, as before, that
\[
\left|\cF_{b}'\right|\ge\left|\cF_{b}\right|-2^{-2\log e\cdot\beta^{2}\cdot\left|I_{1}\right|}\cdot2^{\left|I_{1}\right|}\ge\left(2\cdot2^{-(\gamma+2\log e\cdot\varepsilon)\cdot\left|I_{1}\right|}-2^{-2\log e\cdot\beta^{2}\cdot\left|I_{1}\right|}\right)\cdot2^{\left|I_{1}\right|}.
\]
Again, since we will choose the constants~$\beta,\gamma,\varepsilon$
such that they satisfy that $\gamma+2\log e\cdot\varepsilon\le2\log e\cdot\beta^{2}$,
it follows that 
\[
\left|\cF_{b}'\right|\ge2^{-(\gamma+2\log e\cdot\varepsilon)\cdot\left|I_{1}\right|}\cdot2^{\left|I_{1}\right|}\ge2^{-(\gamma+2\log e\cdot\varepsilon)\cdot m}\cdot2^{\left|I_{1}\right|}.
\]
It now follows by an averaging argument that there exists a set~$I_{g}\subseteq I_{1}$
such that $I_{g}\in\cF_{b}'$ for at least $2^{-(\gamma+2\log e\cdot\varepsilon)\cdot m}$~fraction
of the strings $b\in\cB_{0}$. Let $\cB_{1}$ be the set of those
strings~$b\in\cB_{0}$ , so $\left|\cB_{1}\right|\ge2^{-(\gamma+2\log e\cdot\varepsilon)\cdot m}\cdot\left|\cB_{0}\right|$.
Since the rectangle~$\cA_{1}\times\cB_{0}$ is $\frac{1}{4m}$-fortified
on Alice's side, it holds that
\[
\L(\cA_{1}\times\cB_{1})\ge\frac{1}{4m}\cdot2^{-(\gamma+2\log e\cdot\varepsilon)\cdot m}\cdot\L(\cA_{1}\times\cB_{0})\ge2^{-(2\gamma+4\log e\cdot\varepsilon)\cdot m-2\log m-8}\cdot\L(\cA\times\cB)
\]
Observe that $\cY(b)|_{I_{g}}$ is prefix thick with degree~$\left(\frac{1}{2}+\varepsilon\right)\cdot2^{n-1}$
for every $b\in\cB_{1}$ by definition. Moreover, for every $a\in\cA_{1}$,
the set~$\cX(a)|_{I_{g}}$ is prefix thick with degree~$\left(\frac{1}{2}+\varepsilon\right)\cdot2^{n-1}$,
because $I_{g}\subseteq I_{1}$ and $\cX(a)|_{I_{1}}$ is prefix thick
with degree~$\left(\frac{1}{2}+\varepsilon\right)\cdot2^{n-1}$

Finally, we show that there exist strings~$a_{g}\in\cA_{1}$ and~$b_{g}\in\cB_{1}$
such that $a_{g}|_{\left[m\right]-I_{g}}=b_{g}|_{\left[m\right]-I_{g}}$.
To this end, we will choose the constants~$\beta,\gamma,\varepsilon,\kappa$
later such that 
\[
\L(\cA_{1}\times\cB_{1})>2^{\left|\left[m\right]-I_{g}\right|+\log m}.
\]
We claim that this implies that there exist strings~$a_{g}\in\cA_{1}$
and~$b_{g}\in\cB_{1}$ such that $a_{g}|_{\left[m\right]-I_{g}}=b_{g}|_{\left[m\right]-I_{g}}$.
Indeed, suppose otherwise, namely, assume that for every $a\in\cA_{1}$
and $b\in\cB_{1}$ it holds that $a|_{\left[m\right]-I_{g}}\ne b|_{\left[m\right]-I_{g}}$.
Consider now the following protocol for $\KW_{\cA_{1}\times\cB_{1}}$:
on inputs $a\in\cA_{1}$ and~$b\in\cB_{1}$, Alice sends $a|_{\left[m\right]-I_{g}}$,
and Bob replies with the coordinate $i\in\left[m\right]-I_{g}$ such
that $a_{i}\ne b_{i}$ (which exists by the assumption). This protocol
transmits at most $\left|\left[m\right]-I_{g}\right|+\log m$~bits,
and hence its size is at most $2^{\left|\left[m\right]-I_{g}\right|+\log m}$.
This contradicts, however, our lower bound on $\L(\cA_{1}\times\cB_{1})$
above. It follows that there exist strings~$a_{g}\in\cA_{1}$ and~$b_{g}\in\cB_{1}$
such that $a_{g}|_{\left[m\right]-I_{g}}=b_{g}|_{\left[m\right]-I_{g}}$,
and this concludes the proof.

\paragraph*{Choosing the universal constants.}

We now show how to choose the universal constants $\beta,\gamma,\varepsilon,\kappa$.
We choose~$\kappa=8$. The constants $\beta$, $\gamma$, and~$\varepsilon$,
are required to satisfy the following constraints:
\begin{align*}
\gamma+2\log e\cdot\varepsilon & \le2\log e\cdot\beta^{2}\\
\L(\cA_{1}\times\cB_{1}) & >2^{\left|\left[m\right]-I_{g}\right|+\log m}.
\end{align*}
Observe that by our choice of~$\kappa$ it holds that
\[
\L(\cA_{1}\times\cB_{1})\ge2^{-(2\gamma+4\log e\cdot\varepsilon)\cdot m-2\log m-8}\cdot\L(\cA\times\cB)>2^{(1-3\cdot\gamma-4\log e\cdot\varepsilon)\cdot m+\log m},
\]
and that $\left|I_{g}\right|\ge\left(\frac{1}{2}-\beta\right)^{2}\cdot m$.
Therefore, a sufficient condition for the second constraint above
to be satisfied is
\[
1-3\cdot\gamma-4\log e\cdot\varepsilon\ge1-\left(\frac{1}{2}-\beta\right)^{2}.
\]
By rearranging the equations, it follows that it suffices to choose
$\beta$, $\gamma$, and~$\varepsilon$ such that
\begin{equation}
\gamma\le\min\left\{ 2\log e\cdot\left(\beta^{2}-\varepsilon\right),\frac{1}{3}\left(\frac{1}{2}-\beta\right)^{2}-4\log e\cdot\varepsilon\right\} .\label{eq:choice-of-gamma}
\end{equation}
In fact, it suffices to choose $\beta$ and~$\gamma$ such that
\[
\gamma<\min\left\{ 2\log e\cdot\beta^{2},\frac{1}{3}\left(\frac{1}{2}-\beta\right)^{2}\right\} ,
\]
and then we can choose $\varepsilon$ to be sufficiently small such
that \ref{eq:choice-of-gamma} holds. It can now be checked that $\beta=0.12$
and $\gamma=0.041$ satisfy the last requirement.

\subsection{\label{subsec:Independence-number-of-G'}The independence number
of~$\protect\cG'$}

In this section, we prove \ref{independence-number-of-G'}. Let $\gamma$,
$\varepsilon$, $\cV'$, $I$, $a$, and~$b$, be as in \ref{construction-of-G'},
and let $\cG'$ be the sub-graph of~$\Gp$ induced by $\cV'$. We
prove that the independence number of~$\cV'$ is at most 
\[
2^{-(\frac{\log e}{32}\cdot\varepsilon^{2}\cdot2^{n}-\frac{1}{2}\cdot m\cdot n-4\cdot m-1)}\cdot\left|\cV'\right|.
\]
To this end, let $\cS\subseteq\cV'$ be a subset of functions that
is larger than the above bound. We prove that $\cS$ is not an independent
set of~$\cG'$. 

We start by simplifying the notation. Without loss of generality,
assume that $I=\left\{ 1,\ldots,\left|I\right|\right\} $. For every~$g\in\cV'$,
let us denote $\XI_{g}=\Xp(g,a)|_{I}$ and $\YI_{g}=\Yp(g,b)|_{I}$.
Observe that, for every two functions $g_{A},g_{B}\in\cS$, if the
sets $\XI_{g_{A}}$ and~$\YI_{g_{B}}$ intersect then $g_{A}$ and~$g_{B}$
are neighbors in~$\cG'$: indeed, if the two sets intersect then
there exist matrices $X\in\Xp(g_{A},a)$ and $Y\in\Yp(g_{B},b)$ such
that $X_{i}=Y_{i}$ for every $i\in I$, and this implies that $g_{A}$
and~$g_{B}$ are neighbors because~$a|_{\left[m\right]-I}=b|_{\left[m\right]-I}$.
Thus, it suffices to prove that there exist functions $g_{A},g_{B}\in\cS$
such that $\XI_{g_{A}}\cap\YI_{g_{B}}\ne\emptyset$.

In order to prove it, we consider two random functions~$g_{A}$ and~$g_{B}$
that are uniformly distributed over the entire set of balanced functions~$\cV_{0}$
(rather than just~$\cS$). For every function~$g\in\cV_{0}\backslash\cS$,
we define $\XI_{g}$ and~$\YI_{g}$ to be some arbitrary subsets
of~$g^{-1}(a)|_{I}$ and~$g^{-1}(b)|_{I}$ that are prefix thick
with degree~$\left(\frac{1}{2}+\varepsilon\right)\cdot2^{n-1}$.
We will prove that
\[
\Pr\left[\XI_{g_{A}}\cap\YI_{g_{B}}=\emptyset\right]<\left(\frac{\left|\cS\right|}{\left|\Vb\right|}\right)^{2}=\Pr\left[g_{A},g_{b}\in\cS\right],
\]
and this will imply that there exist some functions $g_{A},g_{B}\in\cS$
such that $\XI_{g_{A}}\cap\YI_{g_{B}}\ne\emptyset$, as required.

To this end, observe that if all the following three events happen
simultaneously, then $\XI_{g_{A}}$ and $\YI_{g_{B}}$ are \emph{not}~disjoint:
\begin{enumerate}
\item \label{independent-set-random-universe-size}For every $i\in I$,
it holds that $\left|g_{A}^{-1}(a_{i})\cap g_{B}^{-1}(b_{i})\right|\le\left(1+\frac{\varepsilon}{2}\right)\cdot2^{n-2}$.
\item \label{independent-set-thick-Alice}The set $\XI_{g_{A}}\cap g_{B}^{-1}(b)|_{I}$
is prefix thick with degree $\left(\frac{1}{2}+\frac{\varepsilon}{2}\right)\cdot2^{n-2}$.
\item \label{independent-set-thick-Bob}The set $\YI_{g_{B}}\cap g_{A}^{-1}(a)|_{I}$
is prefix thick with degree $\left(\frac{1}{2}+\frac{\varepsilon}{2}\right)\cdot2^{n-2}$.
\end{enumerate}
The reason is that if all the above three events happen, then $\XI_{g_{A}}\cap g_{B}^{-1}(b)|_{I}$
and $\YI_{g_{B}}\cap g_{A}^{-1}(a)|_{I}$ are both prefix-thick subsets
of 
\[
g_{A}^{-1}(a)|_{I}\cap g_{B}^{-1}(b)|_{I}=\prod_{i=1}^{\left|I\right|}\left(g_{A}^{-1}(a_{i})\cap g_{B}^{-1}(b_{i})\right)
\]
(where we view the sets $g_{A}^{-1}(a_{i})\cap g_{B}^{-1}(b_{i})$
as alphabets), and hence they intersect by \ref{prefix-thick-sets-intersect}.
Therefore, in order to upper bound the probability that $\XI_{g_{A}}$
and $\YI_{g_{B}}$ are disjoint, it is sufficient to upper bound the
probability that one of the foregoing events does \emph{not} happen.
We will upper bound the probability of each of the three events separately,
and then apply the union bound.

We start with upper bounding the probability that Event~\ref{independent-set-thick-Alice}
does not happen. Fix a function~$g_{A}$. Since $\XI_{g_{A}}$ is
prefix thick, there exists a subset $\cX'\subseteq\XI_{g_{A}}$ whose
a prefix tree~$T$ has minimal degree is greater than $\left(\frac{1}{2}+\varepsilon\right)\cdot2^{n-1}$.
Let $T'$ be the prefix tree of the subset $\cX'\cap g_{B}^{-1}(b)|_{I}$
of~$\XI_{g_{A}}\cap g_{B}^{-1}(b)|_{I}$. We prove that with high
probability over the choice of~$g_{B}$, the minimal degree of~$T'$
is greater than~$\left(\frac{1}{2}+\frac{\varepsilon}{2}\right)\cdot2^{n-2}$,
and this will imply the required bound. Fix an internal vertex~$v$
of~$T$ at depth~$i$. Observe that if $v$~is also a vertex of~$T'$,
then its children in~$T'$ are exactly its children in~$T$ whose
labels belong to~$g_{B}^{-1}(b_{i+1})$. By assumption, $v$~has
at least 
\[
\left(\frac{1}{2}+\varepsilon\right)\cdot2^{n-1}=\left(\frac{1}{4}+\frac{\varepsilon}{2}\right)\cdot2^{n}
\]
children in~$T$. Therefore, by \ref{sampling}, the probability
over the choice of~$g_{B}$ that less than $\left(\frac{1}{4}+\frac{\varepsilon}{4}\right)\cdot2^{n-1}$
of their labels belong to~$g_{B}^{-1}(b_{i+1})$ is less than 
\[
2^{-2\log e\cdot\left(\frac{\varepsilon}{4}\right)^{2}\cdot2^{n-1}}=2^{-\frac{\log e}{16}\cdot\varepsilon^{2}\cdot2^{n}}
\]
(as $g_{B}^{-1}(b_{i+1})$ is a uniformly distributed subset of~$\B^{n}$
of size $2^{n-1}$). There are at most $2^{m\cdot n}$ vertices in~$T$.
By taking a union bound over all of them, we get that the probability
that for some internal vertex~$v$ in~$T$, it is a vertex of~$T'$
and less than 
\[
\left(\frac{1}{4}+\frac{\varepsilon}{4}\right)\cdot2^{n-1}=\left(\frac{1}{2}+\frac{\varepsilon}{2}\right)\cdot2^{n-2}
\]
of its children of~$v$ belong to $g_{B}^{-1}(b_{i+1})$ is at most
$2^{-\left(\frac{\log e}{16}\cdot\varepsilon^{2}\cdot2^{n}-m\cdot n\right)}$.
Now, observe that whenever the latter event does not happen, it holds
that the minimum degree of~$T'$ is greater than~$\left(\frac{1}{2}+\frac{\varepsilon}{2}\right)\cdot2^{n-2}$.
Hence, the probability that Event~\ref{independent-set-thick-Alice}
above does not happen is at most $2^{-\left(\frac{\log e}{16}\cdot\varepsilon^{2}\cdot2^{n}-m\cdot n\right)}$.
Similarly, the probability that Event~\ref{independent-set-thick-Bob}
does not happen is at most $2^{-\left(\frac{\log e}{16}\cdot\varepsilon^{2}\cdot2^{n}-m\cdot n\right)}$.

Finally, we upper bound the probability that Event~\ref{independent-set-random-universe-size}
does not happen. Fix any choice of~$g_{A}$ and an index~$i$, and
observe that $\left|g_{A}^{-1}(a_{i})\right|=2^{n-1}=\frac{1}{2}\cdot2^{n}$.
Therefore, by \ref{sampling}, the probability over the random choice
of~$g_{B}$ that 
\[
\left|g_{A}^{-1}(1)\cap g_{B}^{-1}(0)\right|>\left(1+\frac{\varepsilon}{2}\right)\cdot2^{n-2}=\left(\frac{1}{2}+\frac{\varepsilon}{4}\right)\cdot2^{n-1}
\]
is at most
\[
2^{-2\log e\cdot\left(\frac{\varepsilon}{4}\right)^{2}\cdot2^{n-1}}=2^{-\frac{\log e}{16}\cdot\varepsilon^{2}\cdot2^{n}},
\]
and thus the probability that this happens for any $i\in\left[m\right]$
is at most $m\cdot2^{-\frac{\log e}{16}\cdot\varepsilon^{2}\cdot2^{n}}$.
By the union bound, the probability that one of the three events does
not happen is at most
\begin{align*}
 & m\cdot2^{-\frac{\log e}{16}\cdot\varepsilon^{2}\cdot2^{n}}+2^{-\left(\frac{\log e}{16}\cdot\varepsilon^{2}\cdot2^{n}-m\cdot n\right)}+2^{-\left(\frac{\log e}{16}\cdot\varepsilon^{2}\cdot2^{n}-m\cdot n\right)}\\
 & \le2^{-\left(\frac{\log e}{16}\cdot\varepsilon^{2}\cdot2^{n}-m\cdot n-2\right)}\\
 & =\left(2^{-(\frac{\log e}{32}\cdot\varepsilon^{2}\cdot2^{n}-\frac{1}{2}\cdot m\cdot n-4\cdot m-1)}\cdot2^{-4\cdot m}\right)^{2}\\
 & <\left(\frac{\left|\cS\right|}{\left|\cV'\right|}\cdot\frac{\left|\cV'\right|}{\left|\Vb\right|}\right)^{2} & \text{(assumptions on \ensuremath{\cS} and \ensuremath{\cV'})}\\
 & =\left(\frac{\left|\cS\right|}{\left|\Vb\right|}\right)^{2},
\end{align*}
as required.
\begin{remark}
There is a small subtlety in the application of~\ref{prefix-thick-sets-intersect}
above. Specifically, we applied the proposition to strings in which
different coordinates belong to different alphabets. The proposition,
on the other hand, was only stated for the case of a single alphabet.
As noted in \ref{prefix-thick-different-alphabets}, this is not a
problem if all the alphabets are of the same size, but here the alphabets
may be of different sizes. Nevertheless, since all the alphabets are
of size at most~$\left(1+\frac{\varepsilon}{2}\right)\cdot2^{n-2}$,
we can pretend that they all have exactly that size by adding dummy
symbols, and the argument would proceed without a change.
\end{remark}

\section{\label{sec:barrier}A barrier to improving~$\gamma$}

It would have been nice if we could improve the value of the constant~$\gamma$
in the structure theorem to a larger value. In particular, if we could
prove our structure theorem with~$\gamma=1$, it would have implied
an almost-optimal composition theorem, analogous to the weak KRW conjecture.
Indeed, such structure theorems have been proved for other variants
of the KRW conjecture \cite{EIRS91,DM16,KM18,RMNPR20}. In this section
we show that improving the value of~$\gamma$ even to~$0.64$ would
require significant new ideas. Specifically, recall that proof of
structure theorem works by showing a lower bound on the chromatic
number~$\chi(\Gp)$. We show that such a lower bound cannot be established
for $0.64$-live transcripts:
\begin{proposition}
\label{barrier}For every $n\in\N$, for every sufficiently large
$m\in\N$ such that $m\ge2n$, and for every $f:\B^{m}\to\B$ such
that $\L(f)\ge\frac{2^{m}}{m}$, the following holds: there exists
a (standard) protocol~$\Pi$ that solves $\KcMn$ and a $0.64$-live
transcript~$\po$ of~$\Pi$ such that~$\chi(\Gp)=1$.
\end{proposition}

\begin{remark}
We note that the constraint that $m\ge2n$ in \ref{barrier} is not
very significant, since the interesting regime for deriving formula
lower bounds typically assumes that $m\approx2^{n}$. Furthermore,
we note that the fact that the protocol~$\Pi$ is a standard protocol
rather than a partially half-duplex one strengthens our result, since
standard protocols can be viewed as partially half-duplex protocols.
\end{remark}

The rest of this section is dedicated to proving \ref{barrier}. Let
$m$, $n$, and~$f$ be as in the proposition. Informally, we construct
the transcript~$\po$ such that, for all functions~$g_{A},g_{B}:\B^{n}\to\B$,
the matrices~$X$ and~$Y$ always disagree on almost all the rows,
and the column vectors~$a$ and~$b$ always disagree on many rows.
Such a choice guarantees that there is always a row~$i$ such that
$X_{i}\ne Y_{i}$ and~$a_{i}\ne b_{i}$, so there are no edges in~$\Gp$.
We construct such a transcript~$\po$ as follows:
\begin{itemize}
\item In order to guarantee that the matrices~$X$ and~$Y$ always disagree
on almost all the rows, we force $X$ to always have $0.08\cdot m$~ones
in its first column, and force~$Y$ to always have $0.92\cdot m$~ones
in its first column. A straightforward calculation shows that the
fraction of such matrices is at least~$2^{-0.64\cdot m}$.\\
We still need to guarantee that there are sufficiently many such matrices
in~$(f\d g_{A})^{-1}(1)$ and $(f\d g_{B})^{-1}(0)$. To this end,
we choose the associated set~$\cV\subseteq\Vp$ of balanced functions~$g$
such that the number of matrices in $(f\d g)^{-1}(1)$ and $(f\d g)^{-1}(0)$
with a given first column is always the same. It is not hard to show
that the fraction of such functions~$g$ is at least $2^{-m}$.
\item A natural way to guarantee that the column vectors~$a,b$ disagree
on many rows would be to force $a$~and~$b$ to belong to a code
with a large distance. Nevertheless, it is not clear how to argue
in such a case that the formula complexity $\L(\Ap(g)\times\Bp(g))$
is large. Instead, we force both $a$~and~$b$ to belong to the
same \emph{coset }of a linear code~$C$ with a large distance \emph{and
a large dimension}.\\
It is not hard to see that such a construction guarantees that $a$~and~$b$
disagree on many rows. Moreover, it can be shown that there exists
a choice of a coset of~$C$ for which $\L(\Ap(g)\times\Bp(g))$ is
at least $\L(f)$ divided by the number of cosets. This is done by
using the fact that the cosets of~$C$ form a partition of~$\B^{m}$,
and by applying the sub-additivity of formula complexity. We upper
bound the number of cosets using the assumption that~$C$ has large
dimension, thus obtaining the desired lower bound on the formula complexity.
\end{itemize}
Details follow. We start with setting up some parameters and notation.
By applying Varshamov's bound (\ref{varshamov}) with $\delta=0.161$
and~$\varepsilon=0.001$, it follows that there exists a code $C\subseteq\B^{m}$
with distance $0.161\cdot m$ and dimension at least $0.362\cdot m$.
Recall that $\Vb$ denotes the set of all balanced functions from~$\B^{n}$
to~$\B$.

We choose the protocol~$\Pi$ to be any protocol that solves~$\KcMn$,
such that in the first messages Alice and Bob communicate the numbers
of ones in the first columns of~$X$ and~$Y$, and the cosets of~$C$
to which~$a$ and~$b$ belong. Consider a partial transcript~$\po$
of~$\Pi$ in which Alice and Bob say the following:
\begin{enumerate}
\item The first column of~$X$ has exactly $\left\lfloor 0.08\cdot m\right\rfloor $
ones.
\item The first column of~$Y$ has exactly $\left\lceil 0.92\cdot m\right\rceil $
ones.
\item The string~$a=g_{A}(X)$ belongs to the coset~$W$ of~$C$ such
that the set $\cA_{W}=W\cap f^{-1}(1)$ maximizes the formula complexity
$\L\left(\cA_{W}\times f^{-1}(0)\right)$.
\item The string~$b$~belongs to the same coset~$W$ of~$C$ as~$a$,
so $b\in\cB_{W}=W\cap f^{-1}(0)$.
\end{enumerate}
Observe that~$\Vp$ is the set of all functions from~$\B^{n}$ to~$\B$,
and that for every $g\in\Vp$, it holds that $\Ap(g)=\cA_{W}$ and
$\Bp=\cB_{W}$. Moreover, for every $g\in\Vp$ and every $a\in\cA_{W}$,
it holds that the set $\Xp(g,a)$ is the set of all matrices $X\in g^{-1}(a)$
with exactly $\left\lfloor 0.08\cdot m\right\rfloor $ ones in their
first column, and a similar claim holds with $b\in\cB_{W}$, $\Yp(g,b)$,
and $\left\lceil 0.92\cdot m\right\rceil $.

We show that the graph~$\Gp$ does not contain any edge, and therefore
$\chi(\Gp)=1$. Assume for the sake of contradiction that there were
two neighbors~$g_{A},g_{B}\in\Vp$ in~$\Gp$. By definition, this
implies that $g_{A}$ and~$g_{B}$ satisfy the weak intersection
property: namely, there exist matrices $X\in\cX_{\pi_{1}}(g_{A},a)$
and $Y\in\cY_{\pi_{1}}(g_{B},b)$, such that $X_{i}=Y_{i}$ for every
$i\in\left[m\right]$ for which $a_{i}\ne b_{i}$, where $a=g_{A}(X)$
and~$b=g_{B}(Y)$ (or the same statement holds while exchanging $g_{A}$
and~$g_{B}$; without loss of generality, assume that we are in the
former case). We know that the first columns of~$X$ and~$Y$ have
exactly $\left\lfloor 0.08\cdot m\right\rfloor $ and $\left\lceil 0.92\cdot m\right\rceil $
ones respectively, and hence $X$ and~$Y$ disagree on at least $0.84\cdot m$
rows. This implies that $a$ and~$b$ have to agree on at least~$0.84\cdot m$
coordinates, so the vector~$a-b$ (over~$\F_{2}$) contains at least~$0.84\cdot m$
zeroes. Nevertheless, we assumed that vectors~$a,b$ belong to the
same coset~$W$ of~$C$, and therefore $a-b\in C$. It follows that
$a-b$ contains at least $0.161\cdot m$ ones by the definition of~$C$,
which contradicts the conclusion that $a-b$ contains at least~$0.84\cdot m$
zeroes. We reached a contradiction, and therefore there are no edges
in~$\Gp$, as required.

It remains to show that $\po$ is $\gamma$-alive for $\gamma=0.64$.
To this end, we choose the set~$\cV$ associated with~$\po$ to
be the set of all functions $g\in\Vb$ that satisfy the following
condition: after the first input bit of~$g$ is fixed to either~$0$
or~$1$, the function~$g$ remains balanced. We will show that for
every $g\in\cV$, $a\in\Ap(g)$, and $b\in\Bp(g)$, it holds that
\begin{align*}
\left|\cV\right| & \ge2^{-m}\cdot\left|\Vb\right|\\
\left|\Xp(g,a)\right| & \ge2^{-\gamma\cdot m+1}\cdot g^{-1}(a)\\
\left|\cY(g,b)\right| & \ge2^{-\gamma\cdot m+1}\cdot g^{-1}(b)\\
\log\L\left(\Ap(g)\times\Bp(g)\right) & \ge(1-\gamma)\cdot m+\kappa\log m+\kappa,
\end{align*}
where $\kappa$ is any universal constant. We start with the first
equation. In order to choose a function~$g\in\cV$, it is sufficient
to choose the two balanced functions from~$\B^{n-1}$ to~$\B$ that
are obtained by fixing the first input bit of~$g$ to~$0$ and~$1$
respectively. By \ref{binomial-entropy-approximation}, the number
of balanced functions from $\B^{n-1}$ to~$\B$ is at least
\[
\binom{2^{n-1}}{\frac{1}{2}\cdot2^{n-1}}\ge\frac{1}{2^{n-1}+1}\cdot2^{H_{2}(\frac{1}{2})\cdot2^{n-1}}\ge2^{2^{n-1}-n}.
\]
It follows that
\begin{align*}
\left|\cV\right| & =\binom{2^{n-1}}{2^{n-1}/2}^{2}\\
 & \ge\left(2^{2^{n-1}-n}\right)^{2}\\
 & =2^{2^{n}-2n}.\\
 & \ge2^{-2n}\cdot\left|\Vb\right|\\
 & \ge2^{-m}\cdot\left|\Vb\right|,
\end{align*}
where the last inequality follows from the assumption that $m\ge2n$.

Next, we show that for every $g\in\cV$ and $a\in\Ap(g)$ it holds
that
\[
\left|\Xp(g,a)\right|\ge2^{-\gamma\cdot m+1}\cdot\left|g^{-1}(a)\right|,
\]
and the analogous inequality for $b\in\Bp(g)$ can be proved similarly.
Let $g\in\cV$ and $a\in\Ap(g)$, and recall that $\Xp(g,a)$ is the
set of matrices $X\in g^{-1}(a)$ such that the first column of $X$
has exactly $\left\lfloor 0.08\cdot m\right\rfloor $~ones. By the
assumption that $g\in\cV_{1}$, it follows that for every column vector~$v\in\B^{m}$,
there are exactly $2^{m\cdot(n-2)}$ matrices $X\in g^{-1}(a)$ whose
first column is~$v$. Hence, it is sufficient to estimate the number
of possible first columns of~$X$, namely, the number of binary strings
of length~$m$ with $\left\lfloor 0.08\cdot m\right\rfloor $~ones.
The number of such binary strings is
\begin{align*}
\binom{m}{\left\lfloor 0.08\cdot m\right\rfloor } & \ge\binom{m}{0.079\cdot m} & \text{(for a sufficiently large\,\ensuremath{m})}\\
 & \ge\frac{1}{m+1}\cdot2^{H_{2}(0.079)\cdot m} & \text{(\ref{binomial-entropy-approximation})}\\
 & \ge2^{H_{2}(0.078)\cdot m} & \text{(for a sufficiently large\,\ensuremath{m})}\\
 & =2^{-\left(1-H_{2}(0.078)\right)\cdot m}\cdot2^{m}\\
 & \ge2^{-0.61\cdot m}\cdot2^{m}\\
 & \ge2^{-\gamma\cdot m+1}\cdot2^{m} & \text{(for a sufficiently large\,\ensuremath{m})}.
\end{align*}
It follows that
\[
\left|\Xp(g,a)\right|\ge2^{-\gamma\cdot m+1}\cdot2^{m}\cdot2^{m\cdot(n-2)}=2^{-\gamma\cdot m+1}\cdot2^{m\cdot(n-1)}=2^{-\gamma\cdot m+1}\cdot\left|g^{-1}(a)\right|,
\]
as required.

Finally, we prove that for every $g\in\cV_{1}$ it holds that
\[
\log\L\left(\Ap(g)\times\Bp(g)\right)\ge(1-\gamma)\cdot m+\kappa\log m+\kappa.
\]
Let $g\in\cV_{1}$, and recall that $\Ap(g)=\cA_{W}$ and $\Bp(g)=\cB_{W}$.
We start by lower bounding the complexity~$\L(\cA_{W}\times f^{-1}(0))$.
Let $\cW$ be the set of cosets of~$C$. Since $C$ has dimension
at least $0.362\cdot m$, it follows that $\left|\cW\right|\le2^{0.638\cdot m}$.
For every $W'\in\cW$, let $\cA_{W'}=f^{-1}(1)\cap W'$, and recall
that $W$~was chosen such that $\L(\cA_{W}\times f^{-1}(0))$ is
maximal. Hence, by the sub-additivity of formula complexity, it holds
that
\[
\L(f)=\L\left(f^{-1}(1)\times f^{-1}(0)\right)\le\sum_{W'\in\cW}\L\left(\cA_{W'}\times f^{-1}(0)\right)\le\left|\cW\right|\cdot\L\left(\cA_{W}\times f^{-1}(0)\right).
\]
It follows that
\[
\L\left(\cA_{W}\times f^{-1}(0)\right)\ge\L(f)/\left|\cW\right|,
\]
and therefore
\begin{align*}
\log\L\left(\cA_{W}\times f^{-1}(0)\right) & \ge\log\L(f)-\log\left|\cW\right|\\
 & \ge\log\left(\frac{2^{m}}{m}\right)-\log\left|\cW\right| & \text{(By choice of\,\ensuremath{f})}\\
 & =m-\log(m)-\log\left|\cW\right|\\
 & \ge m-\log(m)-0.638\cdot m & \text{(\ensuremath{\left|\cW\right|\le2^{0.638\cdot m}})}\\
 & =(1-0.638)\cdot m-\log(m).
\end{align*}
Next, let $\cB'=f^{-1}(0)-\cB_{W}$. By the sub-additivity of formula
complexity, it holds that
\[
\L(\cA_{W}\times\cB_{W})\ge\L\left(\cA_{W}\times f^{-1}(0)\right)-\L(\cA_{W}\times\cB').
\]
Hence, we can lower bound the formula complexity of the rectangle
$\cA_{W}\times\cB_{W}$ by upper bounding the complexity of the rectangle~$\cA_{W}\times\cB'$.
In order to bound the latter complexity, recall that since $W$~is
a coset of the linear subspace~$C$, it is the solution space of
a (possibly non-homogeneous) system of linear equations. In particular,
each string~$b\in\cB'$ violates at least one of the equations. Let
$m'=m-\log\left|C\right|$ be the number of the equations in the system,
and for each $i\in\left[m'\right]$, let $\cB_{i}'$ be the set of
strings $b\in f^{-1}(0)$ that violate the $i$-th equation. It holds
that $\cB'=\bigcup_{i=1}^{m'}\cB_{i}'$, and therefore
\[
\L(\cA_{W}\times\cB')\le\sum_{i=1}^{m'}\L(\cA_{W}\times\cB_{i}').
\]
Now, observe that to determine whether a string~$w\in\B^{m}$ belongs
to $\cA_{W}$ or to~$\cB_{i}'$, it suffices to check whether $w$
satisfies the $i$-th equation or not. This, in turn, amounts to computing
the parity function over at most $m$~bits, and therefore has formula
complexity at most $4m^{2}$ by \ref{complexity-parity}. It follows
that $\L(\cA_{W}\times\cB_{i}')\le4m^{2}$ for every $i\in\left[m'\right]$,
and hence
\[
\L(\cA_{W}\times\cB')\le m'\cdot4m^{2}\le4m^{3}.
\]
This implies that
\begin{align*}
\L(\cA_{W}\times\cB_{W}) & \ge\L\left(\cA_{W}\times f^{-1}(0)\right)-\L(\cA_{W}\times\cB')\\
 & \ge\L\left(\cA_{W}\times f^{-1}(0)\right)-4m^{3}\\
 & \ge2^{(1-0.638)\cdot m-\log(m)}-4m^{3}\\
 & \ge2^{(1-0.638)\cdot m-\log(m)-1} & \text{(for a sufficiently large \ensuremath{m}).}
\end{align*}
We conclude that for every universal constant~$\kappa$ and for every
sufficiently large value of~$m$ it holds that
\[
\log\L(\cA_{W}\times\cB_{W})\ge(1-\gamma)\cdot m+\kappa\cdot\log m+\kappa,
\]
where $\gamma=0.64$, as required.
\begin{acknowledgement*}
The author is grateful to Ronen Shaltiel for many useful discussions
and ideas, and to Sajin Koroth for suggesting the question of whether
the KRW conjecture holds for the notion of strong composition. The
author would like to thank Yahel Manor,  {L\^e Minh Qu\'y}, and anonymous referees for
comments that improved the presentation of this paper, and to Noga
Alon, Ishay Haviv, Lianna Hambardzumyan, Alon Orlitzky, Avi Wigderson,
and an anonymous referee, for references on the graph equality problem
and related problems in the literature. The author would also like
to thank Alexander Smal for an in-depth explanation of his joint works
with Artur Ignatiev and Ivan Mihajlin~\cite{MS21,IMS22}, and to
Ben Barber for his kind explanations on the version of Harper's theorem
for the general Hamming graph. Finally, the author is grateful to
Ville Salo for pointing out his joint work with Ilkka \torma~\cite{ST14}
and its connection to prefix-thick sets.
\end{acknowledgement*}

\newcommand{\etalchar}[1]{$^{#1}$}

\end{document}